\documentclass{elsart}
\usepackage{graphicx}
\usepackage{rotating}

\def\dirac{\rlap{\,/}\mathnormal{D}}

\def\su#1{\mathrm{SU}_L(#1)\otimes\mathrm{SU}_R(#1)}
\begin{document}

\begin{frontmatter}

\begin{flushright}
        \begin{minipage}{3cm}
         INPO-DR/00-33\\
         SHEP 00/16\\
        {\tt hep-ph/0012221}
        \end{minipage}
\end{flushright}

\title{Zweig rule violation in the scalar sector\\
and values of low-energy constants}
\author{S.~Descotes}\thanks{e-mail: \texttt{sdg@hep.phys.soton.ac.uk}}
\address{IPN Groupe de Physique Th\'eorique,\\
Universit\'e Paris-Sud, 91406 Orsay Cedex, France\\
and\\
Department of Physics and Astronomy,\\ 
University of Southampton, Southampton SO17 1BJ, UK
}

\begin{abstract}
We discuss the role of the Zweig rule (ZR) violation in the scalar channel for
the determination of low-energy constants and condensates arising in the effective
chiral Lagrangian of QCD. 
The analysis of the Goldstone boson masses and decay constants shows that
the three-flavor condensate and some low-energy constants are very sensitive
to the value of the ZR violating constant $L_6$. A similar study is performed
in the case of the decay constants. A chiral sum rule 
based on experimental data in the $0^{++}$
channel is used to constrain $L_6$, indicating a significant decrease
between the two- and the three-flavor condensates. The analysis of the
scalar form factors of the pion at zero momentum suggests that the pseudoscalar
decay constant could also be suppressed from $N_f=2$ to 3.
\end{abstract}

\end{frontmatter}

\section{Introduction}

The low-energy constants (LEC's) of the effective chiral Lagrangian of QCD
\cite{g-l} are quantities of great theoretical interest, since they reflect the way 
chiral symmetry is spontaneously broken. However, their determination remains a
particularly awkward problem. In most cases \cite{g-l,daphne,satur,largenc,abt}, 
their values have been inferred 
from observables for the pseudoscalar mesons, with the help of two assumptions: 
(1) the quark condensate is the dominant 
order parameter to describe the Spontaneous
Breakdown of Chiral Symmetry (SB$\chi$S) \cite{g-l}, and (2) the pattern of SB$\chi$S
agrees correctly with a large-$N_c$ description of QCD \cite{theonc}, 
in which quantum fluctuations are treated as small perturbations.

If we admit both assumptions, the SU(2)$\times$SU(2) quark condensate
$\Sigma(2)=-\lim_{m_u,m_d\to 0} \langle \bar{u}u \rangle$ should not depend
much on the mass of the strange quark. We could then set the latter to zero
with no major effect on the quark condensate: 
$\Sigma(2)\sim \lim_{m_s\to 0} \Sigma(2)=\Sigma(3)$. We end up with only
one large condensate for SU(2)$\times$SU(2) and SU(3)$\times$SU(3) chiral limits, which is
not very sensitive to $\bar{q}q$ fluctuations. The LEC's
suppressed by the Zweig rule, $L_4$ and $L_6$, are consistently
supposed to be very small when considered at a typical hadronic scale.

However, several arguments may be raised against this "mean-field approximation"
of SB$\chi$S, in which the Zweig rule applies and the chiral structure of QCD vacuum
is more or less independent of the number of massless quarks.
On the one hand,
the scalar sector $0^{++}$ does not comply with large $N_c$-predictions \cite{scalar},
and some lattice simulations with dynamical fermions suggest a strong 
$N_f$-sensitivity of SB$\chi$S signals \cite{lattice}. On the other hand,
the behaviour of the perturbative QCD $\beta$-function indicates that chiral 
symmetry should be restored for large enough values of $N_f$. In the vicinity
ot the corresponding critical point, chiral order parameters should 
strongly vary with $N_f$. Various approaches, based on the investigation of the
QCD conformal window \cite{conf}, gap equations \cite{gapeq}, or the instanton liquid model
\cite{instanton}, have been proposed to investigate the variations of chiral order 
parameters with the number of massless flavors and to
determine the critical value of $N_f$ for the restoration of chiral symmetry.

In Ref.~\cite{paramag}, the $N_f$-sensitivity of
chiral order parameters has been investigated
without relying on perturbative methods, but rather by
exploiting particular properties of vector-like gauge theories. The mechanism of SB$\chi$S
is indeed related to the dynamics of the lowest eigenvalues of the Dirac operator:
$\dirac=\gamma_\mu(\partial_\mu+iG_\mu)$, considered on an Euclidean torus 
\cite{banks-casher,leut-smil,twoalt}.
Two main chiral order parameters can be expressed in this framework.
The quark condensate $\Sigma$ is related to the average density of eigenvalues around
zero \cite{banks-casher} and the pion decay constant in the chiral limit $F$
can be interpreted as a conductivity \cite{twoalt}. The paramagnetic behaviour of
Dirac eigenvalues leads to a suppression of both order parameters when
the number of flavors increases:
\begin{equation}
F^2(N_f+1)<F^2(N_f), \qquad \Sigma(N_f+1)<\Sigma(N_f)
\end{equation}

This sensitivity of chiral order parameters to light-quark loops
is suppressed in the large-$N_c$ limit and is considered as weak for QCD according to the
second hypothesis of the Standard framework. 
However, the $N_f$-dependence of chiral order parameters
can be measured by correlators that violate the Zweig rule in the scalar (vacuum)
channel. For instance, the difference $\Sigma(2)-\Sigma(3)$ (and the LEC $L_6$) is related
to the correlator $\langle\bar{u}u\ \bar{s}s\rangle$ \cite{Bachir1,Bachir2} (this correlator can be 
interpreted as fluctuations of the density of Dirac eigenvalues \cite{paramag}).
The large ZR violations observed in the
$0^{++}$ channel could therefore support a swift evolution in the chiral structure of the 
vacuum from $N_f=2$ to $N_f=3$. The quantum fluctuations of $\bar{q}q$ pairs 
would then play an essential role in the low-energy dynamics of QCD.

Hence, it is worth reconsidering the determination of LEC's without supposing 
(1) the dominance of the quark condensate and (2) the suppression of quantum
fluctuations. This determination starts with quark mass expansion of
measured observables such as $F_\pi^2M^2_\pi$ or $F_K^2M_K^2$,
using Chiral Perturbation Theory ($\chi$PT) \cite{g-l}:
\begin{equation}
F_P^2M_P^2=m_q \Sigma_P + m_q^2 [a_P + b_P \log(M_P)]+  F_P^2\delta_P,
 \label{expfp2mp2}
\end{equation}
where $m_q$ denotes formally light quark masses ($u$, $d$, $s$) and
the remainder $F_P^2 \delta_P$ is of order $m_q^3$.
The coefficients $\Sigma_P$, $a_P$,
$b_P$ are combinations of LEC's. The chiral logarithms
$\log(M_P)$ stem from meson loops. The coefficient of each power of $m_q$
does not depend on the renormalization scale of the effective theory ($F_P^2 M_P^2$
 and $m_q$ are independent of this scale).

Series like Eq.~(\ref{expfp2mp2}) are assumed to converge on the basis of a
genuine dimensional estimate \cite{georgi}. The LEC's involved in the coefficients
are related to Green functions of axial and vector currents, and scalar 
and pseudoscalar densities. The dimensional estimate consists in saturating
the correlators by the exchange of resonances with masses of order 
$\Lambda_\mathrm{QCD}$ \cite{satur}. We obtain coefficients of order
$\sim 1/\Lambda^n_\mathrm{QCD}$ for the power $m_q^n$. The quark mass expansion
would therefore lead to (convergent) series in powers of $m_q/\Lambda_\mathrm{QCD}\ll 1$.

Notice that this genuine estimate cannot be applied to the linear term $\Sigma_P$
corresponding to the quark condensate (there is no colored physical state to saturate
$\langle \bar{q}q \rangle$). 
Moreover, the convergence of the whole
series does not imply that the linear term $\Sigma_P$ is dominant with
respect to the quadratic term. In this article, we will precisely address
(1) the possibility of such a competition between the first two orders in the
quark mass expansions, and (2) the implications of large values for the
ZR-suppressed constants $L_4$ and $L_6$, in particular for the determination
of LEC's.

Unfortunately, the masses and decay constants of the Goldstone bosons do not provide enough
information to estimate the actual size of quantum fluctuations in QCD. To reach this goal,
Refs.~\cite{Bachir1,Bachir2} have proposed a sum rule to estimate $L_6$ [or 
$\Sigma(2)-\Sigma(3)$] from experimental data in the
scalar channel.
Starting with Standard assumptions (two- and three-flavor condensates of large and similar
sizes), Ref.~\cite{Bachir1} ended up with a ratio $\Sigma(3)/\Sigma(2)\sim 1/2$ at 
the Standard $O(p^4)$ order, whereas Ref.~\cite{Bachir2} confirmed a large decrease of the quark
condensate when  Standard $O(p^6)$ contributions were taken into account.
Even though these results suggest a significant variation in the pattern of SB$\chi$S 
from $N_f=2$ to $N_f=3$, it is seems necessary to reevaluate this sum rule without
any supposition about the size of the condensates. This analysis will be performed
in the second part of this article.

We will follow mainly the line of Ref.\cite{fluctu}, which can be considered as
an orientation guide to this article. The first part is devoted to
the determination of the LEC's from the pseudoscalar spectrum. Sec.~2 considers the
role played by $L_6$ for the Goldstone boson masses and the quark condensates, 
whereas the decay constants and $L_4$
are treated in Sec.~3. Sec.~4 deals mainly with the implication of
ZR violation in the $0^{++}$ channel for the determination of LEC's. The second
part of this article focuses on the estimate of $L_6$ from data in the scalar sector.
Sec.~5 introduces the sum rule for $L_6$, and 
sketches the Operator Product Expansion of the involved Green function. 
In Sec.~6, we estimate this sum rule, with a special emphasis on the
the scalar form factors of the pion and the kaon. In Sec.~7, we present
the results obtained for the quark condensates and LEC's from the sum rule, and 
we discuss two other quantities related to the pion scalar form factors: the slope of the
strange form factor and the scalar radius of the pion. Sec.~8 sums up the 
main results of the article. App.~A collects the expansions of pseudoscalar masses
and decay constants in powers of quark masses. App.~B deals with the Operator 
Product Expansion of the correlator $\langle\bar{u}u\ \bar{s}s\rangle$. App.~C 
provides logarithmic derivatives of the pseudoscalar masses with respect to the
quark masses.

\section{Constraints from the pseudoscalar meson masses}

\subsection{Role of $L_6$}

Let us first study the pseudoscalar masses $M_{\pi}$,
$M_K$, $M_{\eta}$, starting from their expansion at the Standard $O(p^4)$
order, Eqs.~(10.7) in Ref.~\cite{g-l}. We reexpress them as:
\begin{eqnarray}\label{pion} 
F_{\pi}^2 M_{\pi}^2 &=& 2m \Sigma (3) +
2m(m_s+2m)Z^S + 4 m^2 A + 4 m^2 B_0^2 L + F_{\pi}^2 \delta_{\pi},\\
\label{kaon}
F_K^2 M_K^2 &=& (m_s+m) \Sigma(3) + (m_s+m)(m_s+2m)Z^S\\
&& \qquad +(m_s+m)^2 A + m(m_s+m)B_0^2 L + F_K^2 \delta_K,\nonumber
\end{eqnarray}
where $m=(m_u+m_d)/2$ and
$Z^S$ and $A$ are scale-independent constants, containing respectively the
LEC's $L_6(\mu)$ and $L_8(\mu)$,
\begin{eqnarray}\label{z}
Z^S &=& 32 B_0^2 \left[L_6(\mu) - \frac{1}{512 \pi^2}\left(\log\frac{M_K^2}{\mu^2}
+\frac{2}{9}\log\frac{M_{\eta}^2}{\mu^2}\right)\right],\\
\label{a}
A &=& 16 B_0^2 \left[ L_8(\mu) - \frac{1}{512 \pi^2}\left(
\log\frac{M_K^2}{\mu^2} +
\frac{2}{3}\log\frac{M_{\eta}^2}{\mu^2}\right)\right],
\end{eqnarray}
with $B_0=\Sigma(3)/F_0^2$ and $F_0\equiv F(3)$. The remaining $O(p^4)$
chiral logarithms are contained in $L$:\footnote{In this article, we use the following
values of masses and decay constants:
$M_\pi=135$ MeV, $M_K=495$ MeV, $M_\eta=547$ MeV, $F_\pi=92.4$ MeV, $F_K/F_\pi=1.22$.}
\begin{equation}
L = \frac{1}{32\pi^2}\left[3 \log\frac{M_K^2}{M_{\pi}^2} +
\log\frac{M_{\eta}^2}{M_K^2}\right]=25.3\cdot 10^{-3}.
\end{equation}

There is a similar equation for $F_{\eta}^2 M_{\eta}^2$:
\begin{eqnarray}
F_\eta^2 M_\eta^2&=&\frac{2}{3}(2m_s+m)\Sigma+\frac{2}{3}(2m_s+m)(m_s+2m)Z^S 
   \label{eta}\\
 &&\quad  +\frac{4}{3}(2m^2_s+m^2)A 
     +\frac{8}{3}(m_s-m)^2 Z^P+\frac{1}{3}B_0^2 L+F_\eta^2\delta_\eta,\nonumber
\end{eqnarray}
with the scale-independent constant $Z^P =16B_0^2L_7$. A 
factor $B_0$ is included in the expression of
$A$, $Z_S$ and $Z_P$ in terms of $L_{i=6,7,8}$,
so that they do not diverge in the limit $\Sigma(3)\to 0$. 
The corresponding equations for the
pseudoscalar decay constants $F_P^2$ ($P$= $\pi$, $K$, $\eta$) will be
treated in Sec.~\ref{secdecayconst}. 

We take
$F_P^2 M_P^2$ and $F_P^2$ as independent observables, in order to separate
in a straightforward way the ``mass'' constants $L_6$, $L_7$, $L_8$ 
from $L_4$, $L_5$ that appear only in the expansion of decay constants
$F_P^2$. There is a second argument supporting the choice of
$F_P^2$ and $F_P^2M_P^2$ as independent observables of the pseudoscalar spectrum.
We expand observables in powers of quark masses, supposing a
good convergence of the series. We have sketched in the introduction
how a naive dimensional estimate justifies this
assumption : LEC's are related to QCD correlation functions, which can be saturated by
massive resonances, leading to series in $(m_q/\Lambda_{QCD})$. 
We should therefore expect good convergence properties for "primary" observables
obtained directly from the low-energy behavior of QCD correlation functions, 
like $F_P^2$ and $F_P^2M_P^2$. For such 
quantities, the higher-order remainders should thus remain small. 
On the other hand, we have to be careful when we deal with "secondary" quantities 
combining "primary" observables. The higher-order remainders may then have a larger influence. 
In particular, ratios of "primary" observables (like $M_P^2=F_P^2M_P^2/F_P^2$)
might be dangerous
if higher-order terms turned out to be sizable [leading to untrustworthy 
approximations like $1/(1+x)\simeq 1-x$ with a large $x$].

In Eqs.~(\ref{pion}), (\ref{kaon}) and (\ref{eta}), all terms linear and
quadratic in quark masses are shown. The remaining contributions, of order
$O(m_\mathrm{quark}^3)$ and higher, are collected in the remainders
$\delta_P$. We can consider that the latter are given to us, so that
Eqs.~(\ref{pion}), (\ref{kaon}) and (\ref{eta}) can be seen as algebraic
identities relating the 3-flavor condensate $\Sigma(3)$
the quark mass ratio
$r=m_s/m$, and the LEC's $F_0$, $L_6(\mu)$, $L_7$ and $L_8(\mu)$.
The three-flavor quark condensate is measured in physical units, using the
Gell-Mann--Oakes ratio:
$X(3)=2m\Sigma(3)/(F_{\pi}M_{\pi})^2$ \cite{gor}.

We are going to assume that the remainders $\delta_P$ are small ($\delta_P \ll M_P^2$), 
and investigate then the consequences of Eqs.~(\ref{pion}) and (\ref{kaon}) for
the values of LEC's. Before
starting, we should comment the status of Eqs.~(\ref{pion}),
(\ref{kaon}) and (\ref{eta}) with respect to Chiral Perturbation Theory ($\chi$PT).
Even if we imposed $\delta_P=0$, we would not work in the frame of
one-loop Standard $\chi$PT \cite{g-l}: we do not suppose that
the condensate $\Sigma(3)$ is dominant in these equations, we do not treat
$1-X(3)$ as a small expansion parameter, and accordingly, we do not replace
(for instance) $2mB_0$ by $M_\pi^2$ in higher-order terms. However, we are
not following Generalized $\chi$PT either \cite{gchipt}, since $B_0$ is not
treated as an expansion parameter: even with
$\delta_P=0$, Eqs.~(\ref{pion}), (\ref{kaon}) and (\ref{eta}) exceed the
Generalized tree level, since these equations include chiral logarithms.
 
It is useful to rewrite Eqs.~(\ref{pion}) and (\ref{kaon}) as:
\begin{eqnarray}\label{condens}
 \frac{2m}{F_{\pi}^2 M_{\pi}^2} [\Sigma(3) + (2m+m_s)Z^S] &=& 1-
 \tilde\epsilon(r)- \frac{4m^2 B_0^2}{F_{\pi}^2 M_{\pi}^2} \frac{rL}{r-1}-\delta,\\
\label{eleight}
\frac{4m^2A}{F_{\pi}^2 M_{\pi}^2} &=& \tilde\epsilon(r) +      
\frac{4m^2B_0^2}{F_{\pi}^2 M_{\pi}^2} \frac{L}{r-1} + \delta',
\end{eqnarray}
with 
\begin{equation}
\tilde\epsilon(r) = 2\frac{\tilde{r}_2-r}{r^2-1}, \qquad
\tilde{r}_2= 2\left(\frac{F_KM_K}{F_{\pi}M_{\pi}}\right)^2 -1\sim 39.
\end{equation}
$\delta$ and $\delta'$ are linear combinations of the remainders $\delta_{\pi}$
and $\delta_K$:
\begin{eqnarray}\label{delta}
\delta &=& \frac{r+1}{r-1}\frac{\delta_{\pi}}{M_{\pi}^2} -
\left(\tilde\epsilon+\frac{2}{r-1}\right)\frac{\delta_K}{M_K^2},\\
\delta' &=& \frac{2}{r-1}\frac{\delta_{\pi}}{M_{\pi}^2}-
\left(\tilde\epsilon + \frac{2}{r-1}\right)\frac{\delta_K}{M_K^2}.
\end{eqnarray}
For large $r$, we expect $\delta' \ll \delta \sim \delta_\pi/M_\pi^2$.
Similarly to Ref.~\cite{abt}, we consider as parameters 
$F_0 = \lim_{m,m_s\to 0} F_{\pi}$
[i.e. $L_4(\mu)$], the ZR violating constant $L_6(\mu)$ and the
quark mass ratio $r=m_s/m$. Eq.~(\ref{condens}) ends up with a non-perturbative
formula (no expansion) for the three-flavor Gell-Mann--Oakes--Renner ratio $X(3)$:
\begin{equation}\label{GOR}
X(3) = \frac{2m\Sigma(3)}{F_\pi^2M_\pi^2}=
\frac{2}{1+[1+\kappa(1-\tilde\epsilon-\delta)]^{1/2}}(1-\tilde\epsilon-\delta),
\end{equation}
where $\kappa$ contains $L_6(\mu)$:
\begin{eqnarray}\label{kappa}
\kappa &=& 64 (r+2) \left(\frac{F_{\pi}M_{\pi}}{F_0^2}\right)^2 
\Bigg\{ L_6(\mu)\\
&&\qquad -\frac{1}{256\pi^2}\left(\log\frac{M_K}{\mu} +
\frac{2}{9}\log\frac{M_{\eta}}{\mu}\right) 
+ \frac{rL}{16(r-1)(r+2)}\Bigg\}.\nonumber
\end{eqnarray}

Eq.~(\ref{GOR}) is an exact identity, useful if the remainder 
$\delta$ in Eq.~(\ref{delta}) is small, i.e. if the expansion of QCD correlators
in powers of the quark masses $m_u$, $m_d$, $m_s$ is globally convergent.
It means that $\delta_P\ll M_P^2$ in Eqs.~(\ref{pion}) and (\ref{kaon}), but the
linear term in these equations (related to the condensate) does not need to dominate.

$\kappa$ describes quantum fluctuations of the condensate, and actually
$\kappa = O(1/N_c)$. $L_6(\mu)$ has to be fixed carefully to keep $\kappa$ small. 
$\kappa$ is equal to zero for $10^3\cdot L_6 = - 0.26$ at the scale
$\mu=M_{\rho}$, which is close to the value usually claimed in Standard
$\chi$PT analysis \cite{daphne}. In this case,
Eq.~(\ref{GOR}) yields $X(3)$ near 1, unless the quark mass ratio $r$ decreases
significantly, leading to $\tilde\epsilon \to 1$. This effect is well-known
in G$\chi$PT \cite{gchipt}: the minimal value of $r$ (corresponding to
$\tilde\epsilon=1$ and $X(3)=0$) is
$\tilde r_1=2(F_K M_K)/(F_{\pi}M_{\pi})-1\sim 8$. Notice that for these very
small values of $r$, the combination of higher-order remainders $\delta$
cannot be neglected any more in Eq.~(\ref{GOR}).

\begin{figure}
\begin{center}
\includegraphics[width=11cm]{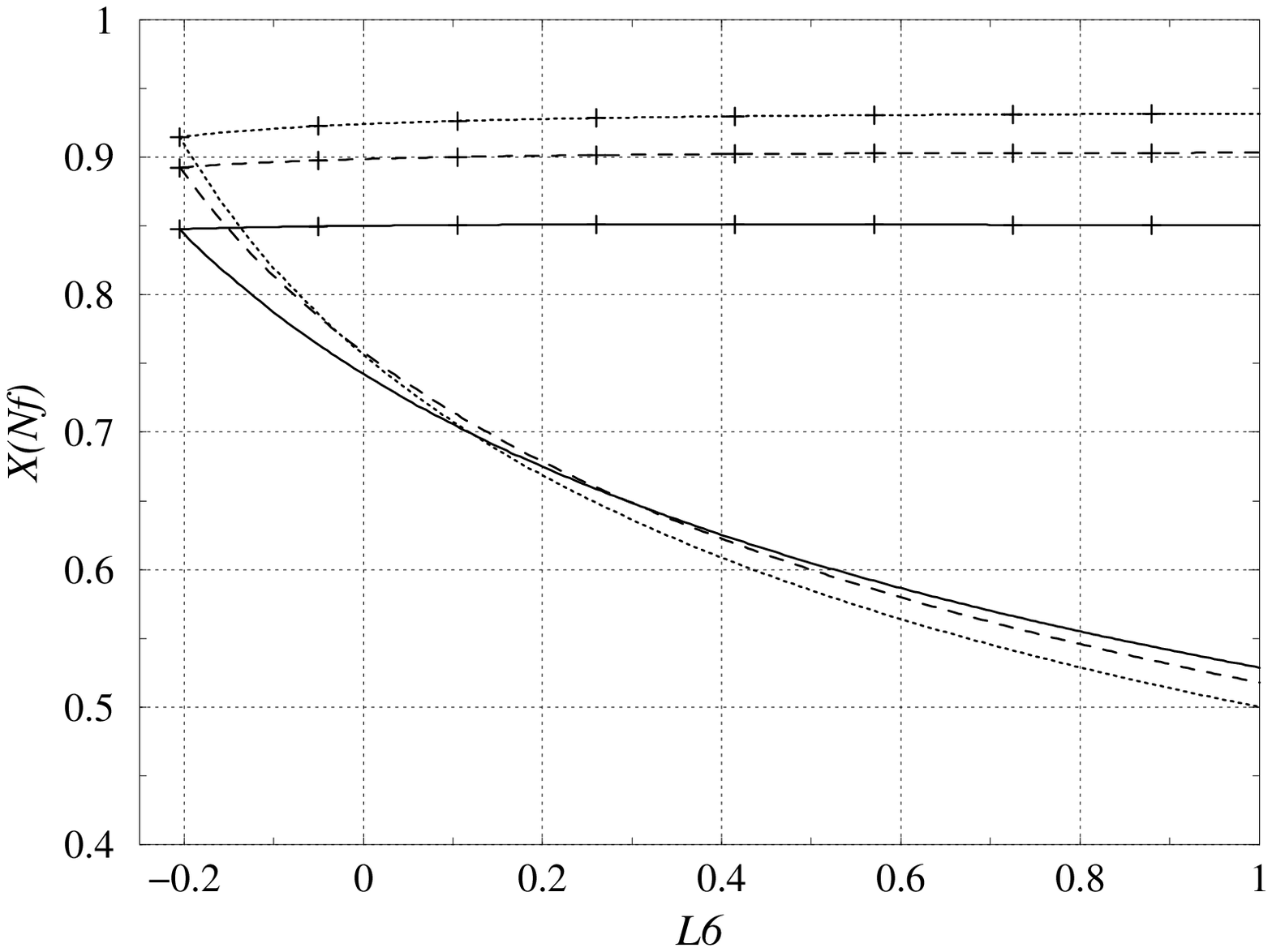}
\includegraphics[width=11cm]{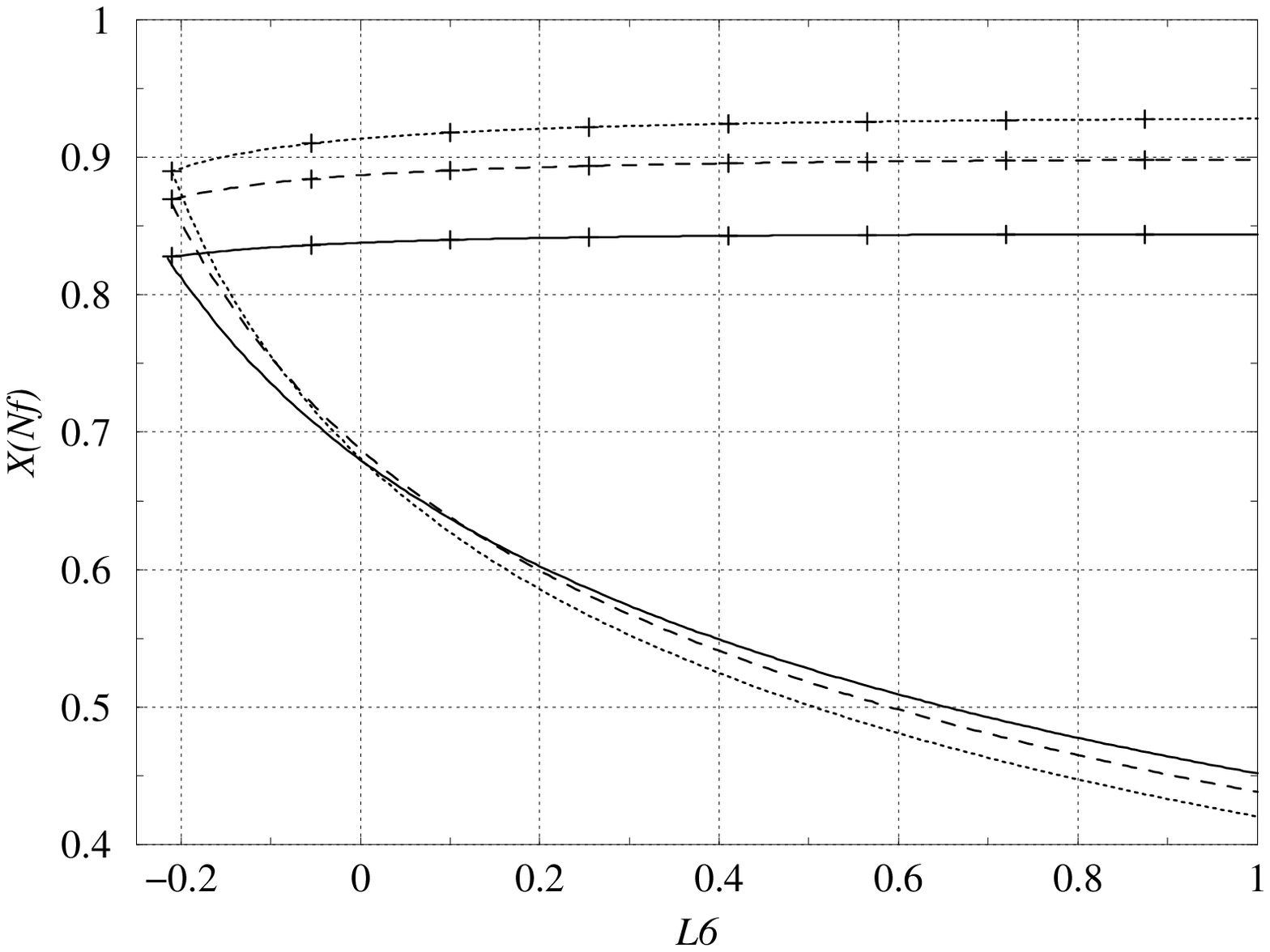}
\caption{$X(2)$ (crosses) and  $X(3)$ (no symbol)
as functions of $L_6(M_\rho)\cdot 10^3$ and $r=m_s/m$ (solid line: $r=20$,
dashed line: $r=25$, dotted line: $r=30$) for $F_0$=85 MeV (upper plot) and 75 MeV
(lower plot).} 
\label{x2x3} 
\end{center} 
\end{figure}

But quantum fluctuations can modify this picture: the number before the
curly brackets in Eq.~(\ref{kappa}) is very large 
($\sim 5340$ for $r=26$ and $F_0$ = 85 MeV).
Hence, even a small positive value of $L_6(M_{\rho})$ can lead to a strong
suppression of $X(3)$, whatever the value of $r=m_s/m$. This effect can be
seen on Fig.~\ref{x2x3},
where $X(3)$ is plotted as a function of $L_6(M_{\rho})$ for $r=20$, $r=25$,
$r=30$ and $F_0$ = 85 and 75 MeV. The decrease of $X(3)$ is slightly
steeper for lower values of $F_0$.

Once $X(3)$ is known, Eq.~(\ref{eleight}) leads to $L_8(\mu)$:
\begin{equation}
L_8(\mu)=\frac{L}{r-1}+\frac{F_0^4}{F_\pi^2M_\pi^2}\frac{\tilde\epsilon+\delta'}{[X(3)]^2}
 +\frac{1}{512\pi^2}
        \left\{\log\frac{M_K^2}{\mu^2}
        +\frac{2}{3}\log\frac{M_\eta^2}{\mu^2}
	\right\}. \label{l8}
\end{equation}
This constant depends on $L_6$ only through $X(3)$. Notice that this
dependence is smaller when $F_0$ decreases ($L_8$ depends actually on $L_6$
through $B_0$).

The ZR violating constant $L_7$ can be obtained from $F_\eta^2M_\eta^2$ Eq.~(\ref{eta}):
\begin{eqnarray}
L_7 &=&
   \frac{1}{32}\frac{F_0^4}{F_\pi^2M_\pi^2}\left\{\frac{1}{(r-1)^2}
\left[3F_\eta^2M_\eta^2+F_\pi^2M_\pi^2-4F_K^2M_K^2\right]
    -F_\pi^2 M_\pi^2\tilde\epsilon(r)\right\}\nonumber\\
&&-\frac{r^2}{32(r-1)^2}\frac{F_0^4}{F_\pi^2M_\pi^2}
  \left[3F_\eta^2 \delta_\eta
    +\frac{8r}{r+1}F_K^2\delta_K
    +(2r-1)F_\pi^2 \delta_\pi\right], \label{l7}
\end{eqnarray}
where $F_\eta$ will be discussed in Sec.~\ref{secdecayconst}. 
The pseudoscalar spectrum satisfies with a
good accuracy the relation $3F_\eta^2M_\eta^2+F_\pi^2M_\pi^2=4F_K^2M_K^2$,
which reduces at the leading order to the Gell-Mann--Okubo formula \cite{GMO}.
This relation leads to a strong correlation between $A$ and $Z_P$: $A+2Z_P\simeq 0$.
This correlation can also be seen in Standard $\chi$PT between $L_7$ and 
$L_8$, and remains to be explained in both frameworks.
No obvious reason forces this particular combination of two
low-energy constants to be much smaller than the typical size of the effective
constants.

\subsection{Paramagnetic inequality for $\Sigma$}

In Ref.~\cite{paramag}, 
$\bar{q}q$ fluctuations were shown to increase the two-flavor condensate 
$\Sigma(2)=-\lim_{m \to 0}\langle\bar uu\rangle|_{m_s\ \mathrm{physical}}$
with respect to $\Sigma(3)$, so that $X(2) > X(3)$.
The two-flavor quark condensate can be obtained through the limit:
\begin{equation}\label{sigma2}
\Sigma(2) = \lim_{m\to 0}\frac{(F_{\pi}M_{\pi})^2}{2m} = \Sigma(3) + 
m_s Z^S|_{m=0} + \delta_2,
\end{equation}
keeping $m_s$ fixed. We have the quantities:
$\delta_2=\lim_{m\to 0}F_\pi^2\delta_\pi/(2m)$ 
and $Z^S|_{m=0}= Z^S + B_0^2 \Delta Z^S$, with:
\begin{equation}
\Delta Z^S=\frac{1}{16\pi^2}
 \left[\log \frac{M_K^2}{\bar M_K^2}+ 
   \frac{2}{9} \cdot \log\frac{M_{\eta}^2}{\bar M_{\eta}^2}\right],
\end{equation}
and $\bar M_P^2 = \lim_{m\to 0} M_P^2$.
The effect of $\Delta Z^S$ is very small\footnote{It can be evaluated
following the procedure of Sec.~\ref{secexpresult}.}. 
$\Delta Z^S$ should be compared to the logarithmic terms included in 
$Z^S$, Eq.~(\ref{z}), at a typical scale
$\mu\sim M_\rho$. $\Delta Z^S$ reaches hardly 10\% of this logarithmic piece.

Once $Z^S$ is eliminated from Eqs.~(\ref{condens}) and (\ref{sigma2}),
we obtain the two-flavor Gell-Mann--Oakes--Renner ratio
$X(2) = 2m\Sigma(2)/(F_{\pi}M_{\pi})^2$:
\begin{eqnarray}\label{xtwo}
X(2)&=& [1-\tilde\epsilon]\frac{r}{r+2} + \frac{2}{r+2}X(3)\\
&&\qquad -\frac{(F_{\pi}M_{\pi})^2}{2 F_0^4}X(3)^2 
\left[\frac{2r^2}{(r-1)(r+2)}L
-r\Delta Z^S\right] + \delta_X,\nonumber
\end{eqnarray}
with:
\begin{eqnarray}
\delta_X&=&\delta_2-\frac{r}{r+2}\delta\\
  &=&\frac{1}{F_\pi^2M_\pi^2}
     \left[m\lim_{m\to 0}\frac{F_\pi^2\delta_\pi}{m}
     -\frac{r(r+1)}{(r+2)(r-1)}F_\pi^2\delta_\pi\right]\\
 &&\qquad+\frac{r}{r+1}\left(\tilde\epsilon+\frac{2}{r-1}\right)
       \frac{\delta_K}{M_K^2}.\nonumber
\end{eqnarray}
In the expression of $\delta_X$, the remainders $\delta_P/M_P^2$ are
suppressed by a factor $m/m_s$: this suppression is obvious for $\delta_K$
[$\tilde\epsilon=O(1/r)$], whereas the operator applied to
$\delta_\pi$ cancels the terms of order $O(mm_s^2)$.
For $r>20$, we expect thus $|\delta_X| \sim |\delta'|$.
The dependence of $X(2)$ on $L_6$ is completely hidden in
$X(3)$, and therefore marginal, as shown in Fig.~\ref{x2x3}.

The paramagnetic inequality $X(2)\geq X(3)$ constrains the maximal value reached by
$X(3)=X(3)|_\mathrm{max}$. If we neglect $\delta\Sigma(2)$ in
Eq.~(\ref{sigma2}), the inequality can be translated into a lower bound for $L_6$:
\begin{equation}
L_6(\mu) \geq \frac{1}{512 \pi^2}\left(\log\frac{\bar{M}_K^2}{\mu^2}
+\frac{2}{9}\log\frac{\bar{M}_{\eta}^2}{\mu^2}\right).
\end{equation}
Fig.~\ref{x2x3} shows clearly the lower bound:
 $L_6(M_{\rho})\geq-0.21\cdot 10^{-3}$.

$X(2)$ is loosely related to $X(3)$, but it is very strongly correlated with
$r$, specially for small values of $r$. Eq.~(\ref{xtwo}) yields the estimate
$X(2)\sim [1-\tilde\epsilon]\cdot r/(r+2)$, up to small correcting terms due
to $X(3)$. We are going to study the effect of these correcting terms.

If we neglect $\Delta Z^S$, $X(2)$ is a quadratic function of
$X(3)$, which is not monotonous when $X(3)$
varies from 0 to $X(3)|_\mathrm{max}$: it first increases, and then decreases
(see Fig.~\ref{x2x3}).
The decrease of $X(2)$ for $X(3)$ close to its upper bound is caused by the
negative term, quadratic in $X(3)$, in Eq.~(\ref{xtwo}). This
decrease of $X(2)$ is more significant for small $F_0$,
because the factor in front of $[X(3)]^2$ in Eq.~(\ref{xtwo}) becomes larger.

\begin{figure}[t]
\begin{center}
\includegraphics[width=11cm]{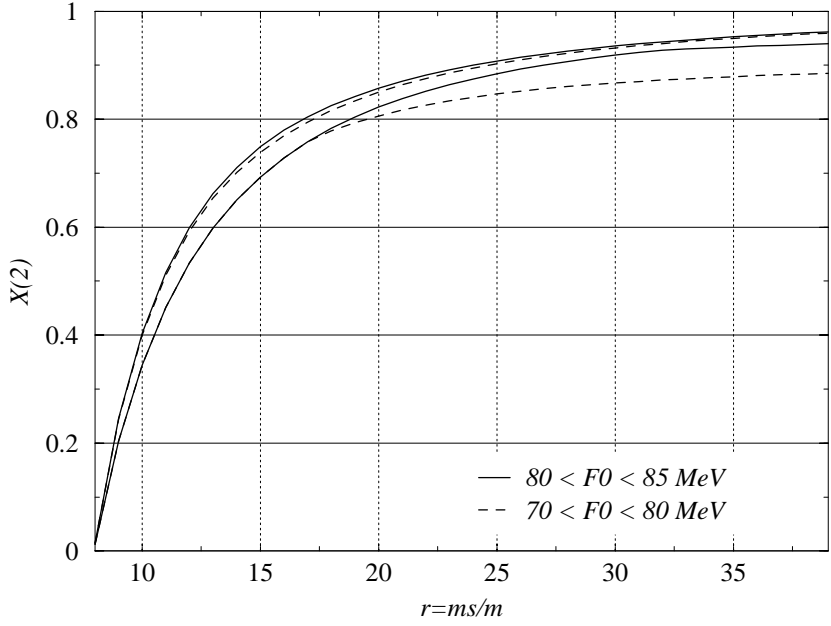}
\caption{$X(2)$ as a function of $r$ for several values of $F_0$}
\label{figx2}
\end{center}
\end{figure}

Therefore, $X(2)$ does not reach its maximum for the paramagnetic bound
$X(2)=X(3)|_\mathrm{max}$, whereas its minimum is the smallest of the two
values obtained for $X(3)=0$ and
$X(3)=X(3)|_\mathrm{max}$\footnote{This updates
Ref.~\cite{paramag}, where the minimum and maximum of $X(2)$ were claimed to
be obtained for $X(3)=0$ and $X(3)=X(3)|_\mathrm{max}$.}. The dependence on $r$
of the minimal value of $X(2)$ can be guessed rather easily. For large $r$, the
term linear in $X(3)$ in Eq.~(\ref{xtwo}) can be neglected: the minimum of
$X(2)$ occurs for  $X(3)=X(3)|_\mathrm{max}$. For small
$r$, $X(2)$ and $X(3)$ tend to 0, and the term quadratic in $X(3)$ should be
small with respect to the linear term. Therefore, $X(2)$ reaches its minimum for
$X(3)=0$ when $r$ is small. 

The numerical analysis of Eq.~(\ref{xtwo}), including $\Delta Z^S$,
supports this intuitive description. In Fig.~\ref{figx2}, the variation ranges of
$X(2)$ are plotted for several values of $F_0$. The curve for the minimum of
$X(2)$ exhibits a cusp when the minimum of $X(2)$
corresponds no more to $X(3)=0$, but to $X(3)|_\mathrm{max}$.
$X(2)$ appears to be strongly correlated to $r$, even though a large value of
$X(2)\sim 0.9$ can be associated to a broad range of $r$.

\section{Constraints from the pseudoscalar decay constants}

\subsection{Role of $L_4$} \label{secdecayconst}

The decomposition used for the Goldstone boson masses can be adapted 
to the decay constants:
\begin{eqnarray}
F_\pi^2&=&F_0^2+2m\xi +2(2m+m_s)\tilde\xi \label{fpi}\\
&&\quad +\frac{1}{16\pi^2}\frac{F_\pi^2M_\pi^2}{F_0^2}X(3)
    \left[2\log\frac{M_K^2}{M_\pi^2}
    +\log\frac{M_\eta^2}{M_K^2}\right]   
     +\varepsilon_\pi,\nonumber\\
F_K^2&=&F_0^2+(m+m_s)\xi +2(2m+m_s)\tilde\xi
     +\frac{1}{2}\frac{F_\pi^2M_\pi^2}{F_0^2}X(3)L
     +\varepsilon_K,\label{fka}
\end{eqnarray}
with the scale-independent constants related to $L_4$ and $L_5$:
\begin{eqnarray}
\xi &=&8B_0\left[
   L_5(\mu)-\frac{1}{256\pi^2}
   \left(\log\frac{M_K^2}{\mu^2}+2\log\frac{M_\eta^2}{\mu^2} \right)
   \right],\label{defl5}\\
\tilde\xi 
  &=&8B_0\left[
   L_4(\mu)-\frac{1}{256\pi^2}\log\frac{M_K^2}{\mu^2}\right], \label{defl4}
\end{eqnarray}
Eqs.~(\ref{fpi}) and (\ref{fka}) contain all the terms constant or linear in
quark masses in the expansion of $F_\pi^2$ and $F_K^2$, whereas
$\varepsilon_P$ denote remainders of order $O(m_\mathrm{quark}^2)$.
There is also a formula for $F_\eta$, which can be written as:
\begin{eqnarray}
F_\eta^2&=&\frac{4}{3}F_K^2-\frac{1}{3}F_\pi^2
      +\frac{1}{24\pi^2}\frac{M_\pi^2
      F_\pi^2}{F_0^2}rX(3)\log\frac{M_\eta^2}{M_K^2}\\
&&\qquad +\frac{1}{48\pi^2}\frac{M_\pi^2
      F_\pi^2}{F_0^2}X(3)
 \left(\log\frac{M_\eta^2}{M_K^2}-\log\frac{M_K^2}{M_\pi^2}\right)
  +\varepsilon_\eta-\frac{4}{3}\varepsilon_K+\frac{1}{3}\varepsilon_\pi.
  \nonumber
\end{eqnarray}

The two scale-independent constants can be
extracted from Eqs.~(\ref{fpi}) and (\ref{fka}):
\begin{eqnarray}
\frac{2m\xi}{F_\pi^2} 
&=&\tilde\eta(r)
+\frac{1}{32\pi^2}\frac{F_\pi^2}{F_0^2}\frac{M_\pi^2}{F_\pi^2}
  \frac{X(3)}{r-1}
\left[5\log\frac{M_K^2}{M_\pi^2}+3\log\frac{M_\eta^2}{M_K^2}\right],
 \label{xi}\\
&&
+\frac{2}{r-1}\left[\frac{\varepsilon_\pi}{F_\pi^2}-\frac{\varepsilon_K}{F_\pi^2}
  \right]\nonumber\\
\frac{2m\tilde\xi}{F_\pi^2} &=&\frac{1}{r+2}
\left\{1-\tilde\eta(r)-\frac{F_0^2}{F_\pi^2}\right.\label{f0l4}\\
&&\qquad\left.-\frac{1}{32\pi^2}\frac{F_\pi^2}{F_0^2}\frac{M_\pi^2}{F_\pi^2}X(3)
  \left[\frac{4r+1}{r-1}\log\frac{M_K^2}{M_\pi^2}+\frac{2r+1}{r-1}
     \log\frac{M_\eta^2}{M_K^2}\right]\right\}\nonumber\\
&&+\frac{1}{r+2}\left[\frac{2}{r-1}
\frac{\varepsilon_K}{F_\pi^2}-\frac{r+1}{r-1}\frac{\varepsilon_\pi}{F_\pi^2}\right],
 \nonumber
\end{eqnarray}
with:
\begin{equation}
\tilde\eta(r)=\frac{2}{r-1}\left(\frac{F_K^2}{F_\pi^2}-1\right)\sim
\frac{0.977}{r-1},
\end{equation}
where the latter estimate is obtained for $F_K/F_\pi=1.22$.

$\xi$ (i.e. $L_5$) turns out to depend essentially on
$X(3)$ and $r$, whereas $\tilde\xi$ (i.e. $L_4$) is related to the
difference between $F(3)$ and $F_\pi$. Eq.~(\ref{f0l4}) leads to a quadratic
equation for $[F(3)/F_\pi]^2$, involving $L_4$:
\begin{equation}
\left(\frac{F(3)}{F_\pi}\right)^4
  -(1-\tilde\eta-\varepsilon)\left(\frac{F(3)}{F_\pi}\right)^2+\lambda X(3)=0,\label{f0fpi}
\end{equation}
with:
\begin{eqnarray}
\lambda &=&8(r+2)\frac{M_\pi^2}{F_\pi^2}
  \left\{L_4(\mu)-\frac{1}{256\pi^2}\log\frac{M_K^2}{\mu^2}\right.\\
&&\left.+\frac{1}{256\pi^2}\frac{1}{(r+2)(r-1)}
\left[(4r+1)\log\frac{M_K^2}{M_\pi^2}+(2r+1)\log\frac{M_\eta^2}{M_K^2}\right]\right\}
\nonumber,\\
\varepsilon&=&\frac{r+1}{r-1}\frac{\varepsilon_\pi}{F_\pi^2}-
 \left(\tilde\eta+\frac{2}{r-1}\right)\frac{\varepsilon_K}{F_K^2}.
\end{eqnarray}

Eq.~(\ref{f0fpi}) has the solution:
\begin{equation}
\left(\frac{F(3)}{F_\pi}\right)^2
 =\frac{1-\tilde\eta-\varepsilon
       +\sqrt{\left(1-\tilde\eta-\varepsilon\right)^2 -4\lambda X(3)}}{2}. \label{solf3}
\end{equation}
Notice that this formula is very close to Eq.~(\ref{GOR}), that relates $X(3)$
to $L_6(\mu)$ through the parameter $\kappa$. For $r=25$, the factor in
front of the curly brackets in the definition of $\lambda$ is of order 
$460$, and $\lambda$ vanishes for
$L_4(M_\rho)=-0.51\cdot 10^{-3}$. 

\begin{figure}
\begin{center}
\includegraphics[width=11cm]{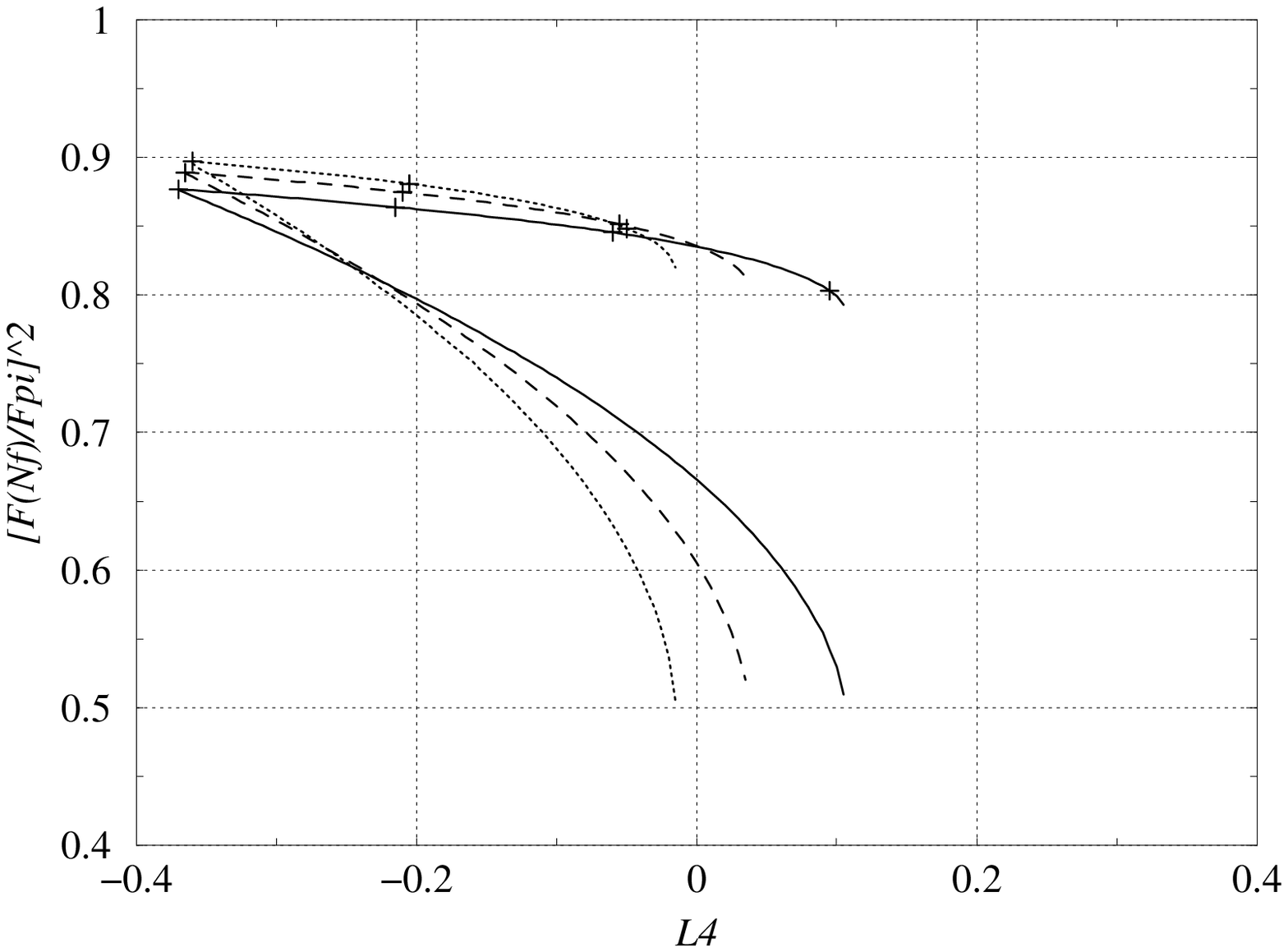}
\includegraphics[width=11cm]{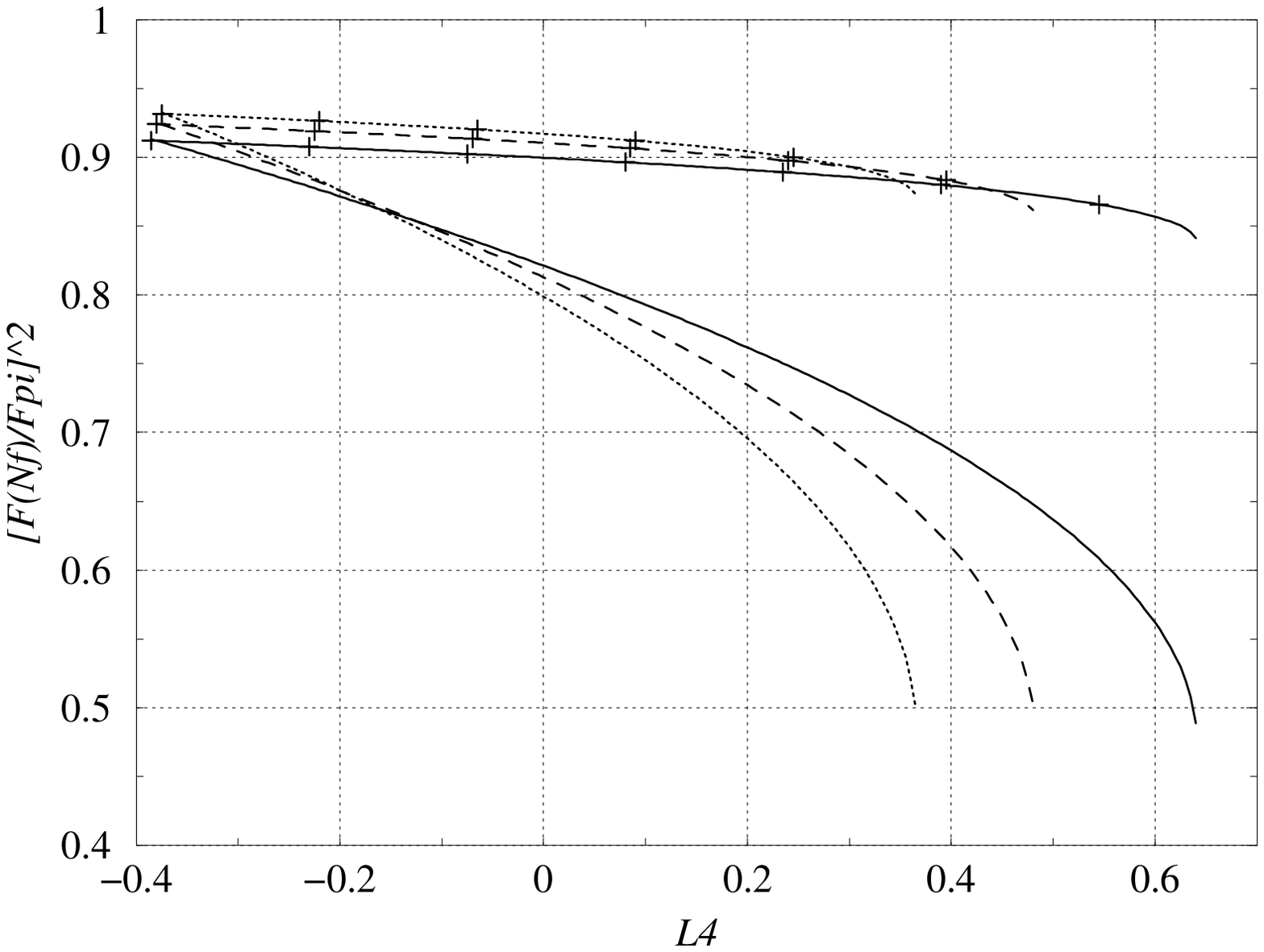}
\caption{$F(2)$ (crosses) and $F(3)$ (no symbol)
as functions of $L_4(M_\rho)\cdot 10^3$ and $r=m_s/m$ (solid lines: $r=20$,
dashed lines: $r=25$, dotted lines: $r=30$) for $X(3)$=0.9 (upper plot) and 0.5
(lower plot).}
\label{f2f3} 
\end{center} 
\end{figure}

Eq.~(\ref{f0fpi}) admits a solution only if
$\lambda X(3)\leq (1-\tilde\eta-\varepsilon)^2/4$.
From Eq.~(\ref{solf3}), we get then a range for $F(3)$:
\begin{equation}\label{ftwoinf}
 \frac{1-\tilde\eta-\varepsilon}{2} \leq 
 \left(\frac{F(3)}{F_\pi}\right)^2\leq 1-\tilde\eta-\varepsilon.
\end{equation}

The variations of $F(3)$ with $L_4(M_\rho)$ are plotted for three values
of $r$ and $X(3)=0.9$ and 0.5 in Fig.~\ref{f2f3}. 
When $X(3)$ decreases, the allowed range for $L_4$ broadens. This is due
to the definition of $L_4$, which relates
$B_0 L_4=\Sigma(3) L_4/F^2(3)$ to the low-energy 
behavior of a QCD Green function. $[F(3)/F_\pi]^2$ starts at
0.9 (for the lowest possible value of $L_4(M_\rho)$) and decreases until 0.5. $F(3)$ 
can thus vary from 87 to 65 MeV, depending on the value of $L_4(M_\rho)$.

\subsection{Paramagnetic inequality for $F^2$}

We obtain $F(2)$ by taking the limit:
\begin{equation}
F^2(2)=\lim_{m\to 0} F_\pi^2
  =F(3)^2+2m_s \tilde\xi|_{m=0}+\varepsilon_2,
\end{equation}
keeping $m_s$ fixed. We have
$\varepsilon_2=\lim_{m\to 0}\varepsilon_\pi$ 
and $\tilde\xi|_{m=0}= \tilde\xi + B_0 \Delta \tilde\xi$, with
\begin{equation}
\Delta \tilde\xi=\frac{1}{32\pi^2}\log \frac{M_K^2}{\bar M_K^2}.
\end{equation}
$\Delta\tilde\xi$ has a tiny effect on the results, similarly to
$\Delta Z^S$ for $X(3)$. We get the equation:
\begin{equation}
\left[\frac{F(2)}{F_\pi}\right]^2
 =\frac{r}{r+2}\left\{[1-\tilde\eta]
     +\frac{2}{r}\left[\frac{F(3)}{F_\pi}\right]^2
     -X(3)\phi\left[\frac{F_\pi}{F(3)}\right]^2-\varepsilon_F
     \right\},     
\label{ftwo}
\end{equation}
with:
\begin{eqnarray}
\varepsilon_F&=&\frac{r+1}{r-1}\frac{\varepsilon_\pi}{F_\pi^2}
   -\frac{r+2}{r}\frac{\varepsilon_2}{F_\pi^2}
  -\left(\tilde\eta+\frac{2}{r-1}\right)\frac{\varepsilon_K}{F_K^2}\\
  &=&\frac{1}{F_\pi^2}
    \left[\frac{r+1}{r-1}\varepsilon_\pi
     -\frac{r+2}{r}\lim_{m\to 0}\varepsilon_\pi\right]
     -\left(\tilde\eta+\frac{2}{r-1}\right)\frac{\varepsilon_K}{F_K^2},\\
\phi&=&\frac{1}{32\pi^2}\frac{M_\pi^2}{F_\pi^2}
\left[ \frac{4r+1}{r-1}\log\frac{M_K^2}{M_\pi^2}
             +\frac{2r+1}{r-1}\log\frac{M_\eta^2}{M_K^2} 
       +(r+2)\log\frac{\bar{M}_K^2}{M_K^2}\right].
\end{eqnarray}
It is interesting to compare the expression of $F(2)$ as a function of
$F(3)$ with Eq.~(\ref{xtwo}) that relates $X(2)$ to $X(3)$. Even though the equations
look rather similar, we can notice that Eq.~(\ref{xtwo}) is a quadratic function of
$X(3)$ whereas Eq.~(\ref{ftwo}) involves
$[F(3)/F_\pi]^2$ and its inverse $[F_\pi/F(3)]^2$. Moreover, Eq.~(\ref{ftwo}) is
an increasing function of $F(3)$, whereas $X(2)$ is not monotonous with
$X(3)$. On the other hand, $\varepsilon_F$ suppresses the remainders
$\varepsilon_P/F_P^2$ by a factor $m/m_s$, in a comparable way to
$\delta_X$.

The paramagnetic inequality $F(2)\geq F(3)$ is translated into an upper
bound for $F(3)$:
\begin{equation}
\left(\frac{F(3)}{F_\pi}\right)^2
 \leq
   \frac{1-\tilde\eta-\varepsilon_F
       +\sqrt{\left(1-\tilde\eta-\varepsilon_F\right)^2 -4\phi X(3)}}{2}.
       \label{ftwosup}
\end{equation}
A quick estimate shows that the condition
$X(3)\leq (1-\tilde\eta-\varepsilon_F)^2/(4\phi)$ is satisfied for any $r$
between $\tilde{r}_1$ and $\tilde{r}_2$. 
This paramagnetic bound corresponds to a lower bound for
$L_4(\mu)$:
\begin{equation}
L_4(\mu) > \frac{1}{256\pi^2}\log\frac{\bar{M}_K^2}{\mu^2}.
\end{equation}
The term $\log(\bar{M}_K^2)$ is very weakly dependent on $r$, $X(3)$ and $F(3)$. 
If we scan the acceptable range of variation for these three parameters, we
obrain the lower bound $L_4(M_\rho)>-0.37\cdot 10^{-3}$. We notice that
the curves for $F(2)$ and $F(3)$ cross each other at this value in 
Fig.~\ref{f2f3}.

Since $\phi>0$, Eqs.~(\ref{ftwoinf}) and (\ref{ftwosup}) lead to the range:
\begin{equation}
\frac{1-\tilde\eta-\varepsilon}{2} \leq\frac{F(3)^2}{F_\pi^2}
  \leq 1-\tilde\eta-\mathrm{max}(\varepsilon_F,\varepsilon).
\end{equation}
The bounds on $F(3)$ are indicated in Tab.~\ref{tabf2} 
(neglecting the remainders $\varepsilon$ and $\varepsilon_F$). 
Table~\ref{tabf2} collects for several values of $r$ the corresponding
bounds for $F(2)$, obtained using Eq.~(\ref{ftwo}). It gives also the values of 
$X(3)$ and $F(3)$ that saturate both paramagnetic bounds: $F(2)=F(3)$ and
$X(2)=X(3)$. In this case, the Zweig rule would be violated neither
for the masses nor for the decay constants.

\begin{table}
\begin{center}
\begin{tabular}{|c|r@{.}lr@{.}l|r@{.}lr@{.}l|r@{.}lr@{.}l|}
\cline{2-13}
\multicolumn{1}{c|}{ }&\multicolumn{4}{c|}{$F(3)$ [MeV]} & 
  \multicolumn{4}{c|}{$F(2)$ [MeV]} & 
  \multicolumn{4}{c|}{No ZR violation}\\
\hline
r & \multicolumn{2}{c}{min} &  \multicolumn{2}{c|}{max}
  & \multicolumn{2}{c}{min} &  \multicolumn{2}{c|}{max}
  & \multicolumn{2}{c}{$F$ [MeV]} &  \multicolumn{2}{c|}{$X$}\\
\hline
10 & 61&69 & 87&24 & 81&31 & 87&24 & 85&57 & 0&403\\
15 & 63&02 & 89&12 & 82&34 & 89&12 & 86&17 & 0&751\\
20 & 63&63 & 89&99 & 83&02 & 89&99 & 86&71 & 0&860\\
25 & 63&99 & 90&50 & 83&38 & 90&50 & 87&08 & 0&905\\
30 & 64&23 & 90&83 & 83&43 & 90&83 & 87&35 & 0&928\\
35 & 64&39 & 91&06 & 83&14 & 91&06 & 87&55 & 0&940\\
\hline
\end{tabular}
\caption{
On the left hand-side, bounds for $F(3)$.
In the middle, corresponding bounds for $F(2)$.
On the right hand-side, values of $F(3)$ and $X(3)$ saturating both
paramagnetic inequalities: $F(2)=F(3)$ and $X(2)=X(3)$.
}
\label{tabf2}
\end{center}
\end{table}

\begin{sidewaystable}
\begin{center}
\begin{tabular}{|r@{.}l|r@{.}lr@{.}lr@{.}lr@{.}l|r@{.}lr@{.}lr@{.}lr@{.}l|
  r@{.}lr@{.}lr@{.}lr@{.}l|}
\cline{3-26}
\multicolumn{2}{c}{}  &   \multicolumn{8}{|c}{$r=20$}  &
\multicolumn{8}{|c}{$r=25$}
    &\multicolumn{8}{|c|}{$r=30$}\\
\hline
\multicolumn{2}{|c}{$L_6$} & \multicolumn{2}{|c}{4} & \multicolumn{2}{c}{5} & 
\multicolumn{2}{c}{7} & \multicolumn{2}{c}{8} & \multicolumn{2}{|c}{4} & 
\multicolumn{2}{c}{5} & \multicolumn{2}{c}{7} & \multicolumn{2}{c}{8} & 
\multicolumn{2}{|c}{4} & \multicolumn{2}{c}{5} & \multicolumn{2}{c}{7} &
\multicolumn{2}{c|}{8}\\
\hline
-0&2  & -0&284 & 2&410 & -1&259 & 2&624 & -0&282 & 1&603 & -0&503 & 0&994 & 
-0&284 & 1&130 & -0&185 & 0&298\\
-0&1  & -0&264 & 2&628 & -1&452 & 3&044 & -0&261 & 1&812 & -0&604 & 1&224 &
-0&262 & 1&332 & -0&233 & 0&416\\
0&     & -0&246 & 2&824 & -1&636 & 3&445 & -0&242 & 1&993 & -0&699 & 1&440 &
-0&243 & 1&501 & -0&276 & 0&525\\
0&2   & -0&215 & 3&167 & -1&986 & 4&207 & -0&210 & 2&306 & -0&880 & 1&848 &
-0&211 & 1&790 & -0&360 & 0&730\\
0&4   & -0&188 & 3&466 & -2&318 & 4&930 & -0&184 & 2&572 & -1&050 & 2&232 & 
-0&184 & 2&036 & -0&439 & 0&925\\
1&     & -0&121 & 4&209 & -3&256 & 6&966 & -0&116 & 3&228 & -1&532 & 3&314 & 
-0&117 & 2&631 & -0&664 & 1&471\\
\hline
-0&2  & 0&068 & 1&833 & -0&816 & 1&653 &-0&013 & 1&174 &-0&325 & 0&586 &
-0&065 & 0&782 & -0&117 & 0&124\\
-0&1  & 0&133 & 2&088 & -0&999 & 2&056 & 0&053 & 1&414 &-0&420 & 0&804 &
0&002 & 1&014 & -0&161 & 0&236\\
0&    & 0&190 & 2&310 & -1&176 & 2&442 & 0&108 & 1&616 &-0&509 & 1&008 & 
0&055 & 1&203 & -0&202 & 0&340\\
0&2   & 0&287 & 2&684 & -1&503 & 3&156 & 0&200 & 1&953 &-0&678 & 1&391 &
0&144 & 1&513 & -0&290 & 0&533\\
0&4   & 0&368 & 3&002 & -1&814 & 3&833 & 0&277 & 2&234 &-0&837 & 1&751 & 
0&217 & 1&770 & -0&354 & 0&715\\
1&    & 0&567 & 3&781 & -2&696 & 5&751 & 0&463 & 2&915 &-1&291 & 2&774 &
0&392 & 2&384 & -0&565 & 1&232\\
\hline
\end{tabular}
\caption{Low-energy constants $L_i(M_\rho)\cdot 10^3$ as functions of
$L_6(M_\rho)\cdot 10^3$ and $r=m_s/m$, for $F_0$=85 MeV (upper part) and 75
MeV (lower part).}\label{LEC}
\end{center}
\end{sidewaystable}

\section{Sensitivity of low-energy constants to ZR violation}

The equations (\ref{l8}), (\ref{l7}),
(\ref{xi}) and (\ref{f0l4}) can be used to extract
the LEC's $L_{i=4,5,7,8}$ as functions of $r$, $F_0$ et
$X(3)$, or equivalently, of $L_6$, $F_0\equiv F(3)$ and $r$ using Eq.~(\ref{GOR}).
Results are shown in Tab.~\ref{LEC} 
as a function of $L_6$, for 2 values of $F_0$ and 3 values of $r$.

This table has been obtained by neglecting the higher-order terms 
$\delta_P$ and $\epsilon_P$, which start at the next-to-next-to-leading order (NNLO). 
If the size of these remainders is large,
the values collected in these tables should be clearly modified.
If we keep considering the low-energy constants as functions of 
$L_6(M_\rho)$ and we change the relative size of the NNLO remainders
within a range of 5\%, the corresponding variations of $X(3)$ are of order
$\pm 0.02$. The impact on the other LEC's is larger for smaller $X(3)$
(of order $\pm 0.3\cdot 10^{-3}$ for $X(3)\sim 0.8$, $r\sim 25$). 

The authors of Ref.~\cite{abt} have estimated the NNLO remainders in the Standard 
Framework. The authors assume first
$L_4(M_\rho)=L_6(M_\rho)=0$ and $r=24$, estimate $O(p^6)$ counterterms (Standard
counting) through a saturation of the associated correlators by resonances,
and fit globally on the available data (masses, decay constants, form factors).
For the decay constants, the obtained NNLO remainders $\epsilon_P$ are less than
5\%. The situation is less clear for the masses, due to a bad convergence of
the series. For instance, Ref.~\cite{abt} has obtained the decomposition: 
$M^2_\pi/(M^2_\pi)_\mathrm{phys}=0.746+0.007+0.247$, where the three terms
correspond respectively to the leading $O(m)$, next-to-leading $O(m^2)$ and 
next-to-next-to-leading $O(m^3)$
orders. Ref.~\cite{abt} suggested that a variation
of $L_4(M_\rho)$ and $L_6(M_\rho)$ could make the NNLO contribution decrease,
but a competition occurs then between the $O(m^2)$ term and the 
leading-order term.

Such a competition could be understood from our analysis as a consequence of
the suppression of the three-flavor condensate $X(3)$.
It would be of great interest to proceed to the same analysis as in Ref.~\cite{abt}, 
and to allow a competition between the terms linear and quadratic in quark masses. 
This might improve the convergence of the expansion of Goldstone boson observables 
in powers of quark masses. Even if the three-flavor condensate $X(3)$ is suppressed 
(first term of the series for pseudoscalar masses), we expect a global convergence of 
the series, i.e. small higher-order remainders $\delta_P$ and $\epsilon_P$.

Standard values of the LEC's at order $O(p^4)$ can be found in 
Ref.~\cite{daphne}, and were derived with the supposition that
the ZR violating LEC's $L_6(\mu)$ and
$L_4(\mu)$ were suppressed at the scale $\mu=M_\eta$. 
The following values have been obtained:
$L_4(M_\rho)\cdot 10^3=-0.3\pm 0.5$,
$L_5(M_\rho)\cdot 10^3=1.4\pm 0.5$,
$L_6(M_\rho)\cdot 10^3=-0.2\pm 0.3$,
$L_7(M_\rho)\cdot 10^3=-0.4\pm 0.2$,
$L_8(M_\rho)\cdot 10^3=0.9\pm 0.3$.
These values are compatible with our analysis: it can be seen on the first lines of
Table~\ref{LEC}, 
with $L_6(M_\rho)\sim -0.2\cdot 10^{-3}$,
$X(2) \sim X(3) \sim 0.9$, $r=25$, $F_0=85$ MeV.
The values of $L_4(M_\rho)$ and $L_6(M_\rho)$ correspond also to the lower
bounds derived from the saturation of the paramagnetic inequalities 
for $F^2$ and $X$: $X(2)=X(3)$ and $F(2)=F(3)$.

We see that $L_4$ is weakly sensitive to $L_6$, in agreement with 
Eqs.~(\ref{defl4}) and (\ref{f0l4}). For large $r$, and $F_0$ close to $F_\pi$, 
$1-\tilde\eta(r)-F_0^2/F_\pi^2$ is nearly vanishing, so that $L_4$ is mainly
fixed by the last term in Eq.~(\ref{f0l4}) with no dependence on $X(3)$.
On the contrary, $L_5$ is clearly dependent on the value of $L_6$.
We can look at Eq.~(\ref{xi}) to understand
this phenomenon : $\tilde\eta(r)$ may be small, but it never vanishes. The first term 
in Eq.~(\ref{xi}) leads therefore to large values of $L_5$ when 
$X(3)$ decreases, whatever values we choose for $r$ and $F_0$ (this is related to the definition
of $L_5$).
The Gell-Mann--Okubo formula Eq.~(\ref{l7}) correlates
strongly $L_7$ and $L_8$, leading to $L_7\simeq -L_8/2$.

From this analysis of the pseudoscalar masses and decay constants, we
cannot conclude whether $\bar{q}q$ fluctuations have important effects
on the chiral structure of QCD vacuum. However, the values of LEC's are
extremely sensitive to the value of $L_6(M_\rho)$. A small shift of $L_6$
towards positive values would immediately split $X(2)$ and $X(3)$, and increase strongly
$L_8(M_\rho)$ and $L_5(M_\rho)$. We will now use additional
information from experimental data in the scalar sector, in order
to constrain $L_6(M_\rho)$.

\section{Sum rule for $X(2)-X(3)$}

\subsection{Correlator of two scalar densities}

We introduce the correlator \cite{Bachir1,Bachir2}:
\begin{equation}
\Pi(p^2)=i\frac{m m_s}{M^2_\pi M^2_K}
   \lim_{m\to 0} \int d^4x \ e^{ip\cdot x}\ 
   \langle 0|T\{\bar{u}u(x)\ \bar{s}s(0)\}|0\rangle,
\end{equation}
that is invariant under the QCD renormalization group, and violates the Zweig
rule in the $0^{++}$ channel. For $m_s\neq 0$, $\Pi$ is a
$\su{2}$ order parameter, related to the derivative of $\Sigma(2)$ with
respect to $m_s$: $mm_s\partial\Sigma(2)/\partial m_s=M_\pi^2 M_K^2 \Pi(0)$.

We can use the relation Eq.~(\ref{xtwo}) between $r$, $X(3)$ and $X(2)$
to compute $\partial\Sigma(2)/\partial m_s$. Eq.~(\ref{z}) can be used to compute
$(\partial Z^S/\partial m_s)_{m=0}$.
This leads to an equation involving $Z^S$ and $\Pi(0)$:
\begin{eqnarray}
X(2)-X(3)&=&\frac{2mm_s}{F_\pi^2M_\pi^2}
  \left.Z^S \right|_{m=0}+\frac{m}{F_\pi^2M_\pi^2}
    \lim_{m\to 0}\frac{F_\pi^2 \delta_\pi}{m}\\
&=&\frac{2M_K^2}{F_\pi^2}\Pi(0)
  +\frac{r[X(3)]^2}{32\pi^2}\frac{F_\pi^2M_\pi^2}{F_0^4}
    \left(\bar\lambda_K+\frac{2}{9}\bar\lambda_\eta\right)\nonumber\\
&&\qquad   +\frac{m}{F_\pi^2M_\pi^2}
   \left(1-m_s\frac{\partial}{\partial m_s}\right)
    \lim_{m\to 0}\frac{F_\pi^2\delta_\pi}{m},
    \label{diffgor}
\end{eqnarray}
with the logarithmic derivatives:
$\bar\lambda_P=m_s\cdot\partial(\log\bar{M}_P^2)/\partial m_s$. This
equation contains the NNLO
remainder $F_\pi^2\delta_\pi$ in the quark mass expansion of
$F_\pi^2M_\pi^2$. Its leading term is
$O(mm_s^2)$, so that the last term in Eq.~(\ref{diffgor}) should be of order
$\sim (-\delta_\pi/M_\pi^2)$.

$L_6$ (or the difference $X(2)-X(3)$) measures the violation of the Zweig
rule in the scalar channel. We are going to exploit experimental information
about this violation and to evaluate $\Pi(0)$ through the sum rule:
\begin{eqnarray}
\Pi(0)&=&
  \frac{1}{\pi}\int_0^{s_1} ds\ \mathrm{Im}\ \Pi(s) 
     \ \frac{1}{s}\left(1-\frac{s}{s_0}\right)\label{regsom}\\
&&\qquad +\frac{1}{\pi}\int_{s_1}^{s_0} ds\ \mathrm{Im}\ \Pi(s)
     \ \frac{1}{s}\left(1-\frac{s}{s_0}\right)
  + \frac{1}{2i\pi}\int_{|s|=s_0} ds\ \Pi(s) 
     \ \frac{1}{s}\left(1-\frac{s}{s_0}\right)\nonumber.
\end{eqnarray}

\begin{figure}[ht]
\begin{center}
\includegraphics[height=6cm]{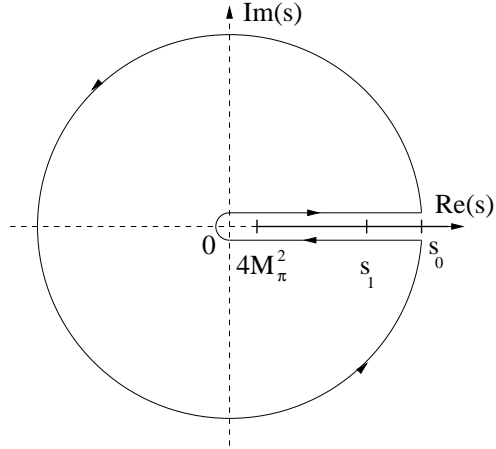}
\caption{Contour of the integral in the sum rule for the correlator
$\langle \bar{u}u\ \bar{s}s\rangle$.}
\end{center}
\end{figure}

The three terms will be estimated in different ways:
\begin{itemize}
\item For $0\leq \sqrt{s} \leq \sqrt{s_1}\sim 1.2\ \mathrm{GeV}$,
the spectral function $\mathrm{Im}\ \Pi$ is obtained by solving
Omn\`es-Muskhelishvili equations for two coupled channels, using several
$T$-matrix models in the scalar sector.
\item For $\sqrt{s_1}\leq \sqrt{s} \leq \sqrt{s_0}\sim 1.5\ \mathrm{GeV}$,
the spectral function under $s_1$ is exploited through another sum rule in order to bound
the contribution of the integral.
\item For $|s|=s_0$, we estimate the integral through Operator Product Expansion (OPE).
\end{itemize}

\subsection{Asymptotic behavior} \label{secopepi}

$\Pi$ can be expanded using OPE:
\begin{eqnarray}
\Pi(p^2)&=&i\frac{m m_s}{M^2_\pi M^2_K}
   \lim_{m\to 0} \int d^4x \ e^{ip\cdot x}\ 
   \langle 0|T[\bar{u}u(x)\ \bar{s}s(0)]|0\rangle\\
   &{\displaystyle \mathop\sim_{P^2\to\infty}}&\frac{m m_s}{M^2_\pi M^2_K}
      \sum_{n\geq 4} \frac{1}{(P^2)^{n/2-1}} C^{(n)}(t)
       \langle 0|\mathcal{O}_n|0\rangle,
\end{eqnarray}
with $P^2=-p^2$, $\mu$ a renormalization scale, $t=\mu^2/P^2$, and
$\mathcal{O}_n$ a combination of $n$-dimensional operators.
$\Pi$ transforms chirally as
$(\bar{u}u) (\bar{s}s)$ and we take the chiral limit $m\to 0$.
Hence, the lowest-dimension operator is
$\mathcal{O}_4=m_s \bar{u}u$, and the contributing diagrams include at least
two gluonic lines \cite{Bachir1}.

\begin{figure}[ht]
\begin{center}
\includegraphics[width=11cm]{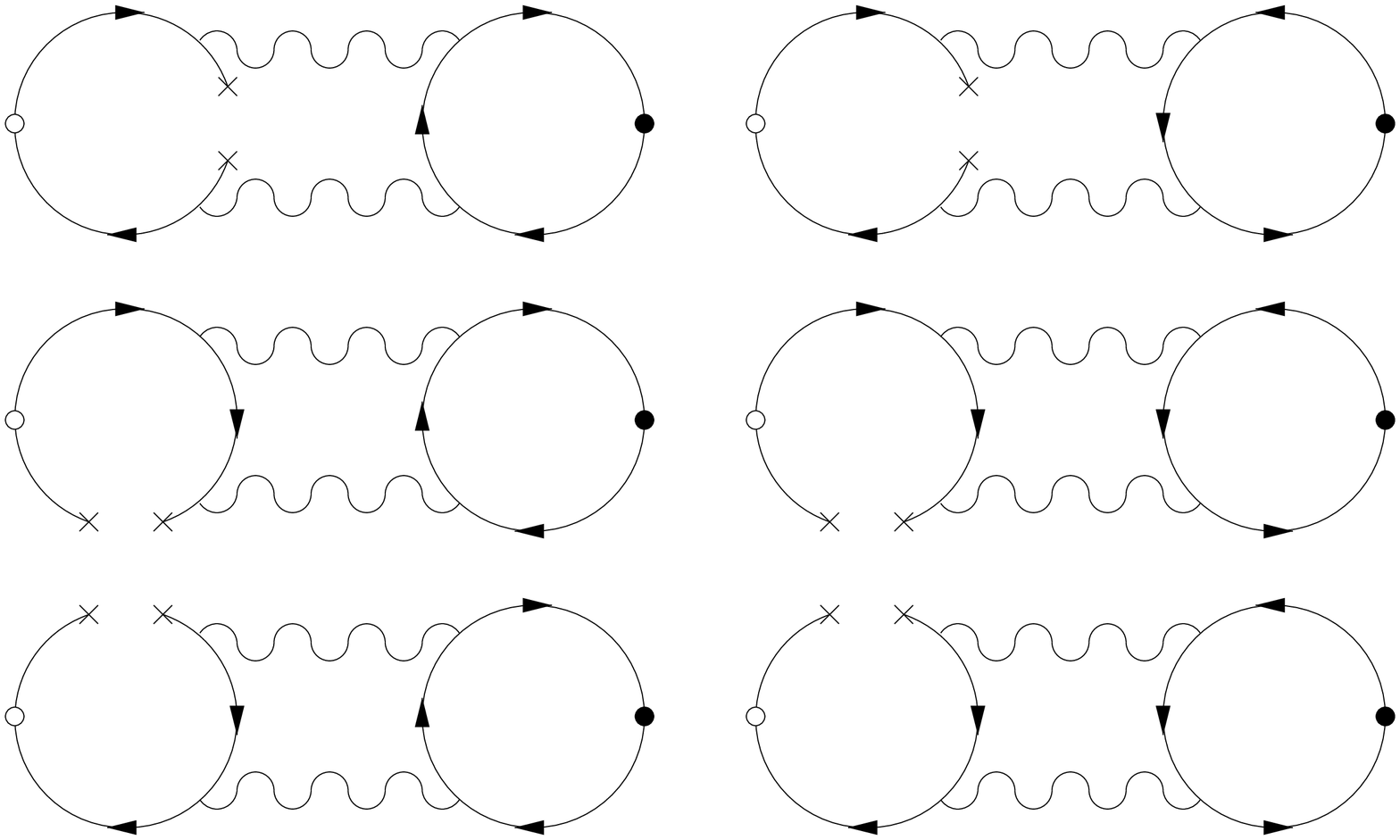}
\caption{Feynman diagrams contributing to $m_s\langle\bar{u}{u}\rangle$
in OPE of $\Pi$ (lowest order in
$\alpha_s$). The white circle is the scalar source $\bar{u}u$,
the black one $\bar{s}s$.}
\label{opefeynman}
\end{center}
\end{figure}

We will work in dimensional regularization ($d=4-2\omega$). In the class of t'Hooft's 
gauges, the gluon propagator reads:
\begin{equation}
\frac{-i}{k^2+i\epsilon} \left(g_{\mu\nu}-(1-\xi)
\frac{k_\mu k_\nu}{k^2+i\epsilon}\right) \delta_{ab},
\end{equation}
with $\xi$ a free real parameter. The Wilson coefficient of $m_s\bar{u}u$
(at the leading order) is obtained by adding 6 two-loop integrals.
It is easy to see that the contributions of 
$\xi$ and $\xi^2$ cancel in this sum of integrals. Hence, we check that the Wilson
coefficient of $m_s\bar{u}u$ at the leading order is independent of the chosen
gauge, in the classe of t'Hooft's gauges.

We want the large-$P^2$ limit of integrals like:
\begin{eqnarray}
&&g_s^4 \mu^{4\omega} m_s\langle\bar{u}u\rangle
 \frac{1}{p^2}
 \int \frac{d^4q}{(2\pi)^{d}}\frac{d^4k}{(2\pi)^{d}} 
   \mathcal{P}(p^2,q^2,k^2,p\cdot q, p\cdot k, q\cdot k,m_s^2)
   \label{forminteg}\\
&&\qquad \times \frac{1}{[(p-q)^2-m_0^2]^{n_2} [q^2-m_0^2]^{n_3}}\nonumber\\
&&\qquad \times 
 \frac{1}{[(k+p)^2-m_s^2]^{n_4} [(k+q)^2-m_s^2]^{n_5} [k^2-m_s^2]^{n_6}},
    \nonumber
\end{eqnarray}
where $\mathcal{P}$ is a polynomial of degree 2. $m_0$ corresponds at the same time
to $m=m_u=m_d$ for fermion propagators in the loop of $u-d$ quarks, and to a 
fictitious mass to regularize infrared gluonic divergences 
(we take at the end the limit $m_0\to 0$).

Using identities like
$2(k\cdot q)=[(k+q)^2+m_s^2]-[k^2+m_s^2]-q^2$, we can reexpress the sum in terms of:
\begin{eqnarray}
&&\frac{1}{p^{2\nu_0}} J(\{\nu_i\},\{m_i\},p)=
\frac{1}{p^{2\nu_0}}
  \int\frac{d^4q\ d^4k}{[q^2-m_0^2]^{\nu_1} [k^2-m_s^2]^{\nu_2}}\\ 
&&\qquad \times
  \frac{1}{[(k+q)^2-m_s^2]^{\nu_3} [(p-q)^2-m_0^2]^{\nu_4}
  [(k+p)^2-m_s^2]^{\nu_5}},
  \nonumber
\end{eqnarray}
with $m_1=m_4=m_0$ and $m_2=m_3=m_5=m_s$. These integrals are formally identical to the
ones arising for two-loop self-energies. The behavior of such integrals at large external
momentum has already been studied. The basic idea consists in following the flux of the
large external momentum through the Feynman diagrams, in order to Taylor expand 
the propagators \cite{largemoment}. This procedure, based on the asymptotic expansion
theorem \cite{asympexp}, is sketched in App.~\ref{appsecopepi}.

Rather lengthy computations lead to the first term arising in the OPE of the correlator.
Some integrals contain poles in $1/\omega$, but these divergences cancel when
all the contributions are summed (this cancellation is a non-trivial check of the
procedure). The first term in OPE is:
\begin{eqnarray}
&&i\frac{mm_s}{M_\pi^2 M_K^2}\lim_{m\to 0}\int d^4x \ e^{ip\cdot x}\ 
   \langle 0|T\{m\bar{u}u(x)\ m_s\bar{s}s(0)\}|0\rangle\label{opepi}\\
&&\quad {\displaystyle \mathop\sim_{P^2\to\infty}}
  -\frac{18[1-2\zeta(3)]}{P^2}\left(\frac{\alpha_s}{\pi}\right)^2
      \frac{m_s^2}{M_\pi^2 M_K^2} m\langle\bar{u}u\rangle.\nonumber
\end{eqnarray}
The involved condensate should be the two-flavor one ($m\to 0$, $m_s\neq 0$).

\subsection{Contribution for $s\leq s_1$ : pion and kaon scalar form factors}

\subsubsection{Omn\`es-Muskhelishvili equations}

In order to compute the integral:
\begin{equation}
\mathcal{I}=
  \frac{1}{\pi}\int_{0}^{s_1} ds\ \mathrm{Im}\ \Pi(s)
     \ \frac{1}{s}\left(1-\frac{s}{s_0}\right),
\end{equation}
we have to know $\mathrm{Im}\ \Pi$ between 0 and $s_1$ ($\sqrt s_1 \sim$ 1.2
Gev). The procedure is explained in detail in Refs.~\cite{Bachir1,Bachir2}, and we shall
merely sketch its main features for completeness. In the range of energy between 0 and $s_1$,
the  $\pi\pi$- and $K\bar{K}$- channels should dominate the spectral function
\cite{Bachir1,Donoghue,Meissner}. If we denote these channels
respectively 1 and 2, the spectral function is:
\begin{equation}
\mathrm{Im}\ \Pi(s) \label{foncspec0}
=\frac{m m_s}{M^2_\pi M^2_K}
\frac{1}{16\pi}\sum_{i=1,2} \sqrt{\frac{s-4M_i^2}{s}} 
  [n_i F_i(s)][n_i G^*_i(s)]
      \theta(s-4M_i^2),
\end{equation}
with the scalar form factors for the pion and the kaon:
\begin{equation}
\vec{F}(s)=
  \left(
  \begin{array}{c}
  \langle 0|\bar{u}u|\pi\pi\rangle\\
  \langle 0|\bar{u}u|K\bar{K}\rangle
  \end{array}
  \right), \qquad
\vec{G}(s)=
  \left(
  \begin{array}{c}
  \langle 0|\bar{s}s|\pi\pi\rangle\\
  \langle 0|\bar{s}s|K\bar{K}\rangle
  \end{array}
  \right),  
\end{equation}
with $M_1=M_\pi$ and $M_2=M_K$. $n_1=\sqrt{3/2}$ and $n_2=\sqrt{2}$ 
are numerical factors related to the normalization of the states
$|\pi\pi\rangle$ and $|K\bar{K}\rangle$.

The form factors are analytic functions in the complex plane, with the
exception of a right cut along the real axis. They should have the asymptotic
behaviour $F_i(s)\sim 1/s$ when $s\to \infty$, and verify a dispersion relation
with no subtraction. Obviously, when $s$ increases,
new channels open, and the two-channel approximation is no more sufficient.
But we need $\vec{F}$ and $\vec{G}$ for $s\leq s_1$, and we are not interested 
in the behaviour of the spectral function at much higher energies. 
We can therefore suppose that the two-channel approximation is valid for any
energies, with the discontinuity along the cut:
\begin{eqnarray}
S_{ij}&=&\delta_{ij}+2i\sigma_i^{1/2} T_{ij} \sigma_j^{1/2}
\theta(s-4M_i^2)\theta(s-4M_j^2),
\label{ffcut}\\
\label{ffimag}
\mathrm{Im}\ F_i(s)&=&\sum_{j=1}^n T_{ij}^*(s)\sigma_j(s) F_j(s)
\theta(s-4M_j^2),\ \sigma_i=\sqrt\frac{s-4M_i^2}{s},
\end{eqnarray}
and we will suppose in addition that the two-channel $T$-matrix model impose
the correct asymptotic behaviour for $\vec{F}$ and $\vec{G}$.

Under these assumptions, $\vec{F}$ and $\vec{G}$ satisfy separately a set of
coupled Omn\`es-Muskhelishvili equations \cite{Bachir1,Donoghue,Omnes,Mush}:
\begin{equation}\label{omeq}
F_i(s)=\frac{1}{\pi}\sum_{j=1}^n\int_{4M_\pi^2}^\infty ds' \frac{1}{s'-s}
   T_{ij}^*(s') \sqrt\frac{s'-4M_j^2}{s'} \theta(s'-4M_j^2) F_j(s'),
\end{equation}
with the condition that the $T$ matrix leads to the expected decrease of
the form factors for $s\to\infty$. Ref.~\cite{Bachir1} has proved a condition of existence
and unicity for the solution of Eq.~(\ref{omeq}): $\Delta(\infty)-\Delta(4M_\pi^2)=2\pi$, 
where $\Delta(s)$ is the sum of the $\pi\pi$ and $K\bar{K}$ phase shifts.
In that case, the set of linear 
equations admits a unique solution, once the values at a given energy
are fixed \cite{Bachir2}. 
All the solutions are thus linear combinations of a basis, for instance
the solutions $\vec{A}(s)$ and $\vec{B}(s)$ such as:
$\vec{A}(0)={1\choose 0}$ and
$\vec{B}(0)={0\choose 1}$.
$\vec{F}$ and $\vec{G}$ can therefore be written as:
\begin{equation}
\vec{F}(s)=F_1(0)\vec{A}(s)+F_2(0)\vec{B}(s),\qquad
\vec{G}(s)=G_1(0)\vec{A}(s)+G_2(0)\vec{B}(s).\label{declin}
\end{equation}
The value of the form factors at zero is related to the derivatives of 
the pseudoscalar masses with respect to the quark masses:
\begin{eqnarray}
F_1(0)=\frac{1}{2}\left(\frac{\partial M_\pi^2}{\partial m}\right)_{m=0},
& \qquad &
F_2(0)=\frac{1}{2}\left(\frac{\partial M_K^2}{\partial m}\right)_{m=0},\\
G_1(0)=\left(\frac{\partial M_\pi^2}{\partial m_s}\right)_{m=0}=0,
& \qquad &
G_2(0)=\left(\frac{\partial M_K^2}{\partial m_s}\right)_{m=0}.
\end{eqnarray}
Up to now, we have followed the same line as Refs.~\cite{Donoghue,Bachir1}. But in these
papers, the value of the scalar form factors at zero was derived using Standard $\chi$PT,
i.e. supposing that the three-flavor quark condensate dominates the expansion of the
pseudoscalar masses.
We are going to allow a competition between the terms linear
and quadratic in quark masses, so that the normalization of the form factors may
become rather different from what is considered in Refs.~\cite{Bachir1,Donoghue}. In a similar
way, the form factors that we will obtain could differ from the ones obtained by a matching
with Standard one-loop expressions \cite{Meissner,gassmeiss}.

We consider here three models of $T$-matrix, proposed respectively by Oller, Oset and
Pelaez in Ref.~\cite{Oset}, by Au, Morgan and Pennington in Ref.~\cite{Au}, and by
Kaminski, Lesniak and Maillet in Ref.~\cite{Loiseau}. These models fit correctly the
available data in the scalar sector under 1.3 Gev, as discussed in 
Refs.~\cite{Bachir1,Donoghue,Oset}. However, they have to be corrected for very low
and very high energies, as discussed in Ref.~\cite{Bachir1}:
chiral symmetry constrains the $\pi\pi$ phase shift near the threshold,
and the asymptotic behavior of the phases shifts has to be changed to
to insure existence and unicity for the solution of Eq.~(\ref{omeq}).

\subsubsection{Contribution of the first integral}

If we put Eq.~(\ref{declin}) into
Eq.~(\ref{foncspec0}), we obtain the spectral function as the sum of two
contributions:
\begin{eqnarray}
\mathrm{Im}\ \Pi(s)
&=&  \gamma_\pi\lambda_K
   \left[\frac{\sqrt{3}}{32\pi}
      \sum_{i=1,2} \sqrt{\frac{s-4M_i^2}{s}} A_i(s) B^*_i(s)
      \theta(s-4M_i^2)\right]\label{foncspec}\\
&& +\gamma_K\lambda_K\frac{M_K^2}{M_\pi^2}
      \left[\frac{1}{16\pi}\sum_{i=1,2} \sqrt{\frac{s-4M_i^2}{s}} B_i(s) B^*_i(s)
      \theta(s-4M_i^2)\right], \nonumber
\end{eqnarray}
where the logarithmic derivatives of the masses are denoted:
\begin{equation}
\lambda_P
   =\frac{m_s}{M^2_P}\left(\frac{\partial M_P^2}{\partial m_s}\right)_{m=0}
   =\frac{m_s}{M^2_P}\frac{\partial \bar{M}_P^2}{\partial m_s},
 \qquad
\gamma_P
   =\frac{m}{M^2_P}\left(\frac{\partial M_P^2}{\partial m}\right)_{m=0}.
\end{equation}
The two bracketed functions in Eq.~(\ref{foncspec}) can be plotted: 
the first one is called "type $AB^*$", the second one "type
$BB^*$". It is also interesting to study how these two contributions cancel
each other inside the spectral function, by taking the Standard tree-level 
estimates: $\gamma_\pi=1$, $\lambda_K=1-M_\pi^2/(2M_K^2)$ and 
$\gamma_K=M_\pi^2/(2M_K^2)$. A peak, corresponding to the narrow 
resonance $f_0(980)$ \cite{scalar}, arises with a height
depending on the models: Ref.~\cite{Oset} leads to a smaller peak than
Refs.~\cite{Au} and \cite{Loiseau}.

\begin{figure}
\begin{center}
\includegraphics[width=11cm]{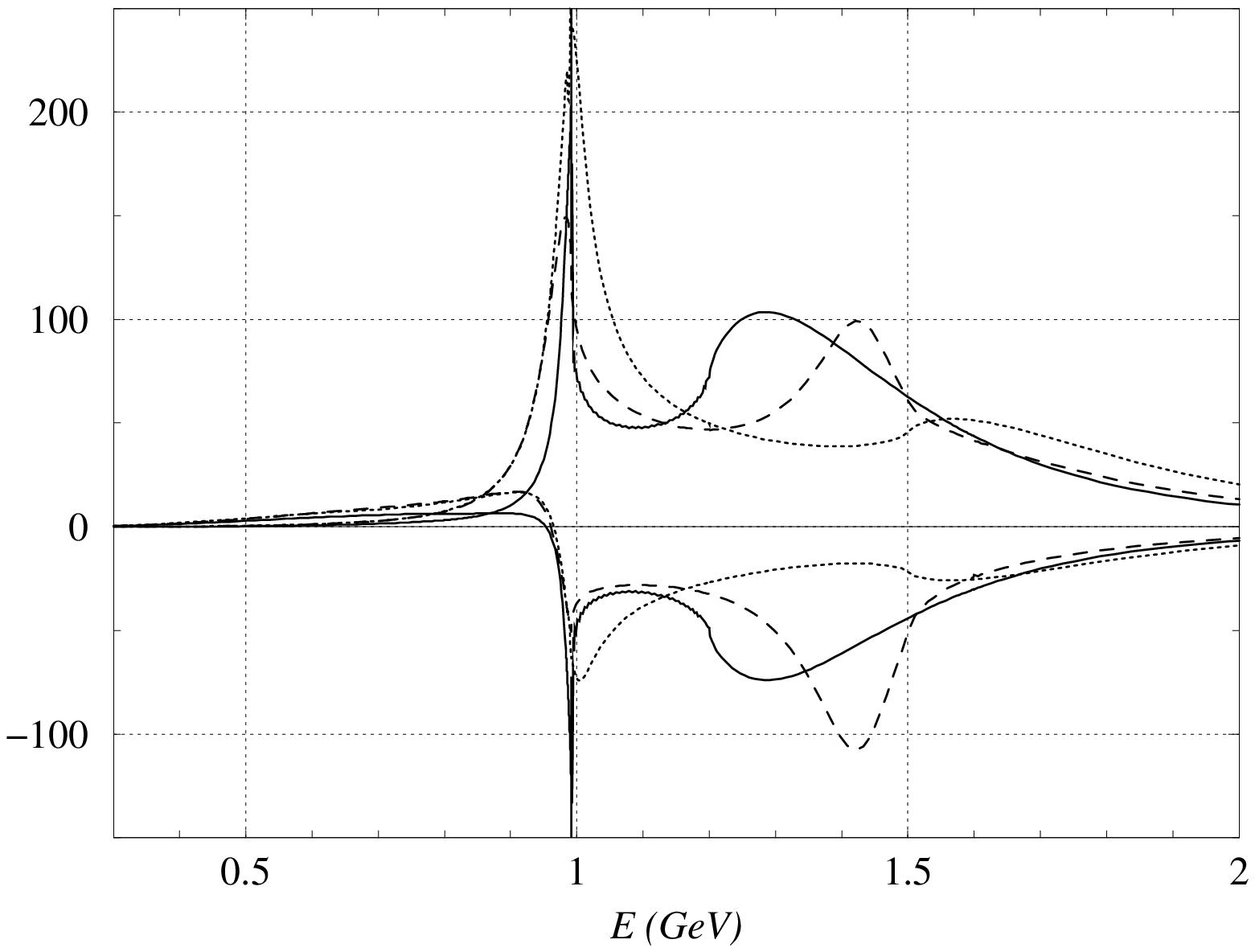}
\includegraphics[width=11cm]{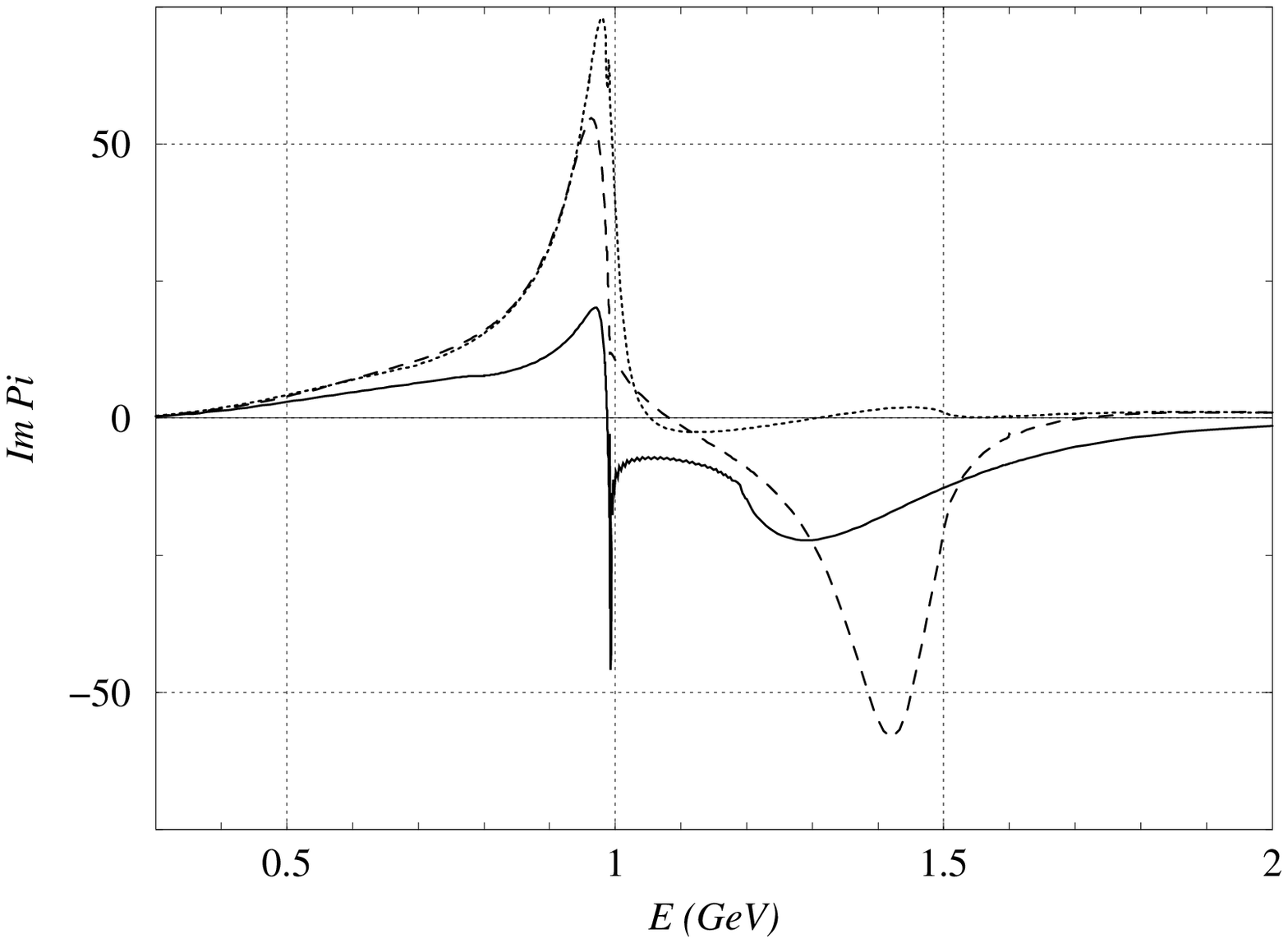}
\caption{Up: Contributions of type $BB^*$ (positive) and type $AB^*$
(negative) to the spectral function. Down:
example of spectral function, obtained with
$\gamma_\pi=1$, $\lambda_K=1-M_\pi^2/(2M_K^2)$ and
$\gamma_K=M_\pi^2/(2M_K^2)$. In both cases, we plot
the results for the $T$-matrix models of Ref.~\cite{Oset} (solid lines),
Ref.~\cite{Au} (dotted) and Ref.~\cite{Loiseau} (dashed).}
\end{center}
\end{figure}

The integral between 0 and $s_1$ in the sum rule
Eq.~(\ref{regsom}) can be written, using Eq.~(\ref{foncspec}):
\begin{equation}
\frac{1}{\pi}\int_0^{s_1} ds\ \mathrm{Im}\ \Pi(s) 
     \ \frac{1}{s}\left(1-\frac{s}{s_0}\right)
   =\gamma_\pi\lambda_K \mathcal{I}_{AB}
   +\gamma_K\lambda_K\frac{M_K^2}{M_\pi^2} \mathcal{I}_{BB},
\end{equation}
where
$\mathcal{I}_{XY}=
   \mathcal{M}_{XY}^{(-1)}-\mathcal{M}_{XY}^{(0)}/s_0$,
involves the moments:
\begin{eqnarray}
\mathcal{M}^{(k)}_{AB}&=&
\frac{\sqrt{3}}{32\pi^2}
\int_0^{s_1} ds
     \ s^k 
   \sum_{i=1,2} \sqrt{\frac{s-4M_i^2}{s}} A_i(s) B^*_i(s)
      \theta(s-4M_i^2),\label{moment1}\\
\mathcal{M}^{(k)}_{BB}&=&
\frac{1}{16\pi^2}
\int_0^{s_1} ds
     \ s^k 
    \sum_{i=1,2} \sqrt{\frac{s-4M_i^2}{s}} B_i(s) B^*_i(s)
      \theta(s-4M_i^2).\label{moment2}
\end{eqnarray}

Notice that we solve Omn\`es-Muskhelishvili equations to obtain the scalar form
factors of the pion and the kaon in the limit $m\to 0$ (and $m_s$ fixed at its physical
value). But we consider $T$-matrix models fitting experimental data, with 
up and down quarks with their physical masses. The limit $m\to 0$ sets
the $\pi\pi$-threshold to
zero\footnote{For $m\to 0$, the cut along the real axis starts
at $s=0$. However, the integral
$\int_0^{s_0} ds\ (1-s/s_0) \cdot \mathrm{Im}\ \Pi(s)/s$
is convergent, since for $s\to 0$:
$F_1(s) \to F_1(0)$ and $G_1(s) \sim G'_1(0)\cdot s$, leading to:
\begin{equation}
\mathrm{Im}\ \Pi(s) \sim 
  \frac{3}{32\pi}\frac{mm_s}{M_\pi^2M_K^2} F_1(0) G'_1(0)\cdot s.
\end{equation}}, changes $\pi\pi$ phase shifts near the threshold and
shifts slightly the $K\bar{K}$ threshold.
Such modifications should not alter significantly the general shape of the
spectral function. In particular, the integral of the spectral function, dominated
by the $f_0(980)$ peak, should be affected only marginally when 
$T_\mathrm{m\neq 0}$ is considered instead of $T_\mathrm{m\to 0}$.

\subsection{Second sum rule : $s_1\leq s\leq s_0$}

The contribution of the integral below $s_1$ is positive and
dominated by the $f_0(980)$ peak. But according to
Sec.~\ref{secopepi}, $\Pi$ is superconvergent, and the integral of the
spectral function from 0 to infinity vanishes. $\mathrm{Im}\ \Pi(s)$
should therefore become negative in some range of energy. In particular, negative
peaks should rather naturally appear in the spectral function, in relation with massive
scalar resonances like $f_0(1370)$ and $f_0(1500)$ \cite{scalar}.

Let us suppose that the spectral function is negative
for $s_1\leq s \leq s_0$.\footnote{If the spectral function is partially
positive in this range, our hypothesis will end up with an estimate for the second
integral that will be smaller than its actual value. In that case, we would 
underestimate the difference $X(2)-X(3)$.} The contribution from the intermediate
region in Eq.~(\ref{diffgor}) can be estimated from:
\begin{equation}
\frac{1}{s_0} \mathcal{J}' \leq 
-\frac{1}{\pi}\int_{s_1}^{s_0} ds\ \mathrm{Im}\ \Pi(s)
     \ \frac{1}{s}\left(1-\frac{s}{s_0}\right)
\leq \frac{1}{s_1} \mathcal{J}', \label{estimate}
\end{equation}
where $\mathcal{J}'$ is the integral:
\begin{equation}
\mathcal{J}'
  =\frac{1}{\pi}\int_{s_1}^{s_0} ds\ \mathrm{Im}\ \Pi(s)
     \ \left(1-\frac{s}{s_0}\right),
\end{equation}
which satisfies the sum rule:
\begin{equation}
\frac{1}{\pi}\int_0^{s_1} ds\ \mathrm{Im}\ \Pi(s) 
     \ \left(1-\frac{s}{s_0}\right)+\mathcal{J}'
+ \frac{1}{2i\pi}\int_{|s|=s_0}\!\!\! ds\ \Pi(s) 
     \ \left(1-\frac{s}{s_0}\right)=0\label{regsom2}
\end{equation}

The first integral in Eq.~(\ref{regsom2}) can be computed from the spectral function
obtained in the previous section:
\begin{equation}
\frac{1}{\pi}\int_{0}^{s_1} ds\ \mathrm{Im}\ \Pi(s)
     \ \left(1-\frac{s}{s_0}\right)
  =\gamma_\pi\lambda_K \mathcal{I}'_{AB}
   +\gamma_K\lambda_K\frac{M_K^2}{M_\pi^2} \mathcal{I}'_{BB},
\end{equation}
with
$\mathcal{I}'_{XY}=\mathcal{M}_{XY}^{(0)}-\mathcal{M}_{XY}^{(1)}/s_0$ involving
the moments Eqs.~(\ref{moment1})-(\ref{moment2}).

The contribution from the complex circle [third integral in Eq.~(\ref{regsom2})]
can be estimated through Operator Product Expansion (OPE), using the method described in
the following section:
\begin{eqnarray}
&&\frac{1}{2i\pi}\int_{|s|=s_0}\!\!\! ds\ \Pi(s) 
     \ \left(1-\frac{s}{s_0}\right)\\
&&=
   9[1-2\zeta(3)]
  \frac{F_\pi^2}{M_K^2} X(2) m^2_s(s_0) a^2(s_0)\nonumber\\
&&\quad \times   \left\{1+\frac{\beta_0\gamma}{2} a(s_0)
         +\left[\frac{\beta_1\gamma}{2}
	     -\frac{\gamma(\gamma+1)}{8}
	     \left(\frac{\pi^2}{3}-2\right)
	      \beta_0^2\right]a^2(s_0)\right\}+\ldots \nonumber\\
&&= 9[1-2\zeta(3)]
  \frac{F_\pi^2}{M_K^2} X(2) m^2_s(s_0) a^2(s_0)\\
&&\qquad \times [1+6.5\cdot a(s_0)-25.125\cdot a^2(s_0)].\nonumber
\end{eqnarray}

\subsection{High-energy contribution : $|s|=s_0$} \label{secbachsr}

We want the contribution of the integral on the large circle:
\begin{equation}
\mathcal{K}=\frac{1}{2i\pi}\!\!\int_{|s|=s_0}\!\!\!\! ds\ \Pi(s) 
     \ \frac{1}{s}\left(1-\frac{s}{s_0}\right)
=\frac{1}{2\pi} \int_{-\pi}^\pi 
         d\theta\ (1+e^{i\theta})\ \Pi(p^2=-s_0 e^{i\theta}).\label{gdcercle}
\end{equation}
The factor $(1-s/s_0)$ suppresses the contribution stemming from the time-like region
around $s_0$, so that we can use in this integral the Operator Product
Expansion of $\Pi$ \cite{Braaten}.
Once Renormalization Group Improvement is applied to
Eq.~(\ref{opepi}), the QCD renormalization group  invariant
$m\langle\bar{u}u\rangle$ gets the coefficient:
\begin{equation}
a^2(P^2)m_s^2(P^2)=a^2(s_0)m_s^2(s_0)\times
\left[\frac{a(P^2)}{a(s_0)}\right]^{8/b_0+2},
\end{equation}
with $a(s)=\alpha_s(s)/\pi$ and $b_0=11-2N_f/3=9$.
The integral Eq.~(\ref{gdcercle}) becomes:
\begin{eqnarray}
\mathcal{K}&=&\frac{9[1-2\zeta(3)]}{2\pi} 
  \frac{F_\pi^2}{M_K^2} X(2) \frac{m^2_s(s_0)}{s_0}
   [a(s_0)]^{-\frac{8}{b_0}}
 \int_{-\pi}^\pi ds\ (1+e^{-i\theta})\ a^\gamma(s_0e^{i\theta}),
\end{eqnarray}
with $\gamma=2+8/b_0=2+8/9$.
To compute this integral, we expand
$a(P^2=s_0e^{i\theta})$ in powers of $a(s_0)$. The behaviour of
$a(t)$ ($t$ complex) depends on the $\beta$ function:
\begin{eqnarray}
t\frac{d}{dt}a(t)=\frac{1}{2\pi}\beta[a(t)],
&\qquad&
\frac{1}{\pi}\beta[a(t)]=-\beta_0a^2-\beta_1a^3+\ldots,\\
\beta_0=\frac{33-2N_f}{6}=\frac{9}{2},&\qquad &
\beta_1=\frac{306-38N_f}{24}=8.
\end{eqnarray}
The expansion of $a(s_0e^{i\theta})$ is:
\begin{equation}
a(s_0e^{i\theta})=a(s_0)-\frac{i}{2}\beta_0 \theta a^2(s_0)
   +\left[\frac{i}{2}\beta_1\theta-\frac{1}{4}\theta\beta_0^2\theta^2\right]
       a^3(s_0)+O(a^4).
\end{equation}

We get:
\begin{eqnarray}
\mathcal{K}&=&9[1-2\zeta(3)]
  \frac{F_\pi^2}{M_K^2} X(2) \frac{m^2_s(s_0)}{s_0} a^2(s_0)\label{valope}\\
&& \quad\times   \left\{1-\frac{\beta_0\gamma}{2} a(s_0)
         -\left[\frac{\beta_1\gamma}{2}
	     +\frac{\gamma(\gamma+1)}{8}
	     \left(\frac{\pi^2}{3}-2\right)
	      \beta_0^2\right]a^2(s_0)\right\}+\ldots \nonumber\\
&=& 
     9[1-2\zeta(3)]
  \frac{F_\pi^2}{M_K^2} X(2) \frac{m^2_s(s_0)}{s_0} a^2(s_0)
    [1-6.5\cdot a(s_0)+48.236\cdot a^2(s_0)].\nonumber
\end{eqnarray}
This negative contribution is strongly suppressed by
$\alpha^2_s$ and $m_s^2/s_0$. We have considered here $m_s\sim 200$ MeV,
but the contribution of this integral is so small that the error due to
$m_s$ and $\alpha_s$ can be neglected. Notice that duality is not supposed to 
arise in the scalar sector for as low energies as in other channels,
due to a probably large contribution
from the direct instantons in this sector \cite{directinst}.

\section{Results}

\subsection{Logarithmic derivatives of pseudoscalar masses}

The logarithmic derivatives of the masses are obtained from the expansions
of $F_P^2$ and $F_P^2M_P^2$:
\begin{equation}
\lambda_P
   =\frac{m_s}{M^2_P}\left(\frac{\partial M_P^2}{\partial m_s}\right)_{m=0}
   =\frac{m_s}{M^2_P}\frac{\partial \bar{M}_P^2}{\partial m_s},
 \qquad
\gamma_P
   =\frac{m}{M^2_P}\left(\frac{\partial M_P^2}{\partial m}\right)_{m=0}.
\end{equation}
The corresponding expressions are given in App.~\ref{secderlog}. We have
$\lambda_\pi=0$ since it is proportional to the derivative of $M^2_\pi$ with
respect to  $m_s$ in the limit $m\to 0$.

\begin{table}
\begin{center}
\begin{tabular}{|r@{.}l|r@{.}lr@{.}lr@{.}l|r@{.}lr@{.}lr@{.}l|r@{.}lr@{.}lr@{.}l|}
\cline{3-20}
\multicolumn{2}{c}{} & \multicolumn{6}{|c}{$r=20$}&\multicolumn{6}{|c}{$r=25$}
   &\multicolumn{6}{|c|}{$r=30$}\\
\hline
\multicolumn{2}{|c}{$X(3)$} & \multicolumn{2}{|c}{$\gamma_\pi$}
  & \multicolumn{2}{c}{$\gamma_K$} & \multicolumn{2}{c|}{$\gamma_\eta$}
  &\multicolumn{2}{|c}{$\gamma_\pi$}
  & \multicolumn{2}{c}{$\gamma_K$} & \multicolumn{2}{c|}{$\gamma_\eta$}
  &\multicolumn{2}{|c}{$\gamma_\pi$}
  & \multicolumn{2}{c}{$\gamma_K$} & \multicolumn{2}{c|}{$\gamma_\eta$}\\
\hline
0&&   0&876 & 0&090 & 0&081 & 0&930 & 0&078 & 0&068 & 0&958 & 0&070 & 0&059\\
0&3 & 0&920 & 0&082 & 0&071 & 0&970 & 0&069 & 0&058 & 0&995 & 0&060 & 0&049\\
0&5 & 0&946 & 0&074 & 0&062 & 0&992 & 0&060 & 0&048 & 1&015 & 0&051 & 0&038\\
0&7 & 0&967 & 0&064 & 0&052 & 1&009 & 0&050 & 0&036 & 1&029 & 0&039 & 0&025\\
0&8 & 0&975 & 0&059 & 0&046 & 1&016 & 0&044 & 0&030 & 1&034 & 0&033 & 0&019\\
0&9 & \multicolumn{2}{c}{-}&
\multicolumn{2}{c}{-} &\multicolumn{2}{c|}{-} & 1&021 & 0&038 & 0&024 & 1&038 & 0&027 & 0&013\\
\hline
0&& 0&892 & 0&080 & 0&072 & 0&943 & 0&070 & 0&060 & 0&970 & 0&063 & 0&053\\
0&3 & 0&941 & 0&072 & 0&062 & 0&990 & 0&061 & 0&050 & 1&011 & 0&053 & 0&041\\
0&5 & 0&965 & 0&063 & 0&051 & 1&009 & 0&051 & 0&038 & 1&029 & 0&042 & 0&029\\
0&7 & 0&980 & 0&052 & 0&039 & 1&020 & 0&038 & 0&025 & 1&037 & 0&029 & 0&015\\
0&8 & 0&983 & 0&046 & 0&032 & 1&022 & 0&032 & 0&018 & 1&037 & 0&022 & 0&009\\
\hline

\end{tabular}
\caption{Logarithmic derivatives with respect to $m$ : $\gamma_\pi$, $\gamma_K$ and $\gamma_\eta$,
as functions of $X(3)$ and $r$, for
$F_0$=85 MeV (upper part) and $F_0$=75 MeV (lower part).}
\label{tablogdeb}
\end{center}
\end{table}

\begin{table}
\begin{center}
\begin{tabular}{|r@{.}l|r@{.}lr@{.}l|r@{.}lr@{.}l|r@{.}lr@{.}l|}
\cline{3-14}
\multicolumn{2}{c}{} & \multicolumn{4}{|c}{$r=20$}&\multicolumn{4}{|c}{$r=25$}
   &\multicolumn{4}{|c|}{$r=30$}\\
\hline
\multicolumn{2}{|c|}{$X(3)$} & 
   \multicolumn{2}{c}{$\lambda_K$} & \multicolumn{2}{c|}{$\lambda_\eta$}&
   \multicolumn{2}{c}{$\lambda_K$} & \multicolumn{2}{c|}{$\lambda_\eta$}&
   \multicolumn{2}{c}{$\lambda_K$} & \multicolumn{2}{c|}{$\lambda_\eta$}\\
\hline
0&& 1&438 & 1&398 & 1&455 & 1&416 & 1&467 & 1&428\\
0&3 & 1&391 & 1&346 & 1&383 & 1&339 & 1&366 & 1&321\\
0&5 & 1&326 & 1&276 & 1&284 & 1&231 & 1&226 & 1&169\\
0&7 & 1&236 & 1&179 & 1&148 & 1&086 & 1&041 & 0&975\\
0&8 & 1&183 & 1&123 & 1&070 & 1&005 & 0&938 & 0&871\\   
0&9 & \multicolumn{2}{c}{-} &
\multicolumn{2}{c|}{-} & 0&986 & 0&921 & 0&833 & 0&768\\
\hline
0&& 1&341 & 1&310 & 1&354 & 1&323 & 1&362 & 1&332\\
0&3 & 1&302 & 1&263 & 1&287 & 1&247 & 1&261 & 1&220\\
0&5 & 1&224 & 1&177 & 1&163 & 1&111 & 1&086 & 1&029\\
0&7 & 1&109 & 1&053 & 0&991 & 0&931 & 0&858 & 0&796\\
0&8 & 1&041 & 0&982 & 0&894 & 0&834 & 0&739 & 0&681\\
\hline
\end{tabular}
\caption{Logarithmic derivatives with respect to $m_s$ : $\lambda_K$ and $\lambda_\eta$,
as functions of $r$ and $X(3)$, for
$F_0$=85 MeV (upper part) and $F_0$=75 MeV (lower part).}
\label{tablogfin}
\end{center}
\end{table}

We would obtain at the one-loop order in the Standard framework:
\begin{eqnarray}
&&\gamma_\pi\sim 1, \qquad \gamma_K \sim \frac{M_\pi^2}{2M_K^2}=0.04,
\qquad \gamma_\eta \sim \frac{M_\pi^2}{3M_\eta^2}=0.02,\label{logstd1}\\
&&\lambda_K \sim 1-\frac{M_\pi^2}{2M_K^2}=0.96, \qquad \lambda_\eta 
  \sim \frac{4M_K^2}{3M_\eta^2}=1.09. \label{logstd2}
\end{eqnarray}
The logarithmic derivatives differ from these values because of the terms quadratic in
the quark masses in the expansions of $F_P^2 M_P^2$. 
The Tables \ref{tablogdeb}-\ref{tablogfin} collect values of these derivatives 
for $r=20,25,30$, and $F_0$=75 MeV and 85 MeV. We note that the values for
$F_0$=85 MeV, $r=25$, $X(3)\sim 0.8$ are in correct agreement with the Standard tree-level
estimates Eqs.~(\ref{logstd1})-(\ref{logstd2}).

We can notice that $\gamma_\pi$ remains close to 1 if we change $F_0$, $r$ and $X(3)$.
$\gamma_\pi$ involves the linear $m$-dependence of the pion mass, which can be written as:
$F_\pi^2M_\pi^2=2m\Sigma(2)+O(m^2)$. Therefore, $\gamma_\pi$ is related to 
the two-flavor quark condensate $\Sigma(2)$, which is very weakly dependent on the values of 
$F_0$ and $X(3)$ [it is affected by the value of $r$, but our tables show only large values of
$r$ where $X(2)$ does not strongly vary]. On the other hand, $\gamma_K$ and $\gamma_\eta$ 
are rather sensitive to $X(3)$. These two
logarithmic derivatives are $1/r$-suppressed when the three-flavor quark condensate is
large. If $X(3)$ decreases, these 2 quantities feel strongly the presence of large
higher-order contribution.

$\lambda_K$ and $\lambda_\eta$ increase from 1 to 1.4 when $X(3)$ decreases down to 0.
If $X(3)$ vanishes, the pseudoscalar masses are dominated by terms quadratic in $m_s$. In
that case, we would naively expect logarithmic derivatives to be twice as large as for
$X(3)\sim 1$. To understand this discrepancy, it is useful to consider the
second kind of logarithmic derivatives arising in Eq.~(\ref{diffgor}):
\begin{equation}
\bar\lambda_P=\frac{m_s}{\bar{M}^2_P}\frac{\partial \bar{M}_P^2}{\partial m_s}
 =\frac{M_P^2}{\bar{M}^2_P}\lambda_P,\qquad P=K,\eta
\end{equation}
$\lambda_P$ and $\bar\lambda_P$ are related, but the first is suppressed with respect 
to the latter by a factor $\bar{M}^2_P/M_P^2$. For instance, in the limit of a
vanishing three-flavor condensate, the expansion of
$\bar{F}_P^2\bar{M}_P^2$ begins with terms quadratic in $m_s$, so that
$\bar\lambda_P$ is of order 2, whereas
$\lambda_P=\bar\lambda_P \cdot \bar{M}^2_P/M_P^2$ is suppressed, and
reaches lower values around 1.4-1.5.

To obtain the values of Tables~\ref{tablogdeb}-\ref{tablogfin}, 
we had to neglect the remainders of higher order
$\epsilon_P$ and $\delta_P$. It is difficult to estimate the size of the resulting
errors for the logarithmic derivatives $\gamma_P$ and $\lambda_P$. Suppose for instance
that $\epsilon_P/F_P^2$ and $\delta_P/M_P^2$ are smaller than 10 \%. To know
the impact on $\gamma_P$ and $\lambda_P$, we would have to calculate the values of the
derivatives of $\epsilon_P$ and $\delta_P$ with respect to $m$ and $m_s$. 
If we know only the size of $\epsilon_P/F_P^2$ and $\delta_P/M_P^2$, 
it is hard to get any information about their derivatives, and  to estimate the resulting
errors on the logarithmic derivatives of the pseudoscalar masses.

\begin{figure}
\begin{center}
\includegraphics[width=11cm]{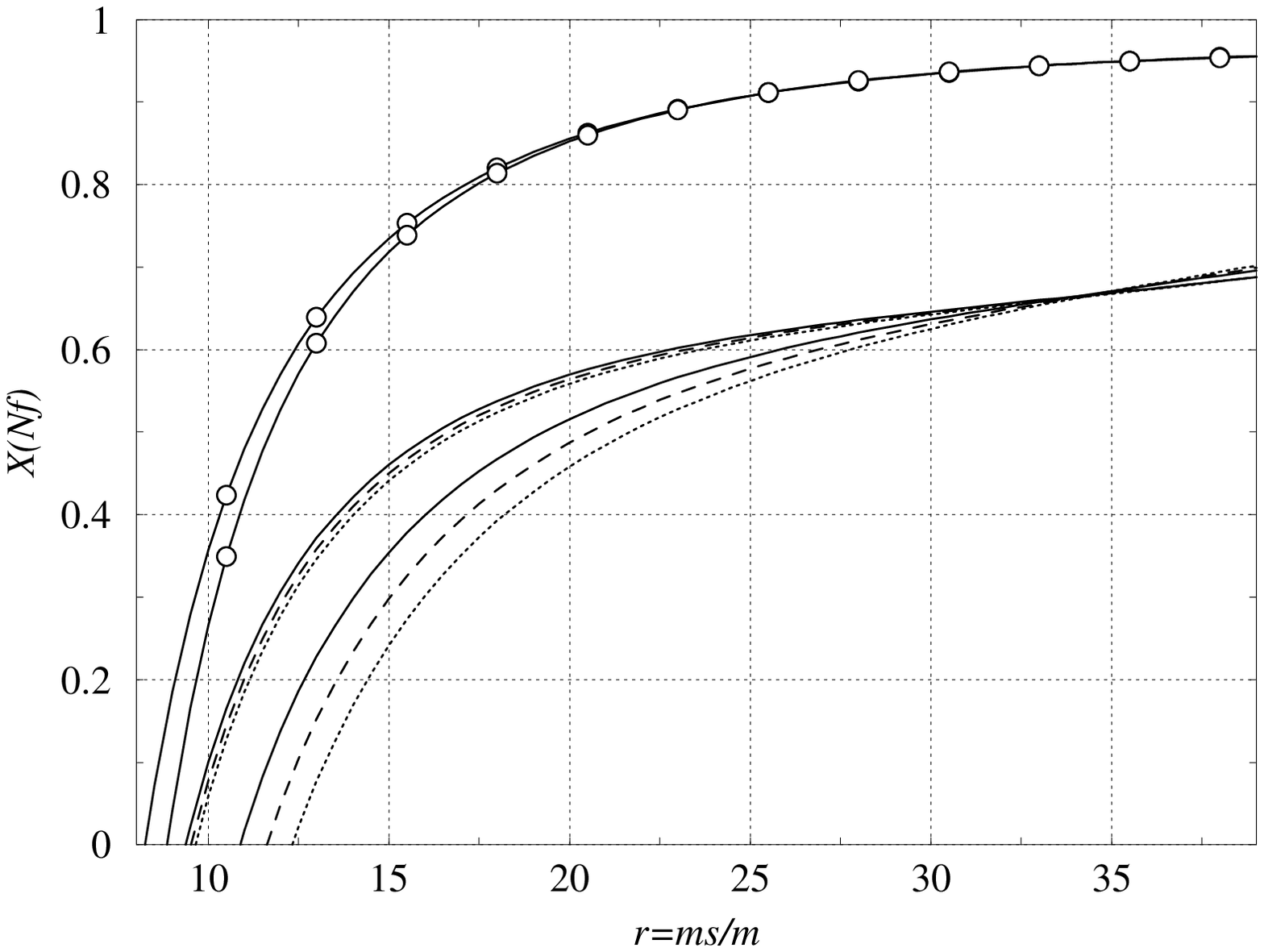}
\includegraphics[width=11cm]{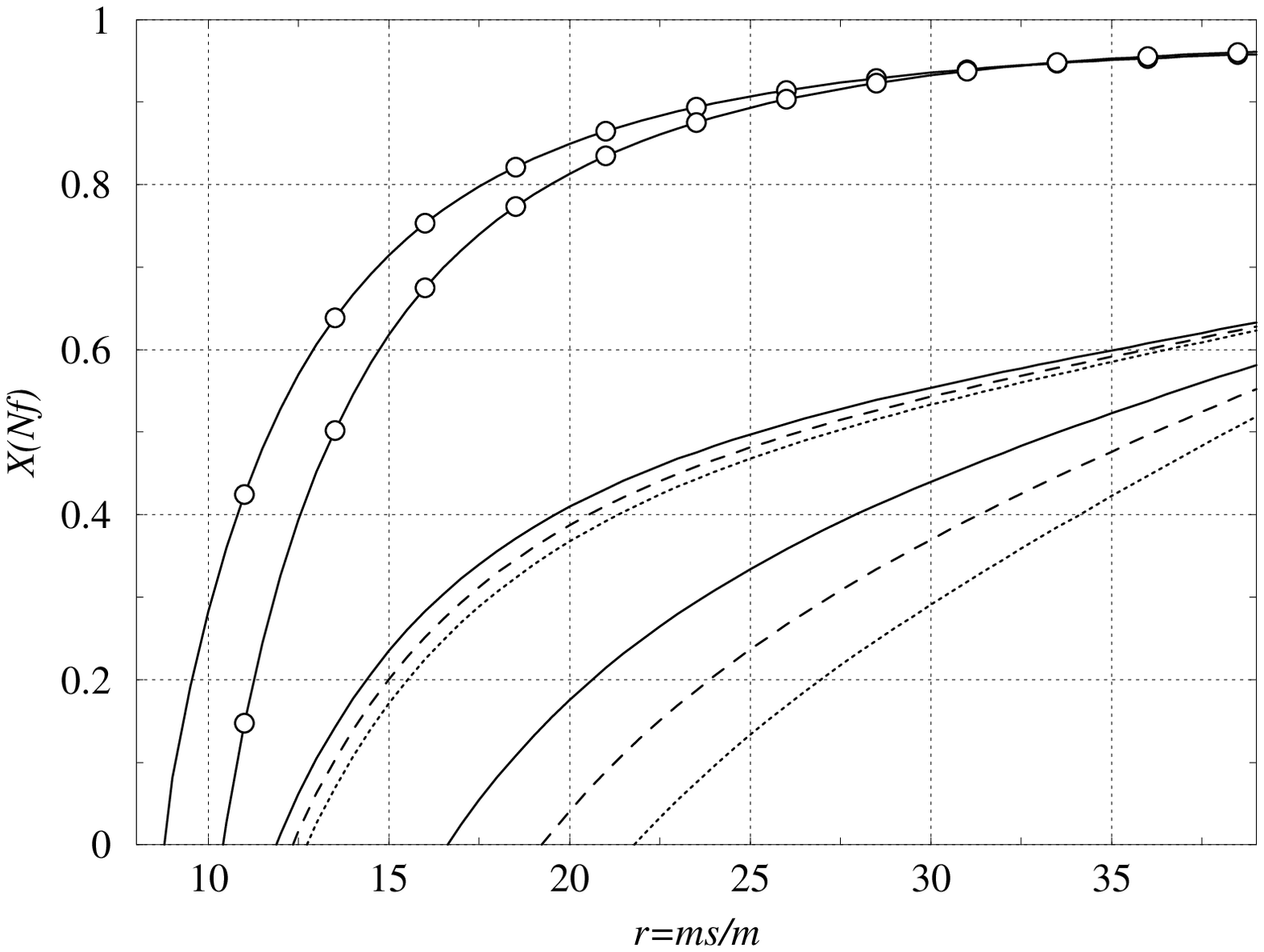}
\caption{Sum rule : Range for $X(3)$ as a function of
$r=m_s/m$ for $F_0$=85 MeV, with the $T$-matrix models of
Refs.~\cite{Oset} (up) and \cite{Au} (down). The results are plotted
for $s_1$=1.2 GeV and $s_0$=1.5
GeV (solid lines), 1.6 GeV (dashed lines) and 1.7 GeV (dotted lines).
The lines with white circles show the corresponding range for $X(2)$.} 
\label{xnf85-1}
\end{center}
\end{figure}

\begin{figure}[t]
\begin{center}
\includegraphics[width=11cm]{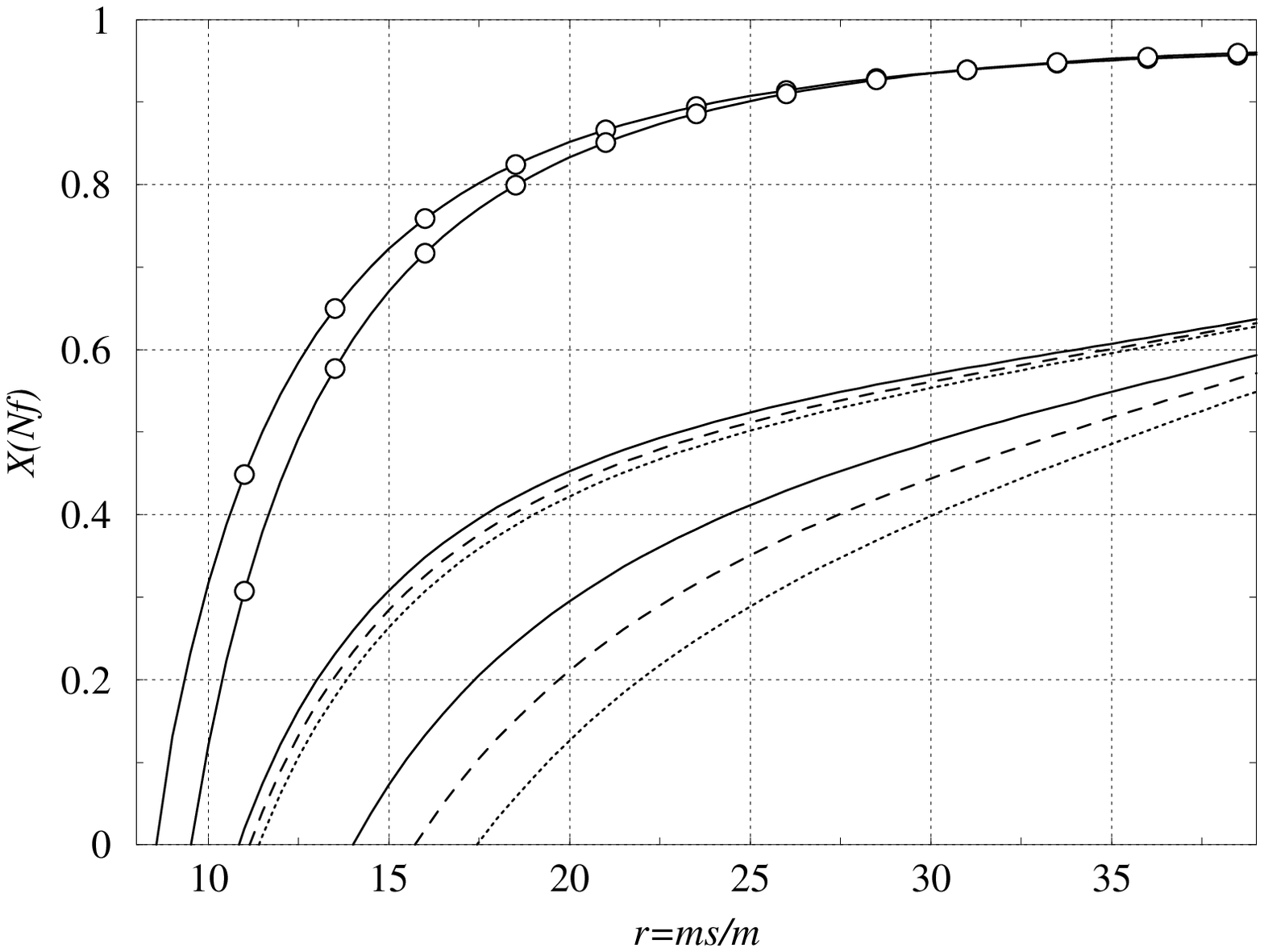}
\caption{Sum rule : Range for $X(3)$ as a function of
$r=m_s/m$ for $F_0$=85 MeV, with the $T$-matrix models of
Ref.~\cite{Loiseau}. The results are plotted
for $s_1$=1.2 GeV and $s_0$=1.5
GeV (solid lines), 1.6 GeV (dashed lines) and 1.7 GeV (dotted lines).
The lines with white circles show the corresponding range for $X(2)$.} 
\label{xnf85-2}
\end{center}
\end{figure}

\begin{figure}
\begin{center}
\includegraphics[width=11cm]{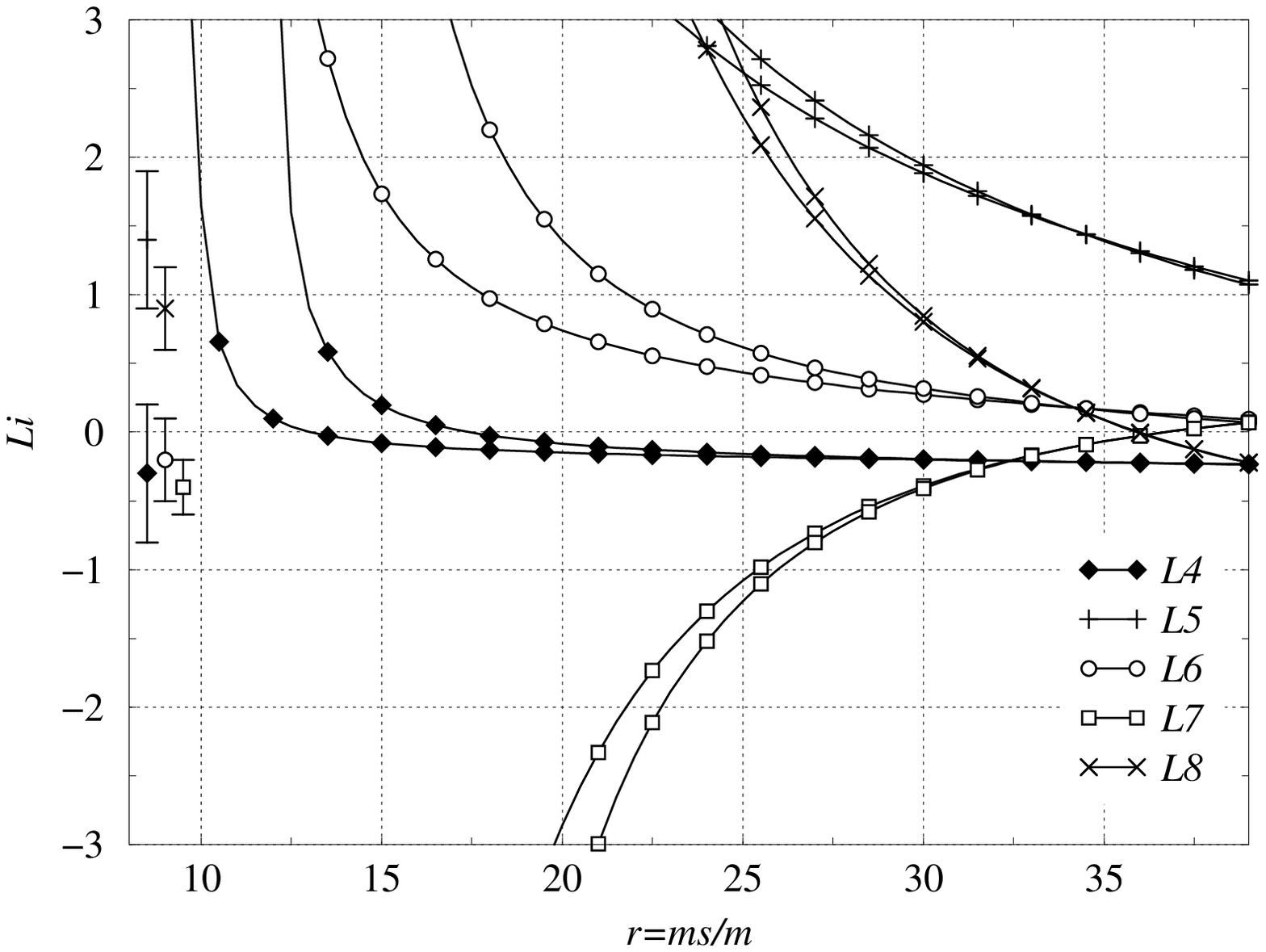}
\includegraphics[width=11cm]{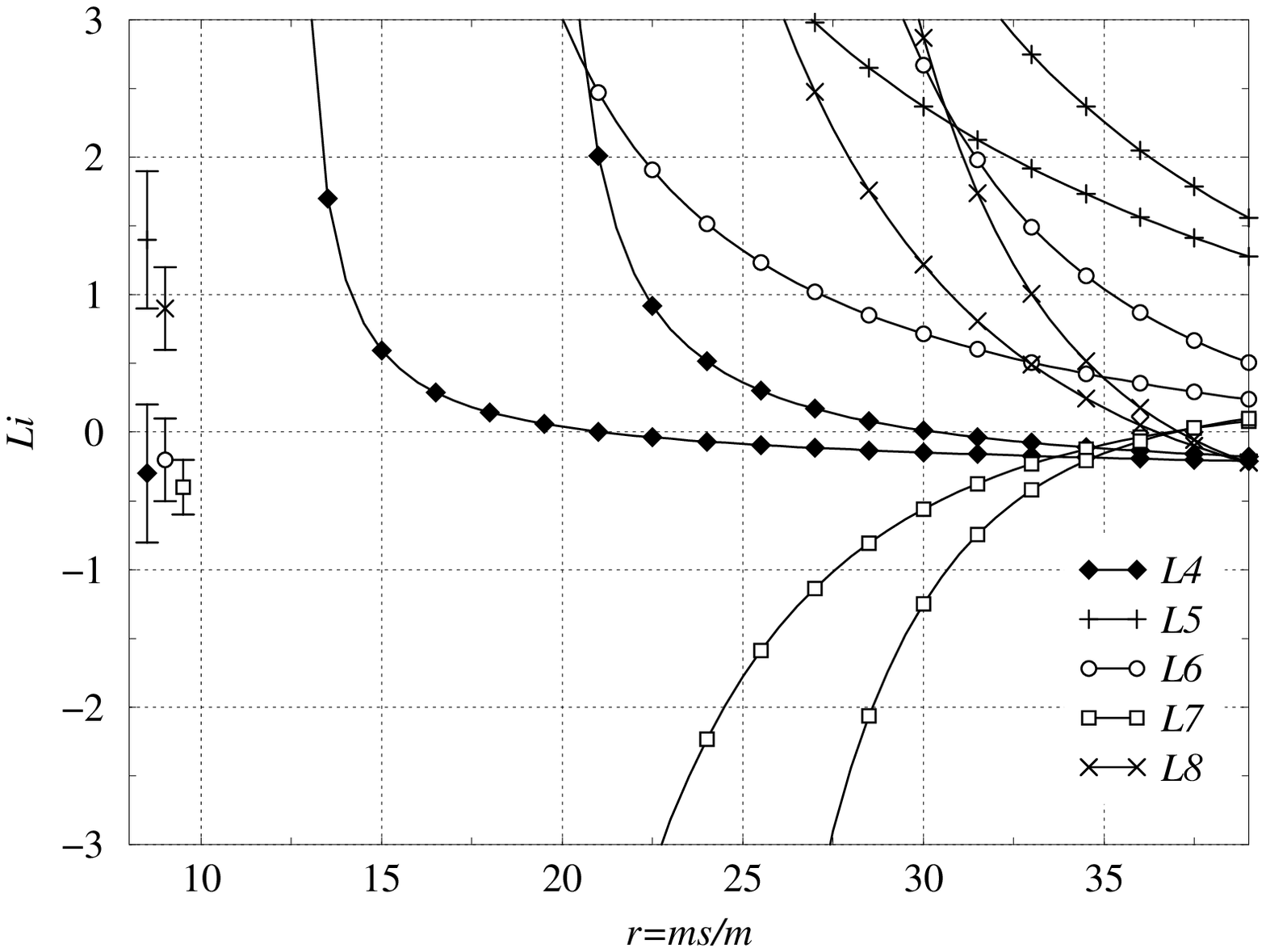}
\caption{Sum rule : Low-energy constants $L_{i=4\ldots 8}(M_\rho)\cdot 10^3$
as functions of $r=m_s/m$ for $F_0$=85 MeV, $s_1$=1.2 GeV and $s_0$=1.6 GeV,
with the $T$-matrix models of Refs.~\cite{Oset} (up) and \cite{Au} (down).
The values plotted on the left, along the vertical axis, are the Standard estimates
stemming from Ref.~\cite{daphne}.}
\label{lec85-1}
\end{center}
\end{figure}

\begin{figure}[t]
\begin{center}
\includegraphics[width=11cm]{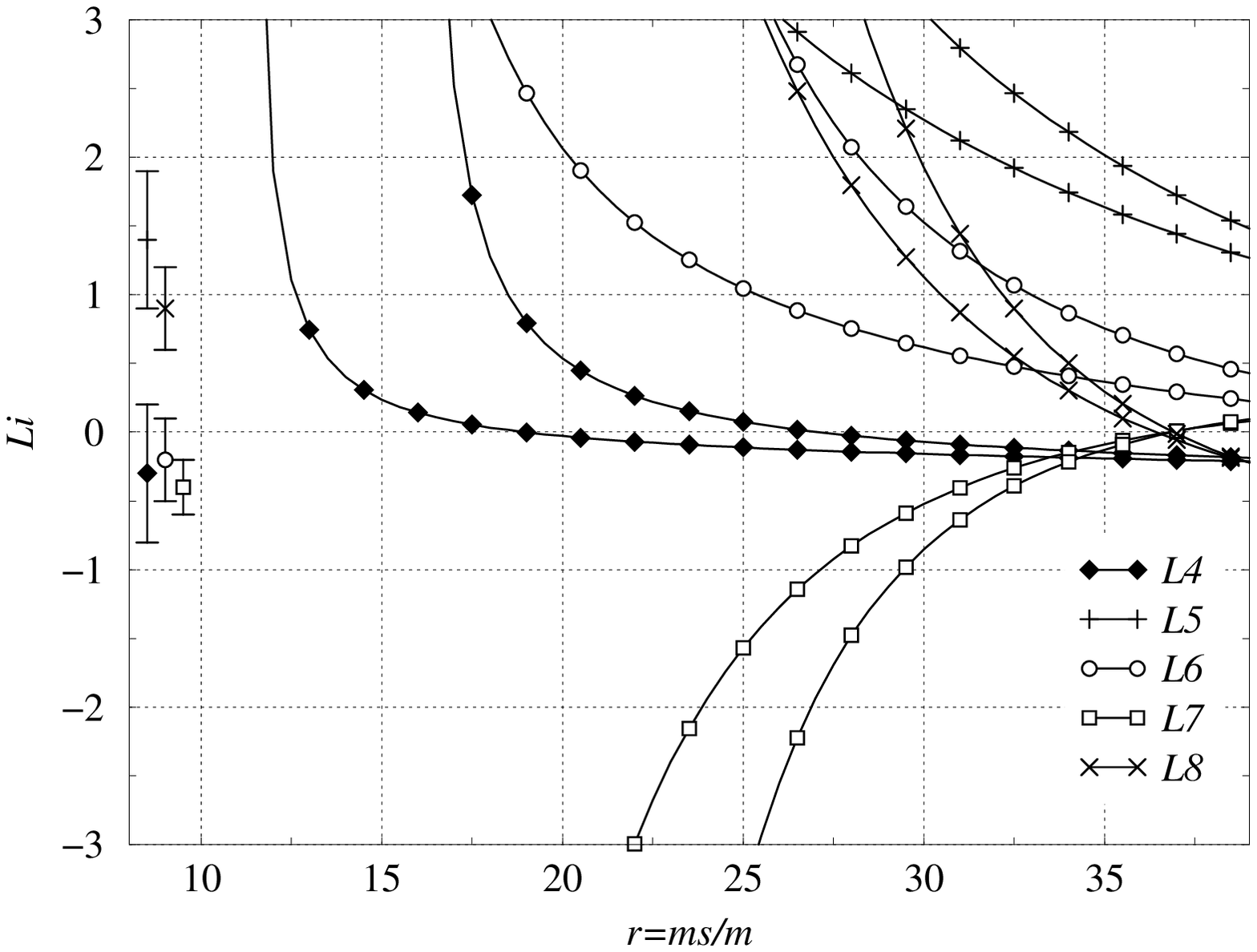}
\caption{Sum rule : Low-energy constants $L_{i=4\ldots 8}(M_\rho)\cdot 10^3$
as functions of $r=m_s/m$ for $F_0$=85 MeV, $s_1$=1.2 GeV and $s_0$=1.6 GeV,
with the $T$-matrix models of Refs.~\cite{Loiseau}.
The values plotted on the left, along the vertical axis, are the Standard estimates
stemming from Ref.~\cite{daphne}.}
\label{lec85-2}
\end{center}
\end{figure}

\subsection{Estimate of $X(3)$ and of LEC's}

Hence, two different estimates of $X(2)-X(3)$ are available: the first one is the
relation between $X(2)$ and $X(3)$ [Eq.~(\ref{xtwo})], the second one consists of
the relation between $X(2)-X(3)$ and $\Pi(0)$
[Eq.~(\ref{diffgor})] and the sum rule for $\Pi(0)$ 
[Eq.~(\ref{regsom})]. In both cases, the difference $X(2)-X(3)$ can be expressed as a function
of the observables and of $F_0,r,X(3)$. This overdertermination can be viewed
as a constraint fixing $X(3)$ in terms of $r$ and $F_0$, see
Figs.~\ref{xnf85-1}-\ref{xnf85-2}.

This analysis contains 3 sources of errors. {\it i)} First, we have neglected
NNLO remainders in the expansions of pseudoscalar masses and decay constants.
Their effect is easy to control in the relations between $X(2)$ and $X(3)$ 
[Eq.~(\ref{xtwo})] or $F(2)$ and $F(3)$ [Eq.~(\ref{ftwo})], but the situation
gets more complicated for the sum rule Eq.~(\ref{diffgor}) and for the logarithmic 
derivatives $\lambda_P$ and $\gamma_P$. In the framework of Standard $\chi$PT,
the authors of Ref.~\cite{abt} noticed that the dependence on $m_s$ of 
$\Sigma(2)$ is not really affected by two-loop effects. In addition, these effects
have the same sign as one-loop contributions: if they were significant, they would
increase (and not decrease) the gap between $X(2)$ and $X(3)$. A similar conclusion
was drawn in Ref.~\cite{Bachir2}. The NNLO remainders are supposed here to be small,
and they are not included in the results.

{\it ii)}
The evaluation of the sum rule Eq.~(\ref{regsom}) relies on an estimate of the integral
between $s_1$ and $s_0$. If we choose a couple $(F_0,r)$, we will not end up
with one value for $X(3)$, but rather a range of acceptable values that will also
depend on the separators $s_1<s_0$. On the Figs.~\ref{xnf85-1}-\ref{xnf85-2}, 
the upper bound for $X(3)$ remains stable for $\sqrt{s_0}>1.5$ GeV, whereas the
lower bound depends strongly on $s_0$. When $s_0$ increases, the lower bound
of Eq.~(\ref{estimate}) is too loose to be saturated. A more stringent lower bound
would be welcome.

{\it iii)}
The third source of error is the $T$-matrix used to build the spectral function
Eq.~(\ref{foncspec}) for $s<s_1$. Three different models of $T$-matrix have been
used \cite{Au,Loiseau,Oset}. The central element is the shape of the $f_0(980)$
peak. Ref.~\cite{Oset} leads to the least pronounced effect. The two other
models \cite{Au,Loiseau} lead to a higher $f_0(980)$ peak, a larger value for
$\Pi(0)$, and a smaller value for $X(3)$.

The range for $X(3)$ is much narrower for large values of $r$, and
can be even reduced to one value in the case of Ref.~\cite{Oset}. 
This range should be broadened if we took into account the errors related to higher orders 
in the expansion of pseudoscalar masses and decay constants. 
The value of $F_0$ has no major influence on the constraint for
$[X(3),r]$. For instance, choosing $F_0$=75 MeV would slightly shift the curves
for $X(3)$ towards the left of the graphs ($r\to r-2$).
Similarly, a change of
$\sqrt{s_1}$ around 1.2 GeV does not affect strongly the results. If we choose $\sqrt{s_1}$=1.3 GeV,
the convergence of the upper bound is slightly less good, but its values
remain very close to Figs.~\ref{xnf85-1}-\ref{xnf85-2}. We should add
a last comment for $r\sim 25$ (commonly used in the Standard framework): the
values of $X(3)$ correspond then to the half of $X(2)$. We end up with a similar
result to the one obtained in 
Refs.~\cite{Bachir1,Bachir2}, but without relying on the hypothesis $X(3) \sim 1$.

The results of the sum rule for $X(3)$ can be converted into bounds
for $L_{i=4\ldots 8}$, plotted on Figs.~\ref{lec85-1}-\ref{lec85-2} as functions of $r$,
for $\sqrt{s_1}=$ 1.2 GeV, $\sqrt{s_0}=$ 1.6 GeV, and $F_0=$ 85 MeV.
For small $r$, the LEC's reach very large values: their definition
from the low-energy behaviour of QCD correlators
includes $1/B_0$ factors that make them diverge when $X(3)\to 0$. We notice also
the large values of
$L_5$, $L_7$ and $L_8$ for $r \sim 25$, because the sum rule pushes
$L_6(M_\rho)$ towards slightly \emph{positive} values. 
The value of $L_4$ is not predicted by the sum rule: it depends essentially on
the value fixed for $F_0$. For instance, choosing $F_0$=75 MeV would yield a slightly positive
value for $L_4$ as $r$ becomes large. 

We have plotted on the left side of Figs.~\ref{lec85-1}-\ref{lec85-2} the values of the
LEC's of Ref.~\cite{daphne}, which were derived assuming that $X(3)$ is of order 1 and
$L_4(M_\eta)=L_6(M_\eta)=0$. Let us remind that the values
obtained for $L_5$, $L_7$, and $L_8$ are strongly dependent on these assumptions.
The values of the LEC's of Ref.~\cite{daphne} hardly agree with the ones obtained 
from the sum rule, because the latter leads to positive values of $L_6(M_\rho)$ and to
a small three-flavor condensate.

\subsection{Slope of the strange scalar form factor of the pion}

Additional information about the decay constants is provided by the 
scalar form factors through a low-energy theorem. Consider the correlator:
\begin{equation}
D^{ij}_{\mu\nu}(p,q)=
  \lim_{m\to 0} \int\!\!\!\int e^{i(p\cdot x-q\cdot y)}
    \langle 0|T\{A^i_\mu(x)\ A^j_\nu(0)\ \bar{s}s(y)| 0 \rangle^{(c)},
\end{equation}
where $i,j=1\ldots 3$, and $(c)$ denotes the connected part of
$(A^i_\mu A^j_\nu)(\bar{s}s)$.
The Ward identities $p^\mu D^{ij}_{\mu\nu}= r^\nu D^{ij}_{\mu\nu}=0$
(with $r=q-p$) yield the Lorentz decomposition:
\begin{eqnarray}
D^{ij}_{\mu\nu}&=&m_s \delta_{ij} 
  \big\{K [r^2 p_\mu p_\nu-(p\cdot r) p_\mu r_\nu 
                 + p^2 r_\mu r_\nu-p^2r^2 g_{\mu\nu}]\nonumber\\
&&\qquad\qquad + L [r_\mu p_\nu -(p\cdot r) g_{\mu\nu}]\big\},
\end{eqnarray}
where $K$ and $L$ are scalar functions of $p^2$, $q^2$ and $r^2$.

On the one hand, we have:
\begin{equation}
D^{ij}_{\mu\nu}(p,0)
  =\frac{\partial}{\partial m_s}
     \left[i\int d^4x e^{ip\cdot x}
     \langle 0|T\{A^i_\mu(x)\ A^j_\nu(0)\}| 0 \rangle \right].
\end{equation}
The correlator $A^i_\mu A^j_\nu$ admits the following decomposition:
\begin{equation}
i\int d^4x e^{ip\cdot x}
     \langle 0|T\{A^i_\mu(x)\ A^j_\nu(0)\}| 0 \rangle
=\delta^{ij} 
  [p_\mu p_\nu-p^2 g_{\mu\nu}] \Phi(p^2).
\end{equation}
$\Phi(p^2)$ contains a pole at $p^2=0$ stemming from one-pion states:
\begin{equation}
\Phi(p^2)=-\frac{F^2(2)}{p^2}+\ldots 
\end{equation}
where the dots denote contributions from the other states. We have therefore:
\begin{equation}
D^{ij}_{\mu\nu}(p,0)=-\delta^{ij} [p_\mu p_\nu -g_{\mu\nu} p^2]
  \left\{\frac{1}{p^2} \frac{\partial F^2(2)}{\partial m_s}
   +\ldots\right\}. \label{T1}
\end{equation}

On the other hand, $T^{ij}_{\mu\nu}$ is dominated at low energy by the exchange of
two pions between $\bar{s}s$ and each of the axial currents:
\begin{equation}
D^{ij}_{\mu\nu}
  =2 \delta^{ij} F^2(2) G_1(q^2) \frac{p_\mu r_\nu} {p^2 r^2 (p\cdot r)}
   +\ldots  
\end{equation}
which contributes to $K$:
\begin{equation}
K(p^2,q^2,r^2)=-2 F^2(2) \frac{G_1(q^2)}{p^2 r^2 (p\cdot r)}+\ldots\label{T2}
\end{equation}
whereas $L(p^2,q^2,r^2)$ receives no contribution.
We compare Eqs.~(\ref{T1}) and (\ref{T2}) for $p, q \to 0$ to obtain:
\begin{equation} \label{lowenergyth}
\frac{\partial F^2(2)}{\partial m_s}
  =2 F^2(2) \lim_{q^2 \to 0} \frac{G_1(q^2)}{q^2}
 \qquad \frac{m_s}{F(2)} \frac{\partial F(2)}{\partial m_s}=m_s G'_1(0).
\end{equation}
This low-energy theorem \cite{Bachir1,Bachir2,Donoghue} provides a relation
between the logarithmic derivative of $F(2)$ with respect to 
$m_s$, and the slope of the strange scalar form factor of the pion for a vanishing
momentum.

We can now exploit the solutions of Omn\`es-Muskhelishvili equations.
According to Eq.~(\ref{declin}), we get for the slope of the form factor:
\begin{equation}
m_s G'_1(0)=m_s G_2(0) B'_1(0)
 =m_s \frac{\partial \bar{M}_K^2}{\partial m_s} B'_1(0)
 =\lambda_K M_K^2 B'_1(0).
\end{equation}
$B'_1(0)$ is computed by taking the derivative with respect to s at
0 of Eq.~(\ref{omeq}):
\begin{equation}\label{bprime1}
B'_1(0)=\frac{1}{\pi}\sum_{j=1}^2\int_{4M_\pi^2}^\infty ds' \frac{1}{s'^2}
   T_{1j}^*(s') \sqrt\frac{s'-4M_j^2}{s'} \theta(s'-4M_j^2) B_j(s').
\end{equation}
The numerical resolution of Omn\`es-Muskhelishvili equations 
Eq.~(\ref{omeq}) yields the values of
$\vec{B}(s)$ at the points of integration used for the 
Gauss-Legendre quadrature \cite{Bachir1}. Hence, we can compute
directly the integral Eq.~(\ref{bprime1}) by the same integration method.

On the other hand, Eq.~(\ref{fpi}) leads to:
\begin{equation}
\frac{m_s}{2F^2(2)} \frac{\partial F^2(2)}{\partial m_s}
=\frac{1}{\bar{F}_\pi^2}
   \left[m_s \tilde\xi - \frac{rX(3)}{64\pi^2} 
     \frac{F_\pi^2 M_\pi^2}{F_0^2} 
     \left(\log \frac{\bar{M}_K^2}{M_K^2}+\frac{M_K^2}{\bar{M}_K^2}
        \lambda_K \right)\right].
\end{equation}

\begin{figure}
\begin{center}
\includegraphics[width=11cm]{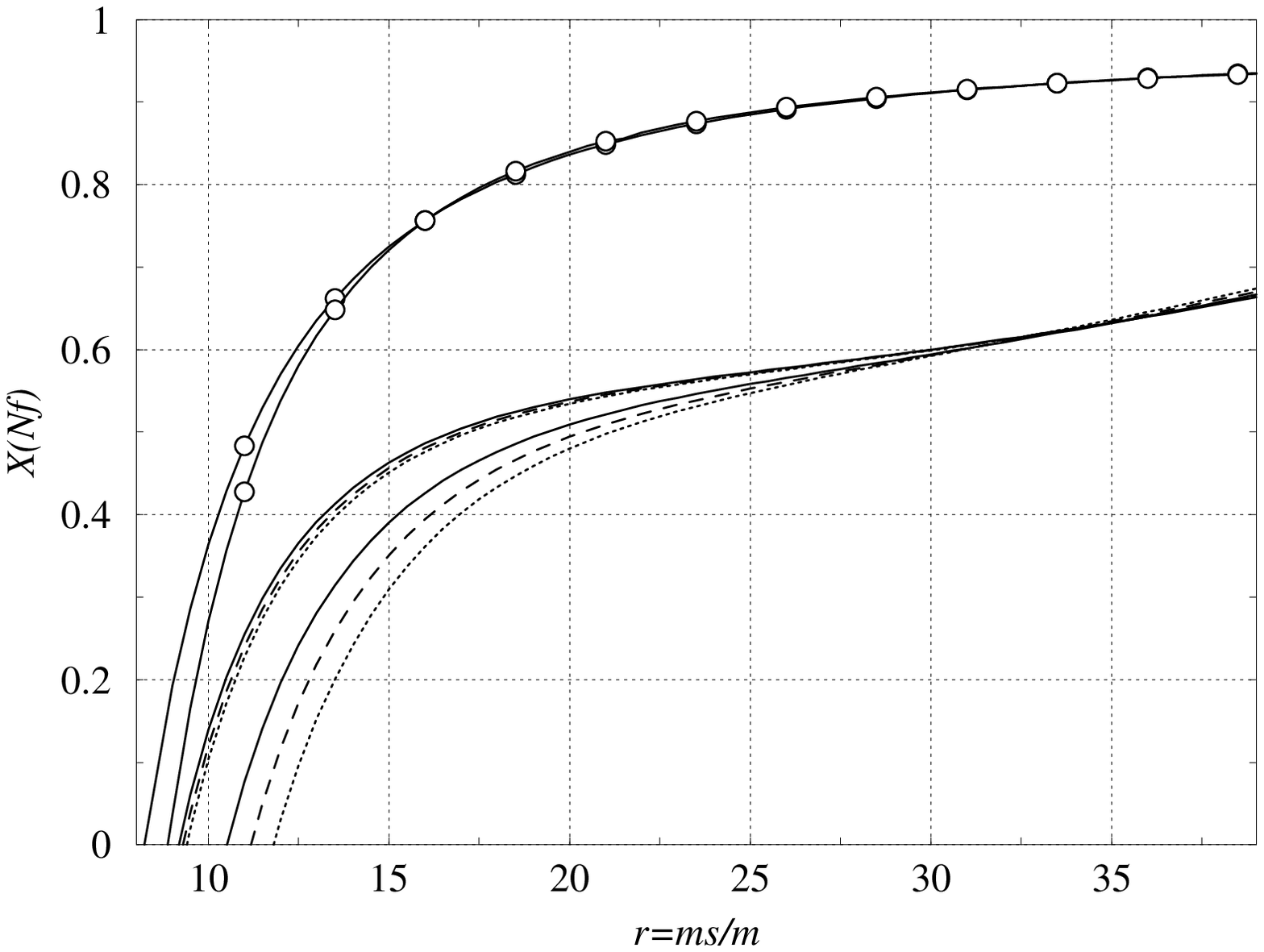}
\includegraphics[width=11cm]{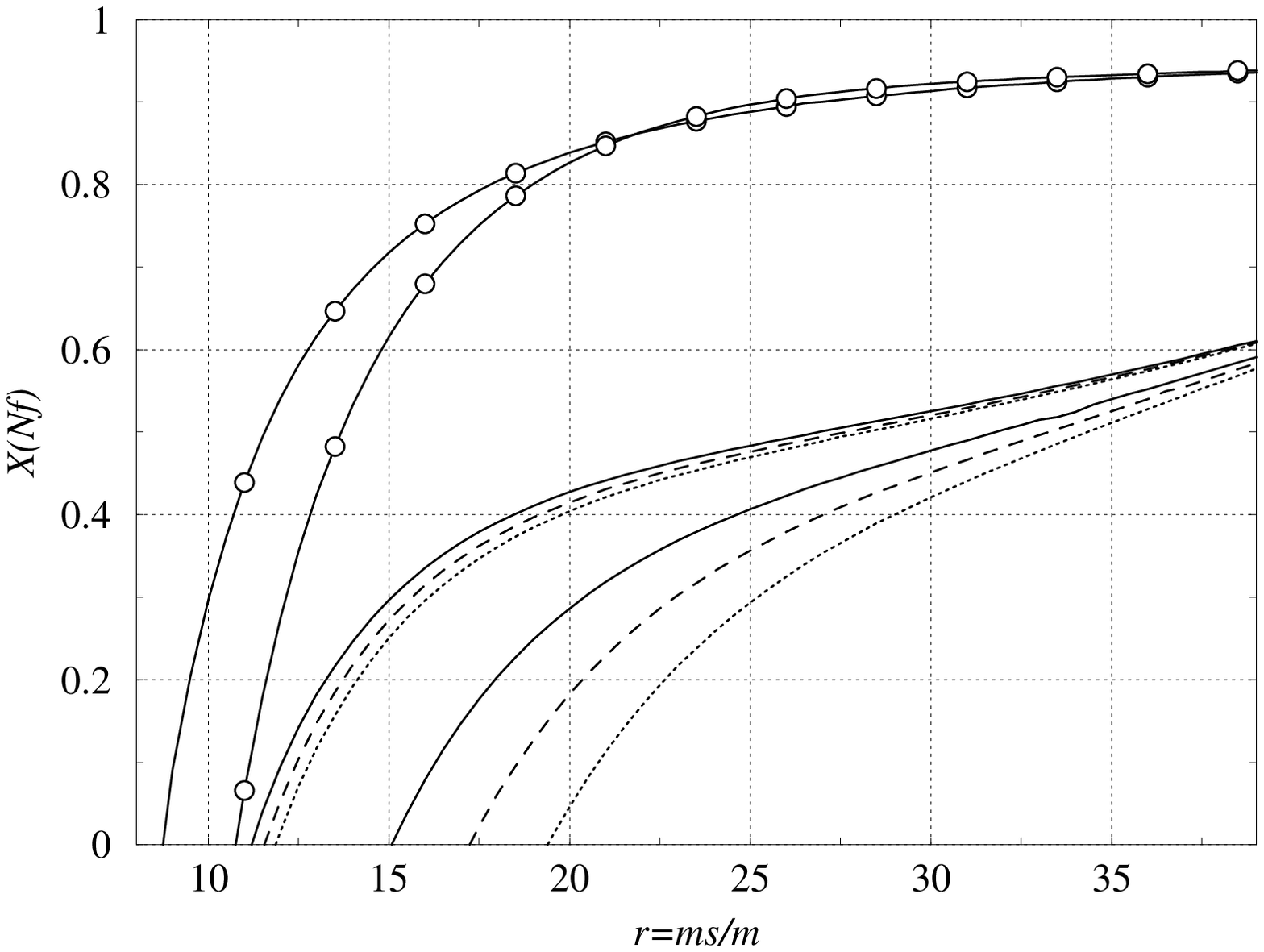}
\caption{Sum rule and slope of the strange form factor of the pion : 
Range for $X(3)$ as a function of
$r=m_s/m$ with the $T$-matrix models of
Refs.~\cite{Oset} (up) and \cite{Au} (down). The results are plotted
for $s_1$=1.2 GeV and $s_0$=1.5
GeV (solid lines), 1.6 GeV (dashed lines) and 1.7 GeV (dotted lines).
The lines with white circles show the corresponding range for $X(2)$.} 
\label{conjx-1}
\end{center}
\end{figure}

\begin{figure}[t]
\begin{center}
\includegraphics[width=11cm]{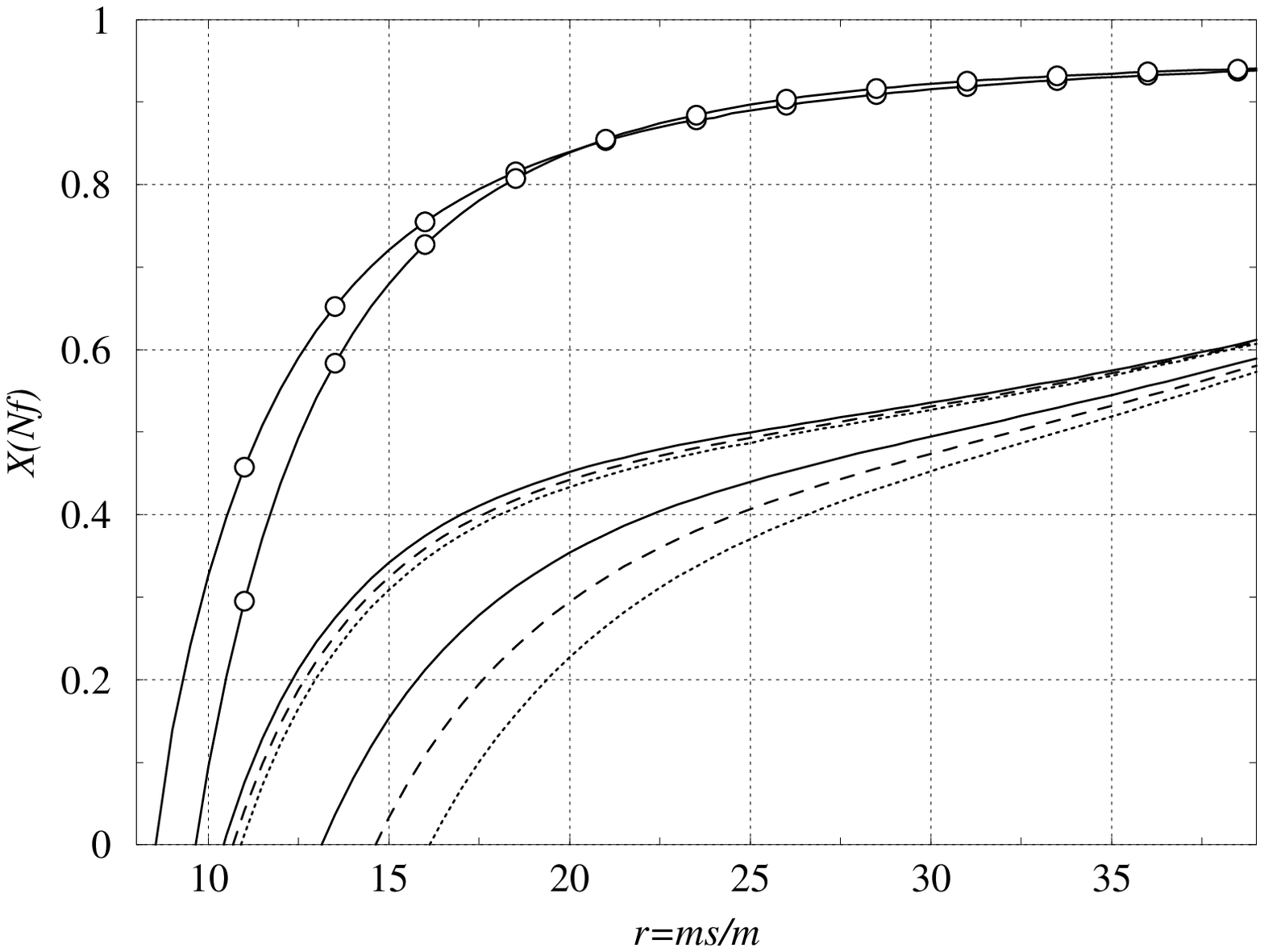}
\caption{Sum rule and slope of the strange form factor of the pion : 
Range for $X(3)$ as a function of
$r=m_s/m$ with the $T$-matrix models of
Ref.~\cite{Loiseau}. The results are plotted
for $s_1$=1.2 GeV and $s_0$=1.5
GeV (solid lines), 1.6 GeV (dashed lines) and 1.7 GeV (dotted lines).
The lines with white circles show the corresponding range for $X(2)$.} 
\label{conjx-2}
\end{center}
\end{figure}

\begin{figure}
\begin{center}
\includegraphics[width=11cm]{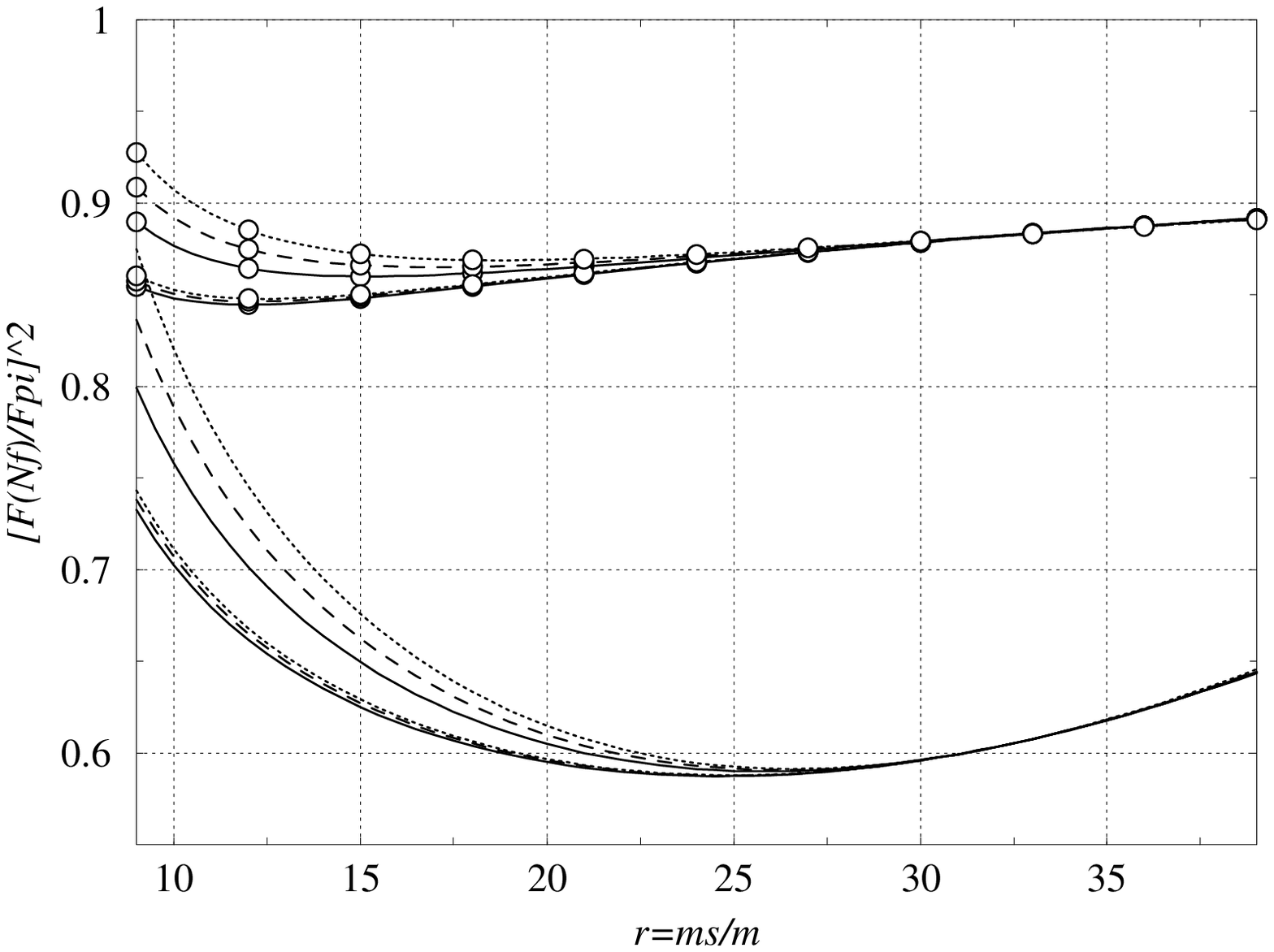}
\includegraphics[width=11cm]{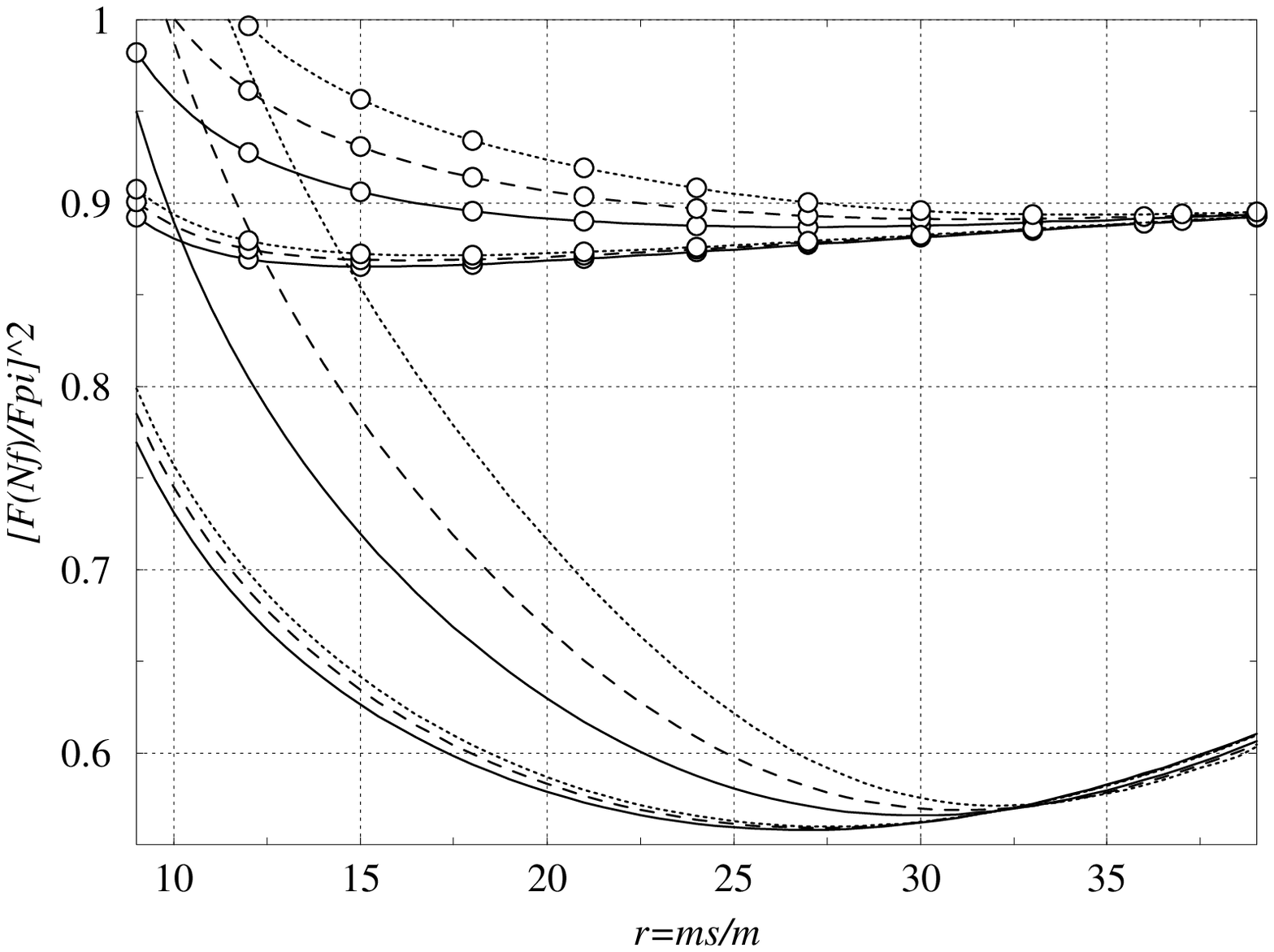}
\caption{Sum rule and slope of the strange form factor of the pion : 
Ranges for $[F(3)/F_\pi]^2$ (no symbol) and $[F(2)/F_\pi]^2$ (white circles)
as functions of $r=m_s/m$ with the $T$-matrix models of
Refs.~\cite{Oset} (up) and \cite{Au} (down). The results are plotted
for $s_1$=1.2 GeV and $s_0$=1.5
GeV (solid lines), 1.6 GeV (dashed lines) and 1.7 GeV (dotted lines).}
\label{conjf-1}
\end{center}
\end{figure}

\begin{figure}[t]
\begin{center}
\includegraphics[width=11cm]{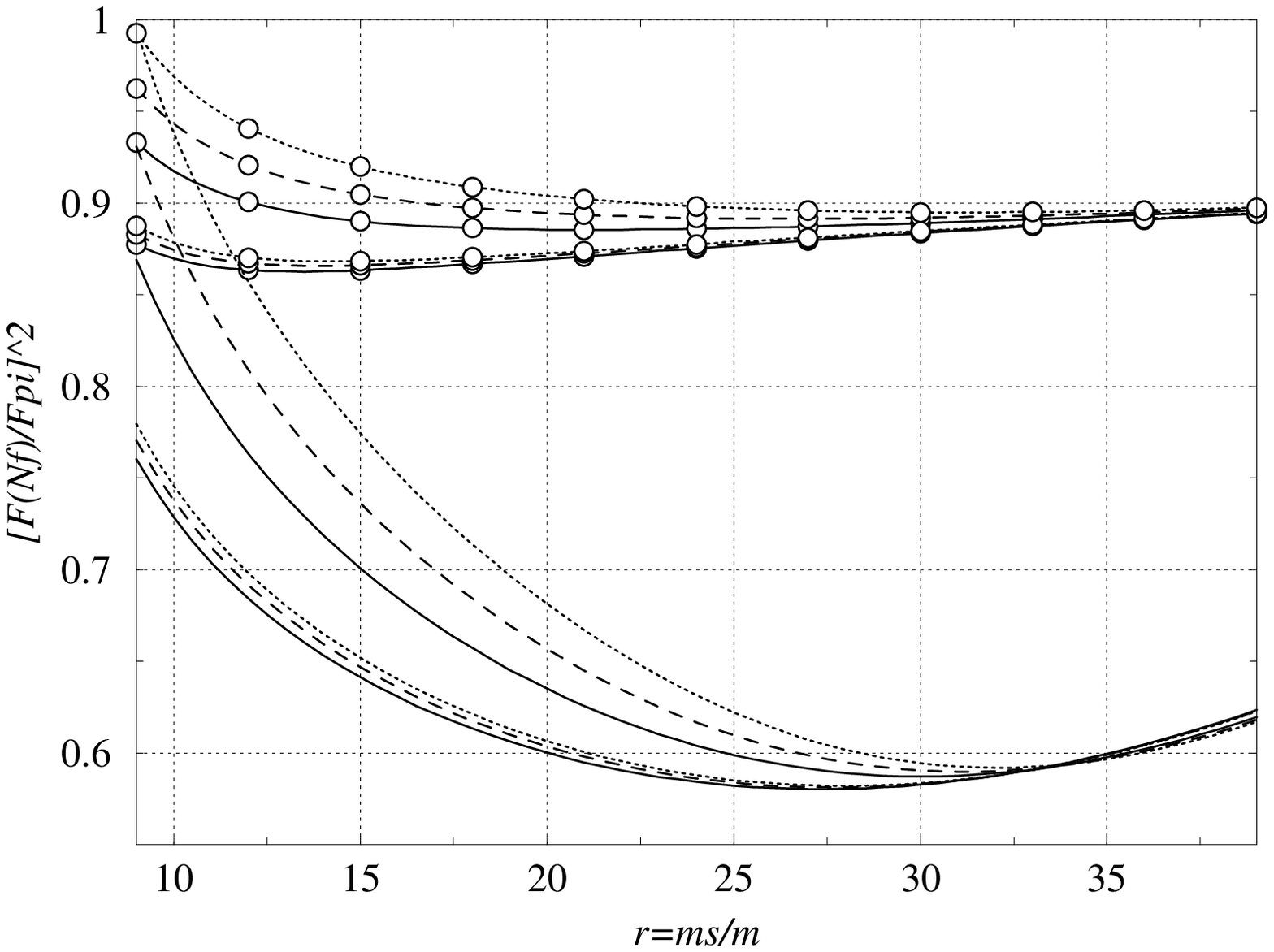}
\caption{Sum rule and slope of the strange form factor of the pion : 
Ranges for $[F(3)/F_\pi]^2$ (no symbol) and $[F(2)/F_\pi]^2$ (white circles)
as functions of $r=m_s/m$ with the $T$-matrix model of
Refs.~\cite{Loiseau}. The results are plotted for $s_1$=1.2 GeV and $s_0$=1.5
GeV (solid lines), 1.6 GeV (dashed lines) and 1.7 GeV (dotted lines).}
\label{conjf-2}
\end{center}
\end{figure}

\begin{figure}
\begin{center}
\includegraphics[width=11cm]{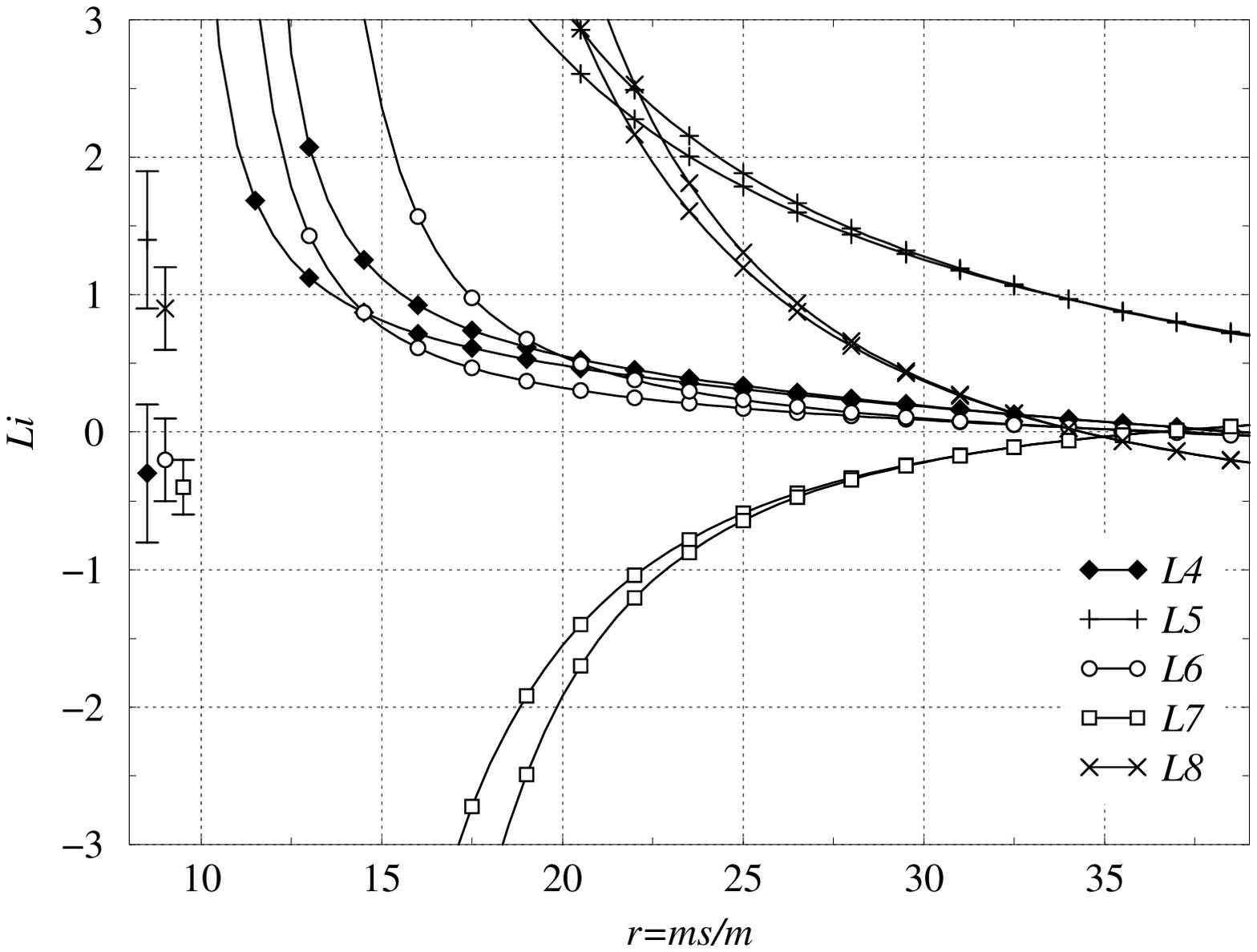}
\includegraphics[width=11cm]{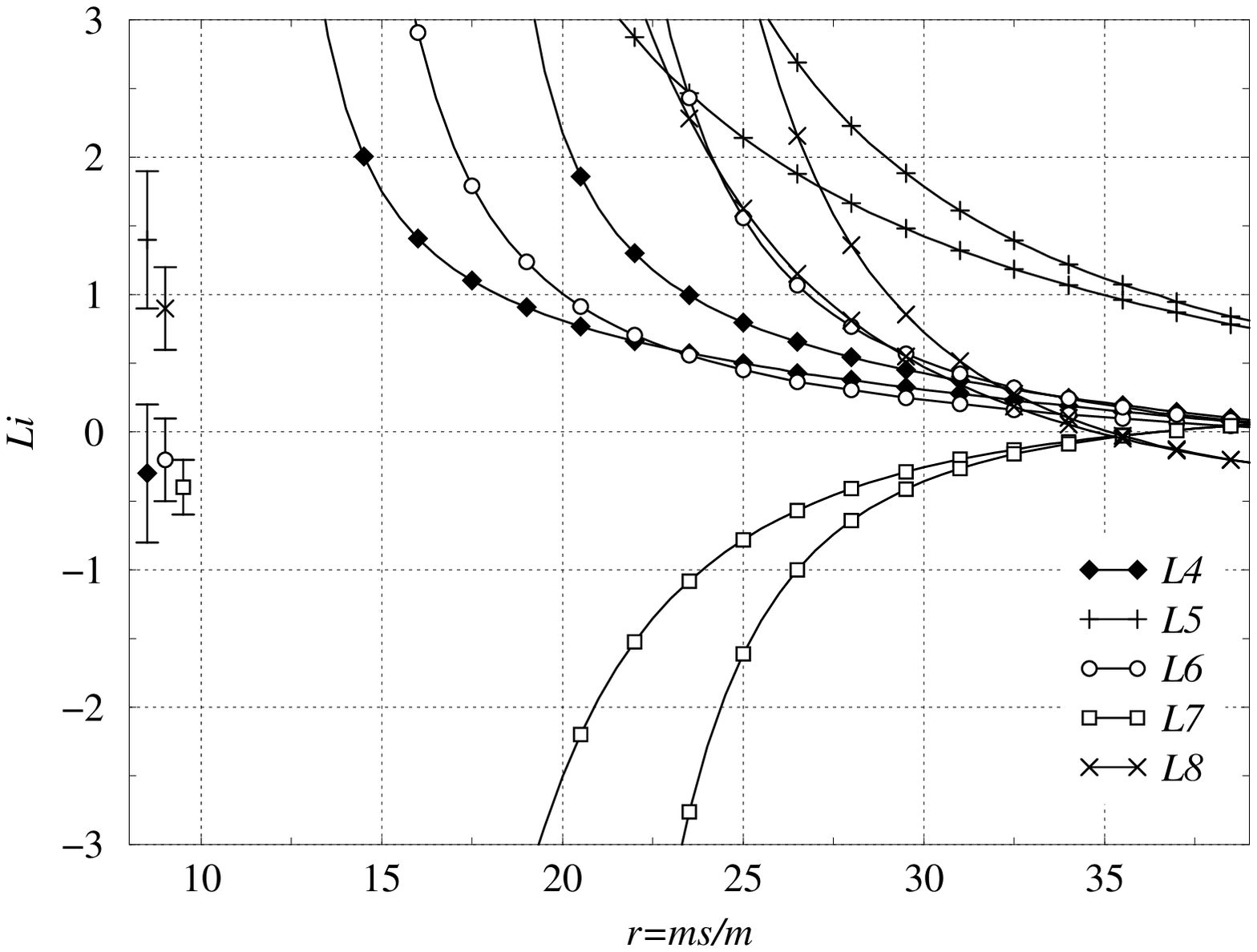}
\caption{Sum rule and slope of the strange form factor of the pion : 
Low-energy constants $L_{i=4\ldots 8}(M_\rho)\cdot 10^3$
as functions of $r=m_s/m$ for $F_0$=85 MeV, $s_1$=1.2 GeV and $s_0$=1.6 GeV,
with the $T$-matrix models of Refs.~\cite{Oset} (up) and \cite{Au} (down).
The values plotted on the left, along the vertical axis, are the Standard estimates
stemming from Ref.~\cite{daphne}.}
\label{conjlec-1}
\end{center}
\end{figure}

\begin{figure}[t]
\begin{center}
\includegraphics[width=11cm]{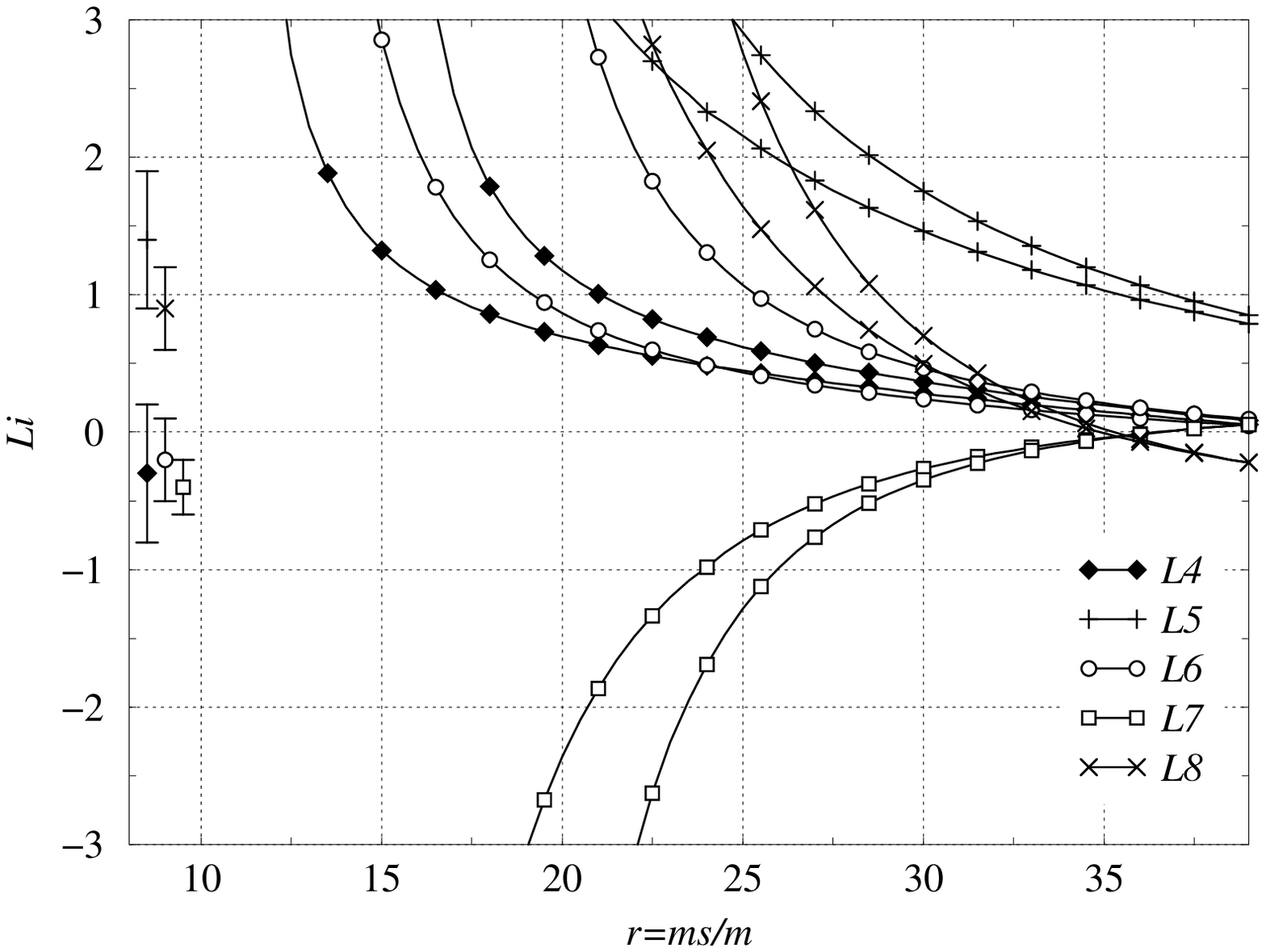}
\caption{Sum rule and slope of the strange form factor of the pion : 
Low-energy constants $L_{i=4\ldots 8}(M_\rho)\cdot 10^3$
as functions of $r=m_s/m$ for $F_0$=85 MeV, $s_1$=1.2 GeV and $s_0$=1.6 GeV,
with the $T$-matrix model of Ref.~\cite{Loiseau}.
The values plotted on the left, along the vertical axis, are the Standard estimates
stemming from Ref.~\cite{daphne}.}
\label{conjlec-2}
\end{center}
\end{figure}

We see that Eq.~(\ref{lowenergyth}) is an additional constraint, different from
the sum rule
Eq.~(\ref{regsom}). From the analysis of the pseudoscalar spectrum, we have concluded
that all the quantities could be expressed (at the NLO) as functions of masses,
decay constants, and 3 parameters $F_0, r, X(3)$. The sum rule
was a first constraint, fixing a range for $X(3)$ depending on 
$r$ and $F_0\equiv F(3)$. If we exploit the second constraint
Eq.~(\ref{lowenergyth}), we can obtain ranges for
$F(3)$ and $X(3)$ as functions of $r$, plotted respectively in 
Figs.~\ref{conjx-1}-\ref{conjx-2} and
Figs.~\ref{conjf-1}-\ref{conjf-2}. These results can also be converted into
values for the low-energy constants (see
Figs.~\ref{conjlec-1} and \ref{conjlec-2}).

The values of $X(3)$ are close to the ones obtained
by the only application of the sum rule Eq.~(\ref{regsom}). 
The results obtained then for $X(3)$
were not very sensitive to the valued chosen for $F_0$. We 
see also that the slope of the strange scalar form factor of the pion 
leads to rather small values for $F(3)\equiv F_0$ (around 70 MeV) for $r\sim 25$.
This result is in agreement with the small \emph{positive} values obtained for 
$L_4(M_\rho)$. This increase of $L_4$ (with respect to the previous analysis)
comes with a decrease of $L_5$.
In Ref.~\cite{Donoghue}, the analysis of the same form factor led to
$d_F=m_s/F(2) \cdot \partial\log F(2)/\partial m_s=0.09$.
In the framework of Standard $\chi$PT, such a value corresponds to 
$L_4(M_\rho)\simeq 0.4\cdot 10^{-3}$, i.e.
$F(3)$ of order 75 MeV. This question has also been discussed in 
Refs.~\cite{Bachir1,Bachir2,Meissner}.

However, these two constraints do not demand the same accuracy for the
scalar form factors. The sum rule involves the integral of the spectral function
$\mathrm{Im}\ \Pi$ up to 1.2 GeV, which is dominated by the 
$f_0(980)$ peak. The global shape of the spectral function
(and more precisely around 1 GeV) is the crucial element.
For the low-energy theorem, we are interested in the slope of 
a form factor at zero, i.e. low-energy details. The resulting constraint may 
be less stable than the sum rule. It seemed therefore preferable to split the
analysis in two parts: the first one dealing only with the sum rule, the second
one exploiting both constraints at the same time.

\subsection{Scalar radius of the pion}

The scalar radius of the pion $\langle r^2\rangle^\pi_s$ can also be obtained
from the scalar form factors of the pion, considered out of the chiral
limit [i.e. with the physical masses $m_s$ and $m=(m_u+m_d)/2$]:
\begin{equation}
F_1(s)=F_1(0)\left[1+\frac{1}{6}\langle r^2\rangle^\pi_s s+c_\pi s^2+\ldots
  \right],
\end{equation}
If we project $\vec{F}$ on the two solutions $\vec{A}$ and $\vec{B}$, we obtain:
\begin{equation}\label{rayscal}
\langle r^2\rangle^\pi_s=6\frac{F'_1(0)}{F_1(0)}=6\left[A'_1(0)+
  \frac{M_K^2}{M_\pi^2}\frac{\tilde\gamma_K}{\tilde\gamma_\pi} B'_1(0)\right],
\end{equation}
where a third kind of logarithmic derivatives is involved 
(considered out of the chiral limit):
$\tilde\gamma_P=\partial[\log M_P^2]/\partial[\log m]$.
App.~\ref{tildegamma} collects their expressions in terms of the low-energy constants.

We are interested in a quantity describing
the non-strange pion form factor around the threshold. It should be possible to neglect
the $K\bar{K}$ channel with no major change in the results. This point of view is
supported by a numerical estimate:
$B'_1(0)/A'_1(0)\sim 0.1$ and
$(M_K^2/M_\pi^2)\times (\tilde\gamma_K/\tilde\gamma\pi)\sim 1/2$.
If we restricted our analysis to the $\pi\pi$ channel, only the first term
(the solution $\vec{A}$) would appear on the right side of
Eq.~(\ref{rayscal}). The scalar radius of the pion would be independent of
$r$, $X(3)$ and $F_0$ in that case. Actually,
the second term on the right side of Eq.~(\ref{rayscal}), related to
the $K\bar{K}$ channel, is 
responsible for a weak dependence of $\langle r^2\rangle^\pi_s$
on $r$, $X(3)$, $F_0$. We can use the previous results, where $X(3)$ and $F_0$ are functions
of $r$, in order to study the range of variation for the pion scalar radius:
\begin{center}
\begin{tabular}{llll}
0.537 -- 0.588 $\mathrm{fm}^2$ &\qquad& Oller-Oset-Pelaez & Ref.~\cite{Oset},\\
0.567 -- 0.630 $\mathrm{fm}^2$ &\qquad& Au-Morgan-Pennington &
Ref.~\cite{Au},\\
0.592 -- 0.650 $\mathrm{fm}^2$ &\qquad& Kaminski-Lesniak-Maillet
 & Ref.~\cite{Loiseau},
\end{tabular}
\end{center}
to be compared to the estimates:
$0.6\pm 0.2\  \mathrm{fm}^2$ \cite{pipigl},
$0.55\pm 0.15 \ \mathrm{fm}^2$ \cite{g-l},
0.55 to 0.61 $\mathrm{fm}^2$ \cite{gassmeiss},
0.57 to 0.61 $\mathrm{fm}^2$ \cite{bijcoltal}
and $0.61\pm 0.04 \ \mathrm{fm}^2$ \cite{royeq}. Notice that the matching of
Roy equations with Standard $\chi$PT \cite{royeq} relies strongly on the value
of the scalar radius of the pion.

Information about the scalar radius of the pion could be seen as an additional
constraint on our system, since 
$\langle r^2\rangle^\pi_s$ is related to $2L_4+L_5$. The situation is similar to
$d_F$: this kind of constraint could rather easily be affected by higher-order corrections.
We are also obliged to consider it out of the chiral limit $m\to 0$. It seems therefore
wiser not to use this constraint, 
until a new analysis would treat less crudely NNLO remainders.

\section{Conclusions}

The LEC's of the effective chiral Lagrangian should be determined as accurately as 
possible in order to know and understand the pattern of SB$\chi$S. 
These constants have generally been estimated
from the expansion of Goldstone boson observables in powers of quark masses,
supposing (1) a dominance of the quark condensate and (2) an agreement with the large-$N_c$
picture of QCD. But such determinations of the LEC's could be modified
if quantum fluctuations turned out to be significant. A symptom of large quantum fluctuations
could be seen in the large violation of the Zweig rule in the scalar channel and in
large variations of chiral order parameters (e.g. the quark condensate) 
from $N_f=2$ to $N_f=3$.

First we have studied how the relaxation of the Standard assumptions (1) and (2) could affect
the determination of the LEC's. To reach this goal,
we have studied the expansion in quark masses of the Goldstone boson masses and decay
constants. We have truncated these expansions to keep the first two powers in quark masses
and we have supposed that higher-order remainders [$O(m_\mathrm{quark}^3)$ for $F_P^2M_P^2$
and $O(m_\mathrm{quark}^2)$ for $F_P^2$] are small. These expansions can be written
using ``effective'' scale-independent constants that combine chiral logarithms and
LEC's. $F_P^2M_P^2$ involves $\Sigma(3)$,
$F^2(3)$ and constants related to $L_6$, $L_7$ and $L_8$, whereas
$F_P^2$ is expressed in terms of $F^2(3)$ and constants corresponding to $L_4$ and
$L_5$.

We have not considered these expansions in one-loop Standard $\chi$PT, since the three-flavor
quark condensate $\Sigma(3)$ is not supposed to dominate the expansion of pseudoscalar masses.
We have not worked either in tree-level Generalized $\chi$PT, since we have included
 chiral logarithms.
These relations between LEC's and experimental quantities (masses and decay constants)
can be inverted to express $L_{i=4\ldots 8}$ [and therefore $X(2)$ and $F(2)$] as functions of
$F(3)$, $r=m_s/m$ and $X(3)$. We have then studied a possible competition 
between the first two orders of the quark mass expansions, by admitting large values for
the ZR violating constants $L_4$ and $L_6$ (larger than the expected values on a basis of
large-$N_c$ arguments).

The variation of the Gell-Mann--Oakes--Renner ratio $X$ from $N_f=2$ to
$N_f=3$ is governed by $L_6$. The equality $X(2)=X(3)$ (saturation of the paramagnetic
bound) is realized for $L_6(M_\rho)=-0.21\cdot 10^{-3}$. 
A three-flavor GOR ratio $X(3)$ much smaller than 1 could be obtained for two different reasons.
On the one hand,
the ratio of quark masses $r$ may be smaller than 25 ($r<20$), which leads to small values
of $X(2)$, and then of $X(3)$ [due to the paramagnetic 
inequality $X(2)\geq X(3)$]. On the other hand, $L_6(M_\rho)$ may be larger than the value
$-0.26\cdot 10^{-3}$ saturating the paramagnetic bound for $X$. 
A slight shift of $L_6(M_\rho)$ towards positive values
leads to a significant decrease of $X(3)$, whereas $X(2)$ remains almost constant and unsuppressed.
$X(2)$ could thus be of order 1 and $X(3)$ much smaller than 1, provided 
that $r$ is large ($r\sim 25$) and the Zweig rule is strongly violated for 
the correlator defining $L_6$.

A similar analysis has been performed for the decay constants. $L_4$ tunes 
the difference between $F^2(2)$ and $F^2(3)$: the equality is obtained for
$L_4(M_\rho)=-0.37\cdot 10^{-3}$. If $L_4(M_\rho)$ is heading for positive values,
$F^2(2)$ and $F^2(3)$ split swiftly. For $r=25$, the saturation of both 
paramagnetic inequalities [for $F^2$ and $X$] yields $X(2)=X(3)=0.9$ and
 $F(2)=F(3)=87$ MeV. This ``ultra-Standard'' scenario
corresponds to the minimal values of $L_4$ and $L_6$ (no ZR violation).
A slight drift towards positives values could lead to very different chiral 
structures of the vacuum for $N_f=2$ and $N_f=3$, corresponding to a significant role of
quantum fluctuations in SB$\chi$S.

The pseudoscalar spectrum (masses and decay constants) by itself does not
contain enough information to pin down the size of these fluctuations. This effect can 
however be estimated from experimental data in the scalar channel, through a sum rule.
The difference between $X(2)$ and
$X(3)$ is related to the correlator $\Pi$ of two scalar densities
$\bar{u}u$ and $\bar{s}s$ at vanishing momentum. $\Pi(0)$ can be expressed 
in terms of a sum rule constituted of three distinct integrals.
{\bf i)} We compute the first one, involving the spectral function
$\mathrm{Im}\ \Pi$ up to energies around 1.2 GeV, by solving
coupled Omn\`es-Muskhelishvili equations for the scalar form factors
of the pion and the kaon. The solutions depend on the $T$-matrix model
used to describe the interactions between $\pi\pi$-
and $\bar{K}K$-channels, and on a normalization of the form factors related to the
derivatives of $M_\pi$ and $M_K$ with respect to
$m$ and $m_s$. {\bf ii)} The second integral corresponds to the contribution
of the spectral function $\mathrm{Im}\ \Pi$ between 1.2 and 1.6
GeV, where we cannot trust the two-channel approximation anymore.
A second sum rule is used to estimate roughly this integral.
{\bf iii)} The third integral is performed on a large complex circle, with
a large enough radius to rely on the Operator Product Expansion (OPE) of $\Pi$.

The most significant contribution stems from the first integral: the $f_0(980)$-peak
leads to a large value for $\Pi(0)$, and therefore to an important splitting 
between $X(2)$ and $X(3)$. If we fix $X(3)$, $r$ and $F(3)$, we know 
$X(2)$ and the LEC's $L_{i=4\ldots 8}$, using
our previous analysis of the pseudoscalar spectrum.
The derivatives of $M_\pi$ and $M_K$ with respect to $m$ and $m_s$
can then be directly computed, since they involve $X(3)$, $r$, $F(3)$ and LEC's.
The sum rule Eq.~(\ref{regsom}) can therefore be seen as a constraint, giving
$X(3)$ as a function of $r$ and $F(3)$. Several sources of errors could
affect this sum rule : the higher-order remainders in the expansions of
$F_P^2M_P^2$ and $F_P^2$, the rough estimate of the integral in the intermediate
energy range, the $T$-matrix model. The three models considered here
support nevertheless a large decrease of $X(3)$ with respect to $X(2)$, corresponding
to positive values of $L_6(M_\rho)$. The size of the splitting between the quark condensates
depends on the height of  the $f_0(980)$ peak in the spectral function.
In the particular case of "Standard" inputs $r\sim 25$, $F_0=85$ MeV, 
the results of Ref.~\cite{Bachir1} are
confirmed: $X(3)$ can hardly reach more than one half of $X(2)$ for the three considered
models\footnote{We remind however that this result is barely consistent with 
the Standard hypothesis of a three-flavor condensate dominating the description of SB$\chi$S.}.

The scalar form factors of the pion and the kaon can be exploited in several different ways.
For instance, $L_4$ [i.e. $F(3)$] is related to the slope of the scalar form factor
of the pion at zero. This second constraint may be used to fix
$X(3)$ and $F_0$ as functions of $r$. If the conclusions for $X(3)$ remain unchanged,
positive values of $L_4(M_\rho)$ are obtained, leading to a significant decrease
from $F(2)$ to $F(3)$ (20 to 30\%). The Zweig rule would be violated strongly
for $L_4$ and $L_6$. However, this second constraint is sensitive to fine details
of a form factor (slope at zero), whereas the sum rule depends on the general shape of
the spectral function $\mathrm{Im}\ \Pi$ [and especially on 
the presence of a high peak corresponding to the $f_0(980)$ resonance]. The
scalar radius of the pion has also been computed, in agreement with former estimates.

A large decrease of the quark condensate from 2 to 3 flavors could be understood in terms
of chiral phase transitions \cite{paramag}.
One of these transitions could be triggered by a vanishing quark
condensate. If the corresponding critical value $n_\mathrm{crit}(N_c)$ turned out to be close
to 2-3, we should expect significant variations of the quark condensate with $N_f$ in the vicinity
of the critical point. Moreover, in terms of eigenmodes of the Dirac operator, the quark
condensate can be interpreted as a density of eigenvalues, whereas $L_6$ corresponds to 
fluctuations of this density. Near the critical point where the first vanishes,
the latter is expected to increase significantly. Let us remind that this scenario is only 
a possible explanation for a large difference between $X(2)$ and $X(3)$. The large value of
ZR violating LEC's might be caused by another (and unrelated) mechanism.

Forthcoming experiments \cite{exper} on $\pi\pi$ scattering should
pin down the value of $X(2)$, which is strongly correlated to $r$.
If $X(2)$ turned out to be close to 1, they could also measure
low-energy constants of the SU(2)$\times$SU(2) Lagrangian, $l_3$ and $l_4$ \cite{g-l,pipi}.
However, these experimental values could not be used to fix SU(3)$\times$SU(3) LEC's without
assumptions on the size of the ZR violating LEC's $L_4$ and $L_6$ \cite{paramag}. 
It would be possible to constrain more tightly $L_6$ through a more sophisticated analysis of
the sum rule including bounds on $X(2)$ (or equivalently $r$). However, this remains a very 
indirect determination of the three-flavor condensate. Direct
experimental tests are necessary to investigate closely the chiral structure of 
QCD vacuum for three massless quarks, and to understand the role of quantum fluctuations
in the pattern of SB$\chi$S.

\ack

I thank J.~Stern for suggesting this problem and for constant help and support,
B.~Moussallam for many useful explanations and for providing his program solving numerically
Omn\`es-Muskhelishvili equations, P.~Talavera for a careful reading of the manuscript,
and M.~Knecht, U.-G. Mei{\ss}ner,
E.~Oset and  H.~Sazdjian for various discussions and comments. 
Work partly supported by the EU, TMR-CT98-0169, EURODA$\Phi$NE 
network.

\appendix

\section{Spectrum of pseudoscalar mesons}\label{appcomp23}

\subsection{Decay constants}

The decay constants are \cite{g-l,vus} :
\begin{eqnarray}
F_\pi^2&=&F_0^2+2m\xi +2(2m+m_s)\tilde\xi \\
&&\quad +\frac{1}{16\pi^2}\frac{F_\pi^2M_\pi^2}{F_0^2}X(3)
    \left[2\log\frac{M_K^2}{M_\pi^2}
    +\log\frac{M_\eta^2}{M_K^2}\right]   
     +\varepsilon_\pi,\nonumber\\
F_K^2&=&F_0^2+(m+m_s)\xi +2(2m+m_s)\tilde\xi \\
&&\quad +\frac{1}{2}\frac{F_\pi^2M_\pi^2}{F_0^2}X(3)L
     +\varepsilon_K,\nonumber\\
F_\eta^2&=&F_0^2+\frac{2}{3}(m+2m_s)\xi +2(2m+m_s)
     \tilde\xi \\
&&\quad  +\frac{1}{48\pi^2}\frac{F_\pi^2M_\pi^2}{F_0^2}(2r+1)X(3)
       \log\frac{M_\eta^2}{M_K^2}
  +\varepsilon_\eta,\nonumber
\end{eqnarray}
with the scale-independent low-energy constants:
\begin{eqnarray}
\xi &=&F_0^2\xi(\mu)-\frac{B_0}{32\pi^2}
   \left(\log\frac{M_K^2}{\mu^2}+2\log\frac{M_\eta^2}{\mu^2} \right)\\
   &=&8B_0\left[
   L_5(\mu)-\frac{1}{256\pi^2}
   \left(\log\frac{M_K^2}{\mu^2}+2\log\frac{M_\eta^2}{\mu^2} \right)
   \right],\\
\tilde\xi &=&F_0^2\tilde\xi(\mu)
   -\frac{B_0}{32\pi^2}\log\frac{M_K^2}{\mu^2}\\
  &=&8B_0\left[
   L_4(\mu)-\frac{1}{256\pi^2}\log\frac{M_K^2}{\mu^2}\right],
\end{eqnarray}
and:
\begin{equation}
L=\frac{1}{32\pi^2}
 \left[3\log\frac{M_K^2}{M_\pi^2}+\log\frac{M_\eta^2}{M_K^2}\right].
\end{equation}
The higher-order contributions are denoted by
$\delta_2 F_P^2$.
The effective constants are related to $F_0$ through the relations:
\begin{eqnarray}
m_s\xi &=& \frac{r}{r-1}\left\{F_K^2-F_\pi^2\right.\\
&&\qquad\left.+
  \frac{1}{64\pi^2}\frac{F_\pi^2M_\pi^2}{F_0^2}X(3)
\left[5\log\frac{M_K^2}{M_\pi^2}+3\log\frac{M_\eta^2}{M_K^2}\right]\right\}\nonumber\\
&&+\frac{r}{r-1}\left[\varepsilon_\pi-\varepsilon_K\right],\nonumber\\
m_s\tilde\xi &=&\frac{r}{2(r+2)}
\left\{\frac{r+1}{r-1}F_\pi^2-\frac{2}{r-1}F_K^2-F_0^2\right.\\
&&\qquad\left.-\frac{1}{32\pi^2}\frac{F_\pi^2M_\pi^2}{F_0^2}X(3)
  \left[\frac{4r+1}{r-1}\log\frac{M_K^2}{M_\pi^2}+\frac{2r+1}{r-1}
     \log\frac{M_\eta^2}{M_K^2}\right]\right\}\nonumber\\
&&+\frac{r}{2(r+2)}\left[\frac{2}{r-1}\varepsilon_K
   -\frac{r+1}{r-1}\varepsilon_\pi\right].\nonumber
\end{eqnarray}

The decay constants fulfill the relation:
\begin{eqnarray}
F_\eta^2&=&\frac{4}{3}F_K^2-\frac{1}{3}F_\pi^2
      +\frac{1}{24\pi^2}\frac{M_\pi^2
      F_\pi^2}{F_0^2}rX(3)\log\frac{M_\eta^2}{M_K^2}\\
&&\qquad +\frac{1}{48\pi^2}\frac{M_\pi^2
      F_\pi^2}{F_0^2}X(3)
 \left(\log\frac{M_\eta^2}{M_K^2}-\log\frac{M_K^2}{M_\pi^2}\right)
  \nonumber\\
&&\qquad 
  +\varepsilon_\eta-\frac{4}{3}\varepsilon_K+\frac{1}{3}\varepsilon_\pi.
  \nonumber
\end{eqnarray}

\subsection{Masses}

The pseudoscalar masses are \cite{g-l,vus}:
\begin{eqnarray}
F_\pi^2 M_\pi^2&=&2m\Sigma+(2mm_s+4m^2)Z^S +4m^2 A 
  \label{masspi}\\
&&\quad  +\frac{F_\pi^4M_\pi^4}{F_0^4}[X(3)]^2 L
  +F_\pi^2 \delta_\pi,\nonumber\\
F_K^2 M_K^2&=&(m_s+m)\Sigma+(m_s+m)(m_s+2m)Z^S 
  +(m_s+m)^2 A \\
&&\quad +\frac{1}{4}\frac{F_\pi^4M_\pi^4}{F_0^4}(r+1)[X(3)]^2 L
 +F_K^2 \delta_K,\nonumber\\
F_\eta^2 M_\eta^2&=&\frac{2}{3}(2m_s+m)\Sigma+\frac{2}{3}(2m_s+m)(m_s+2m)Z^S 
   \label{masseta}\\
 &&\quad  +\frac{4}{3}(2m^2_s+m^2)A 
     +\frac{8}{3}(m_s-m)^2 Z ^P\nonumber\\
 &&\quad  +\frac{1}{3}\frac{F_\pi^4M_\pi^4}{F_0^4}[X(3)]^2 L+F_\eta^2\delta_\eta.\nonumber
\end{eqnarray}
with the scale-independent low-energy constants:
\begin{eqnarray}
Z^S &=& 2F^2_0Z_0^S(\mu)-
   \frac{B_0^2}{32\pi^2}
     \left\{2\log\frac{M_K^2}{\mu^2}
        +\frac{4}{9}\log\frac{M_\eta^2}{\mu^2}\right\}\\
 &=&32 B_0^2\left[L_6(\mu)-\frac{1}{512\pi^2}
     \left\{\log\frac{M_K^2}{\mu^2}
        +\frac{2}{9}\log\frac{M_\eta^2}{\mu^2}\right\}\right],\\
A &=& F^2_0 A_0(\mu)-\frac{B_0^2}{32\pi^2}
        \left\{\log\frac{M_K^2}{\mu^2}
        +\frac{2}{3}\log\frac{M_\eta^2}{\mu^2}
	\right\}\\
  &=&16B_0^2\left[L_8(\mu)-\frac{1}{512\pi^2}
        \left\{\log\frac{M_K^2}{\mu^2}
        +\frac{2}{3}\log\frac{M_\eta^2}{\mu^2}
	\right\}
    \right],\\
Z^P &=& F^2_0Z_0^P=16B_0^2L_7.
\end{eqnarray}
The higher-order remainders are denoted by $\delta M_P^2$. The low-energy constants can
be estimated using the relations:
\begin{eqnarray}
m_s^2 Z^S &=& 
  F_\pi^2M_\pi^2\frac{r^2}{2(r+2)}\\
&&\quad \times \left\{
       1-\tilde\epsilon(r)-X(3)
       -\frac{1}{32\pi^2}\frac{F_\pi^2M_\pi^2}{F_0^4}
       \frac{r[X(3)]^2}{r-1}L
       \right\}\nonumber\\
&&     +\frac{r^2}{2(r+2)}\left[\frac{4}{r^2-1} F_K^2 \delta_K
       -\frac{r+1}{r-1} F_\pi^2 \delta_\pi\right],
       \nonumber\\
m_s^2 A &=&F_\pi^2 M_\pi^2
     \frac{r^2}{4}\left\{\tilde\epsilon(r)
+\frac{1}{32\pi^2}\frac{F_\pi^2M_\pi^2}{F_0^4}\frac{[X(3)]^2}{r-1}L\right\}\\
&& +\frac{r^2}{2(r-1)}F_\pi^2 \delta_\pi
     -\frac{r^2}{r^2-1}F_K^2 \delta_K\nonumber,\\
m_s^2 Z^P &=&
   \frac{r^2}{8}\left\{\frac{1}{(r-1)^2}
\left[3F_\eta^2M_\eta^2+F_\pi^2M_\pi^2-4F_K^2M_K^2\right]
    -F_\pi^2 M_\pi^2\tilde\epsilon(r)\right\}\nonumber\\
&&-\frac{r^2}{8(r-1)^2}
  \left[3F_\eta^2 \delta_\eta
    +\frac{8r}{r+1}F_K^2\delta_K
    +(2r-1)F_\pi^2 \delta_\pi\right],
\end{eqnarray}
with:
\begin{equation}
\tilde\epsilon(r)=2\frac{\tilde{r}_2-r}{r^2-1}
  \qquad \tilde{r}_2=2\frac{F_K^2M_K^2}{F_\pi^2M_\pi^2}-1.
\end{equation}

\subsection{Pseudoscalar masses for $m\to 0$}\label{secexpresult}

From the previous relations, one can derive low-energy constants from experimental
data (pseudoscalar masses, $F_\pi$ and $F_K$) and 3 parameters: $r$, $X(3)$ and $F_0$.
\begin{equation}
\begin{array}{rcccl}
F_\pi, F_K &\to& F_\eta, m_s\xi , m_s\tilde\xi 
  &\to& \bar{F}_\pi, \bar{F}_K, \bar{F}_\eta\\
M_\pi, M_K, M_\eta &\to& m_s^2Z^S ,
m_s^2A , m_s^2Z^P 
&\to&\bar{M}_K, \bar{M}_\eta
\end{array}
\end{equation}

In the chiral limit $m\to 0$, we will have to know the effective constants:
\begin{equation}
\lim_{m\to 0} X_i =X_i  + \sum_P C_P\cdot \log\frac{\bar{M}_P^2}{M_P^2}. \label{ceffatm0}
\end{equation}
To compute $\bar{M}_P$ in this expression, we take the chiral limit of the mass expansions
Eqs.~(\ref{masspi})-(\ref{masseta}). But these expansions involve the effective constants at
$m=0$, which leads to corrections containing logarithms of $\bar{M}_Q/M_Q$:
\begin{equation}\label{mesonatm0}
\bar{M}_P=\sum_i a_i X_i + \sum_Q D_Q\cdot \log\frac{\bar{M}_Q^2}{M_Q^2}.
\end{equation}

The equations Eq.~(\ref{mesonatm0}) could be solved iteratively.
Actually, $\bar{M}_Q/M_Q$ remains very close to 1. The calculation is simplified (and
still accurate) if we compute in a slightly different way $\bar{M}_Q$ in the logarithmic
piece of Eq.~(\ref{mesonatm0}). We start from Eq.~(\ref{mesonatm0}), and we neglect the second
(logarithmic) term:
\begin{equation}
\bar{M}_Q=\sum_i a_i X_i.
\end{equation}
$\bar{M}_Q$ is then directly computed from observables and 
$F_0,r,X(3)$. We put then these values of $\bar{M}_Q$ in the logarithmic term of
Eq.~(\ref{mesonatm0}). We end up with values of $\bar{M}_P$ very close to the ones
computed iteratively. These values will be used to compute the low-energy constants
in the chiral limit $X_i|_{m=0}$ using Eq.~(\ref{ceffatm0}).

\section{Operator Product Expansion for $\Pi$} \label{appsecopepi}

Six integrals contribute to the Wilson coefficient of
$m_s\langle\bar{u} u\rangle$ at the leading order in the strong coupling constant.
The corresponding Feynman diagrams are drawn on
Fig.~\ref{opefeynman}. On each line, the left and right diagrams correspond to each other
by crossing the gluonic lines. A simple change of variables in the integrals shows
that the diagrams on the same line contribute identically to the Wilson coefficient.

\begin{figure}
\begin{center}
\includegraphics[height=3cm]{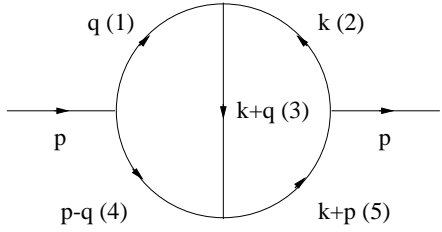}
\caption{Self-energy diagram, leading to the same kind of integrals as
in the OPE of $\Pi$ at the lowest order.}
\label{selfen}
\end{center}
\end{figure}

We want to consider the large-$p^2$ behavior of integrals like:
\begin{eqnarray}
&&J(\{\nu_i\},\{m_i\},p)=
  \int\frac{d^4q\ d^4k}{[q^2-m_1^2]^{\nu_1} [k^2-m_2^2]^{\nu_2}}\\ 
&&\qquad \times
  \frac{1}{[(k+q)^2-m_3^2]^{\nu_3} [(p-q)^2-m_4^2]^{\nu_4}
  [(k+p)^2-m_5^2]^{\nu_5}}.
  \nonumber
\end{eqnarray}
These integrals are formally identical to the integrals arising in two-loop computations
of self-energies, see Fig.~\ref{selfen}.

\begin{figure}[t]
\begin{center}
\includegraphics[width=13cm]{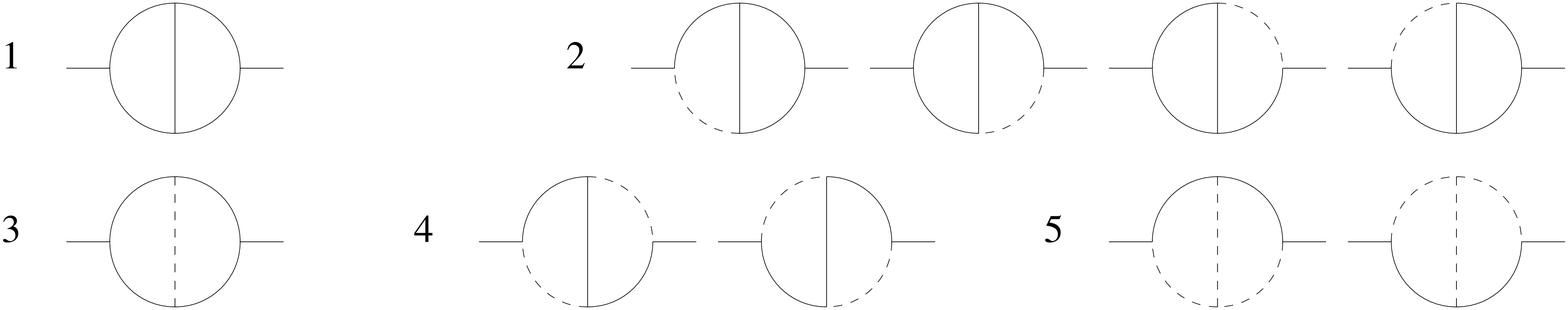}
\caption{Subgraphs involved in the asymptotic expansion of the two-loop Feynman integrals. 
The solid lines
constitute the subgraphs, the dashed lines correspond to the excluded propagators 
\cite{largemoment}.}
\label{taylorexp}
\end{center}
\end{figure}

The behavior of such integrals at large external momentum is known. The basic idea
is to follow the flow of this large external momentum through the Feynman diagram,
in order to Taylor expand correctly the propagators \cite{largemoment}.
This procedure relies on the asymptotic expansion theorem
\cite{asympexp} and can be formally expressed as:
\begin{equation}
J_\Gamma {\displaystyle \mathop\sim_{p^2\to\infty}}
    \sum_\gamma J_{\Gamma/\gamma} \circ \mathcal{T}_{\{m_i\},\{q_i\}} J_\gamma.
\end{equation}
$\Gamma$ denotes the whole graph, $\gamma$ are subgraphs into which the large external 
momentum may flow and $\Gamma/\gamma$ is the complementary 
graph of $\gamma$. For each subgraph $\gamma$
(see Fig.~\ref{taylorexp}), we write the corresponding Feynman integral $J_\gamma$. 
We perform then a Taylor expansion $\mathcal{T}_{\{m_i\},\{q_i\}}$ with respect to 
the masses and the small momenta (external to $\gamma$ and not containing $p$).
We combine the resulting ``expanded'' integral with the remaining graph $\Gamma/\gamma$ and
integrate over internal momenta. The asymptotic behavior of the whole integral $J_\Gamma$ is
obtained by considering all the possible flows $\gamma$ for the large external momentum.

We look for the leading order in $1/p^2$ of the Wilson coefficient.
All the subraphs do not contribute with the same power of $1/p^2$. In particular,
the diagrams of type 5 do not appear in the Wilson coefficient of 
$m_s\langle\bar{u}u\rangle$ at the leading order in $1/p^2$.
Gathering all the contributions, we obtain:
\begin{equation}
\Pi(p^2) {\displaystyle \mathop\sim_{P^2\to\infty}}
  \frac{\alpha_s^2 m_s^2}{M_\pi^2 M_K^2}
  \frac{2}{\pi^2 p^2} \{[5-6\zeta(3)]+[5 -6\zeta(3)]-[1+6\zeta(3)]\},
\end{equation}
where the bracketed terms correspond respectively to the contributions
of the first, second and third lines in Fig.~\ref{opefeynman}. The $\zeta(3)$ terms
are related to subdiagrams of type 1 (see Fig.~\ref{taylorexp}), corresponding to two-loop
massless integrals $J(\{\nu_i\},\{0\},p)$.

\section{Logarithmic derivatives} 

\subsection{Logarithmic derivatives for $m\to 0$} \label{secderlog}

To compute the logarithmic derivatives $\gamma_P$ and
$\lambda_P$:
\begin{equation}
\gamma_P=\frac{m}{M_P^2}
  \left(\frac{\partial M_P^2}{\partial m}\right)_{m=0},
\qquad
\lambda_P=\frac{m_s}{M_P^2}
  \left(\frac{\partial M_P^2}{\partial m_s}\right)_{m=0},
\end{equation}
we use the relations:
\begin{eqnarray}
\gamma_P &=&
   \frac{m}{\bar{F}_P^2 M_P^2} 
      \left(\frac{\partial[F_P^2 M_P^2]}{\partial m}\right)_{m=0}
 \!\!\!
  -\frac{\bar{M}_P^2}{M_P^2}
       \cdot  \frac{m}{\bar{F}_P^2}
        \left(\frac{\partial F_P^2}{\partial m}\right)_{m=0},\\
\lambda_P &=&
   \frac{m_s}{\bar{F}_P^2 M_P^2} 
      \frac{\partial[\bar{F}_P^2 \bar{M}_P^2]}{\partial m_s}
  -\frac{\bar{M}_P^2}{M_P^2}
       \cdot  \frac{m_s}{\bar{F}_P^2}
       \frac{\partial \bar{F}_P^2}{\partial m_s},
\end{eqnarray}
where $\bar{X}=\lim_{m\to 0} X$.

The logarithmic derivatives with respect to $m$ are:
\begin{eqnarray}
\gamma_\pi&=&\frac{1}{\bar{F}_\pi^2M_\pi^2}\left\{
   F_\pi^2M_\pi^2 X(3) +\frac{2}{r}m_s^2 Z^S \right.\\
&&\quad \left.-\frac{1}{32\pi^2}\frac{F_\pi^4 M_\pi^4}{F_0^4}r[X(3)]^2 
     \left(\log\frac{\bar{M}_K^2}{M_K^2}+
\frac{2}{9}\log\frac{\bar{M}_\eta^2}{M_\eta^2}\right)\right\}\nonumber\\
&&   +\frac{m}{\bar{F}_\pi^2 M_\pi^2} 
      \left(\frac{\partial[F_\pi^2 \delta_\pi]}{\partial m}\right)_{m=0}
       ,\nonumber
\end{eqnarray}

\begin{eqnarray}
&&\gamma_K
   \left[1
   -\frac{3}{64\pi^2}\frac{F_\pi^2M_\pi^2}{F_0^2\bar{F}_K^2} rX(3)
   +\frac{3}{128\pi^2}\frac{M_\pi^4F_\pi^4}{F_0^4}\frac{1}{\bar{F}_K^2\bar{M}_K^2} [rX(3)]^2  
   \right]\nonumber\\
&& \quad +\gamma_\eta
   \left[
   -\frac{1}{32\pi^2} \frac{F_\pi^2M_\pi^2}{F_0^2\bar{F}_K^2}
   \frac{M_\eta^2\bar{M}_K^2}{\bar{M}_\eta^2M_K^2} rX(3)    
   +\frac{5}{576\pi^2} \frac{M_\pi^4F_\pi^4}{F_0^4}
    \frac{1}{\bar{F}_K^2 M_K^2} \frac{M_\eta^2}{\bar{M}_\eta^2} [rX(3)]^2     
    \right]\nonumber\\
&& =\frac{1}{\bar{F}_K^2M_K^2}
 \left\{\frac{F_\pi^2M_\pi^2}{2}X(3)+
   \frac{1}{r}\left[3m_s^2Z^S +2m_s^2A 
-\bar{M}_K^2\left(m_s\xi +4m_s\tilde\xi \right)\right]\right.\nonumber\\
&&\qquad \qquad -\frac{1}{16\pi^2}\frac{F_\pi^4 M_\pi^4}{F_0^4} r[X(3)]^2
         \left[\log\frac{\bar{M}_K^2}{M_K^2}+
	    \frac{1}{3}\log\frac{\bar{M}_\eta^2}{M_\eta^2}\right]\nonumber\\
&&\qquad \qquad\left. +\frac{1}{64\pi^2}\frac{F_\pi^2 M_\pi^2}{F_0^2}\bar{M}_K^2  X(3)
         \left[5\log\frac{\bar{M}_K^2}{M_K^2}+
	       2\log\frac{\bar{M}_\eta^2}{M_\eta^2}\right]\right\}\nonumber\\
&&\quad+   \frac{m}{\bar{F}_K^2 M_K^2} 
      \left(\frac{\partial[F_K^2 \delta_K]}{\partial m}\right)_{m=0}
 \!\!\!
  -\frac{\bar{M}_K^2}{M_K^2}
       \cdot  \frac{m}{\bar{F}_K^2}
        \left(\frac{\partial \varepsilon_K}{\partial m}\right)_{m=0}
,
\end{eqnarray}  

\begin{eqnarray}
&&\gamma_K \left[ 
  -\frac{3}{32\pi^2}\frac{F_\pi^2M_\pi^2}{F_0^2\bar{F}_\eta^2}
     \frac{M_K^2\bar{M}_\eta^2}{\bar{M}_K^2M_\eta^2}rX(3) 
  +\frac{1}{24\pi^2} \frac{F_\pi^4M_\pi^4}{F_0^4}
   \frac{1}{\bar{F}_\eta^2 M_\eta^2} \frac{M_K^2}{\bar{M}_K^2}[rX(3)]^2\right]
    \nonumber\\
&&\quad +\gamma_\eta
 \left[1 +\frac{1}{54\pi^2} \frac{F_\pi^4M_\pi^4}{F_0^4}\frac{1}{\bar{F}_\eta^2\bar{M}_\eta^2} 
     [rX(3)]^2 \right]\nonumber\\
&&= \frac{1}{\bar{F}_\eta^2M_\eta^2}\left\{
     \frac{F_\pi^2M_\pi^2}{3}X(3)\right.\nonumber\\
&&\qquad \qquad +\frac{1}{r}\left[\frac{10}{3}m_s^2Z^S +
        \frac{16}{3}m_s^2Z^P 
-\bar{M}_\eta^2\left(\frac{2}{3}m_s\xi +4m_s\tilde\xi \right)
   \right] \nonumber\\
&&\qquad\qquad
  -\frac{1}{32\pi^2}\frac{F_\pi^4M_\pi^4}{F_0^4} 
      r[X(3)]^2\left[\frac{5}{3}\log\frac{M_K^2}{\bar{M}_K^2}+
	    \frac{10}{27}\log\frac{M_\eta^2}{\bar{M}_\eta^2}\right]\nonumber\\
&&\left.\qquad\qquad
  -\frac{1}{32\pi^2}\frac{F_\pi^2M_\pi^2}{F_0^2}\bar{M}_\eta^2 X(3)
         \left[7\log\frac{\bar{M}_K^2}{M_K^2}+
	       2\log\frac{\bar{M}_\eta^2}{M_\eta^2}
	      +2\log\frac{\bar{M}_K^2}{\bar{M}_\eta^2}\right]\right\}\nonumber\\
&&\quad+\frac{m}{\bar{F}_\eta^2 M_\eta^2} 
      \left(\frac{\partial[F_\eta^2 \delta_\eta]}{\partial m}\right)_{m=0}
 \!\!\!
  -\frac{\bar{M}_\eta^2}{M_\eta^2}
       \cdot  \frac{m}{\bar{F}_\eta^2}
        \left(\frac{\partial \varepsilon_\eta}{\partial m}\right)_{m=0}.
\end{eqnarray}

The logarithmic derivatives with respect to $m_s$ are:
\begin{equation}
\lambda_\pi=0,
\end{equation}

\begin{eqnarray}
&&\lambda_K\left[1
  -\frac{3}{64\pi^2} \frac{F_\pi^2M_\pi^2}{F_0^2\bar{F}_K^2} rX(3) 
  +\frac{3}{128\pi^2} \frac{F_\pi^4M_\pi^4}{F_0^4}\frac{1}{\bar{F}_K^2\bar{M}_K^2}[rX(3)]^2
  \right]\\
&&\quad +\lambda_\eta\left[
  -\frac{1}{32\pi^2} 
    \frac{F_\pi^2 M_\pi^2}{F_0^2 \bar{F}_K^2}\frac{\bar{M}_K^2M_\eta^2}{M_K^2\bar{M}_\eta^2}rX(3) 
  +\frac{5}{576\pi^2} \frac{F_\pi^4 M_\pi^4}{F_0^4}\frac{1}{\bar{F}_K^2 M_K^2}
    \frac{M_\eta^2}{\bar{M}_\eta^2} [rX(3)]^2
       \right]\nonumber\\
&&= \frac{1}{\bar{F}_K^2M_K^2}\left\{
    \frac{F_\pi^2M_\pi^2}{2} rX(3)+
       2m_s^2Z^S +2m_s^2A 
       -\bar{M}_K^2(m_s\xi +2m_s\tilde\xi )\right.
  \nonumber\\
&&\qquad\qquad -\frac{1}{128\pi^2}\frac{F_\pi^4M_\pi^4}{F_0^4} [rX(3)]^2
         \left[5\log\frac{\bar{M}_K^2}{M_K^2}+
	       \frac{20}{9}\log\frac{\bar{M}_\eta^2}{M_\eta^2}\right]\nonumber\\
&&\left.\qquad \qquad +\frac{1}{64\pi^2}\frac{F_\pi^2M_\pi^2}{F_0^2}\bar{M}_K^2rX(3)
         \left[3\log\frac{\bar{M}_K^2}{M_K^2}+
	       2\log\frac{\bar{M}_\eta^2}{M_\eta^2}\right]\right\}\nonumber\\
&&\quad
+   \frac{m_s}{\bar{F}_K^2 M_K^2} 
      \left(\frac{\partial[F_K^2 \delta_K]}{\partial m_s}\right)_{m=0}
 \!\!\!
  -\frac{\bar{M}_K^2}{M_K^2}
       \cdot  \frac{m_s}{\bar{F}_K^2}
        \left(\frac{\partial \varepsilon_K}{\partial m_s}\right)_{m=0}
,\nonumber
\end{eqnarray}

\begin{eqnarray}
&&\lambda_K\left[
-\frac{3}{32\pi^2} \frac{F_\pi^2M_\pi^2}{F_0^2\bar{F}_\eta^2}
     \frac{M_K^2}{\bar{M}_K^2}\frac{\bar{M}_\eta^2}{M_\eta^2} rX(3) 
+ \frac{1}{24\pi^2} \frac{F_\pi^4 M_\pi^4}{F_0^4}
    \frac{1}{\bar{F}_\eta^2 M_\eta^2} \frac{M_K^2}{\bar{M}_K^2} [rX(3)]^2
    \right]\nonumber\\
&&\quad +\lambda_\eta\left[1
+\frac{1}{54\pi^2} \frac{F_\pi^4M_\pi^4}{F_0^4}
    \frac{1}{\bar{F}_\eta^2\bar{M}_\eta^2}[rX(3)]^2\right]\nonumber\\
&& =\frac{1}{\bar{F}_\eta^2M_\eta^2}\left\{
     \frac{4F_\pi^2M_\pi^2}{3}rX(3)\right.\nonumber\\
&&\qquad \qquad +\frac{8}{3}\left[m_s^2 Z^S +2m_s^2A 
        +2m_s^2 Z^P \right]
  -\bar{M}_\eta^2\left(\frac{4}{3}m_s\xi +2m_s\tilde\xi \right)
      \nonumber\\
&&\qquad \qquad  -\frac{1}{12\pi^2}\frac{F_\pi^4 M_\pi^4}{F_0^4} 
   [rX(3)]^2\left[\log\frac{\bar{M}_K^2}{M_K^2}+
	    \frac{4}{9}\log\frac{\bar{M}_\eta^2}{M_\eta^2}\right]\nonumber\\
&&\left.\qquad \qquad +\frac{1}{96\pi^2}\frac{F_\pi^2M_\pi^2}{F_0^2}
   \bar{M}_\eta^2rX(3)
         \left[5\log\frac{\bar{M}_K^2}{M_K^2}+
	  4\log\frac{\bar{M}_\eta^2}{M_\eta^2}+
	  4\log\frac{\bar{M}_K^2}{\bar{M}_\eta^2}\right]\right\}\nonumber\\
&&\quad
+   \frac{m_s}{\bar{F}_\eta^2 M_\eta^2} 
      \left(\frac{\partial[F_\eta^2 \delta_\eta]}{\partial m_s}\right)_{m=0}
 \!\!\!
  -\frac{\bar{M}_\eta^2}{M_\eta^2}
       \cdot  \frac{m_s}{\bar{F}_\eta^2}
        \left(\frac{\partial \varepsilon_\eta}{\partial m_s}\right)_{m=0}
.
\end{eqnarray}

\subsection{Logarithmic derivatives for $m\neq 0$} \label{tildegamma}

The same method can be used to compute the logarithmic derivatives involved
in the scalar radius of the pion:
\begin{equation}
\tilde\gamma_P=\frac{m}{M_P^2}
  \frac{\partial M_P^2}{\partial m}.
\end{equation}

We obtain:
\begin{eqnarray}
\tilde\gamma_\pi&=&
  \frac{1}{F_\pi^2M_\pi^2}
  \left\{F_\pi^2M_\pi^2 X(3)
     +\frac{2}{r^2}\left[(r+4)m_s^2Z^S+4m_s^2A\right]\right.\\
&&\qquad\qquad \left.-\frac{2M_\pi^2}{r}[m_s\xi+2m_s\tilde\xi]
    \right\}\nonumber\\
&&\quad +\frac{F_\pi^2M_\pi^2}{F_0^4}\frac{[X(3)]^2}{32\pi^2}
  \left[6\log\frac{M_K^2}{M_\pi^2}+2\log\frac{M_\eta^2}{M_K^2}\right.\nonumber\\
&&\qquad\qquad \left.-3\tilde\gamma_\pi-(r+1)\tilde\gamma_K
    -\frac{1}{9}(2r+1)\tilde\gamma_\eta\right]\nonumber\\
&&\quad -\frac{M_\pi^2}{F_0^2}\frac{X(3)}{32\pi^2}
  \left[4\log\frac{M_K^2}{M_\pi^2}+2\log\frac{M_\eta^2}{M_K^2}
    -4\tilde\gamma_\pi-(r+1)\tilde\gamma_K\right]\nonumber\\
&&\quad
  +\frac{m}{F_\pi^2M_\pi^2}\frac{\partial[F_\pi^2\delta_\pi]}{\partial m}
  -\frac{m}{F_\pi^2}\frac{\partial\varepsilon_\pi}{\partial m},
\end{eqnarray}

\begin{eqnarray}
\tilde\gamma_K&=&
  \frac{1}{F_K^2M_K^2}
  \left\{\frac{F_\pi^2M_\pi^2}{2} X(3)
     +\frac{3r+4}{r^2}m_s^2Z^S+\frac{2(r+1)}{r^2}m_s^2A\right.\nonumber\\
&&\qquad\qquad\left. -\frac{M_K^2}{r}[m_s\xi+4m_s\tilde\xi]
    \right\}\nonumber\\
&&\quad +\frac{F_\pi^4M_\pi^4}{F_0^4F_K^2M_K^2}\frac{[X(3)]^2}{128\pi^2}
  \left[3(r+2)\log\frac{M_K^2}{M_\pi^2}+(r+2)\log\frac{M_\eta^2}{M_K^2}
  \right.\nonumber\\
&&\qquad\qquad \left. -3\tilde\gamma_\pi-3(r+1)^2\tilde\gamma_K
    -\frac{5}{9}(2r+1)(r+1)\tilde\gamma_\eta\right]\nonumber\\
&&\quad -\frac{F_\pi^2M_\pi^2}{F_K^2F_0^2}\frac{X(3)}{64\pi^2}
  \left[3\log\frac{M_K^2}{M_\pi^2}+\log\frac{M_\eta^2}{M_K^2}\right.\nonumber\\
&&\qquad\qquad 
  \left.-3\tilde\gamma_\pi-3(r+1)\tilde\gamma_K-(2r+1)\tilde\gamma_\eta
      \right]\nonumber\\
&&\quad
  +\frac{m}{F_K^2M_K^2}\frac{\partial[F_K^2\delta_K]}{\partial m}
  -\frac{m}{F_K^2}\frac{\partial\varepsilon_K}{\partial m},
\end{eqnarray}

\begin{eqnarray}
\tilde\gamma_\eta&=&
  \frac{1}{F_\eta^2M_\eta^2}
  \left\{\frac{F_\pi^2M_\pi^2}{3} X(3)
     +\frac{2(4+5r)}{3r^2}m_s^2Z^S+\frac{8}{3r^2}m_s^2A
     -\frac{16(r-1)}{3r^2}m_s^2Z^P\right.\nonumber\\
&&\qquad\qquad\left. -\frac{M_\eta^2}{r}\left[\frac{2}{3}m_s\xi+4m_s\tilde\xi\right]
    \right\}\nonumber\\
&&\quad +\frac{F_\pi^4M_\pi^4}{F_0^4F_\eta^2M_\eta^2}\frac{[X(3)]^2}{128\pi^2}
  \left[8\log\frac{M_K^2}{M_\pi^2}+\frac{8}{3}\log\frac{M_\eta^2}{M_K^2}
  \right.\nonumber\\
&&\qquad\qquad \left.-4\tilde\gamma_\pi-\frac{4}{3}(4r+1)(r+1)\tilde\gamma_K
    -\frac{4}{27}(16r^2+10r+1)\tilde\gamma_\eta\right]\nonumber\\
&&\quad -\frac{F_\pi^2M_\pi^2}{F_\eta^2F_0^2}\frac{X(3)}{64\pi^2}
  \left[\frac{4}{3}\log\frac{M_\eta^2}{M_K^2}
   -6(r+1)\tilde\gamma_K\right]\nonumber\\
&&\qquad
  +\frac{m}{F_\eta^2M_\eta^2}\frac{\partial[F_\eta^2\delta_\eta]}{\partial m}
  -\frac{m}{F_\eta^2}\frac{\partial\varepsilon_\eta}{\partial m}.
\end{eqnarray}

This linear system of three equations and three variables is easily solved to
compute  $\tilde\gamma_\pi$ and
$\tilde\gamma_K$ as functions of $F_0$, $r$ and $X(3)$.

\end{document}